# The IceCube Neutrino Observatory Part IV: Searches for Dark Matter and Exotic Particles

THE ICECUBE COLLABORATION

## Contents

## IceCube Collaboration Member List

M. G. Aartsen[2], R. Abbasi[27], Y. Abdou[22], M. Ackermann[41], J. Adams[15], J. A. Aguilar[21], M. Ahlers[27], D. Altmann[9], J. Auffenberg[27], X. Bai[31,a], M. Baker[27], S. W. Barwick[23], V. Baum[28], R. Bay[7], J. J. Beatty[17,18], S. Bechet[12], J. Becker Tjus[10], K.-H. Becker[40], M. Bell[38], M. L. Benabderrahmane[41], S. BenZvi[27], P. Berghaus[41], D. Berley[16], E. Bernardini[41], A. Bernhard[30], D. Bertrand[12], D. Z. Besson[25], G. Binder[8,7], D. Bindig[40], M. Bissok[1], E. Blaufuss[16], J. Blumenthal[1], D. J. Boersma[39], S. Bohaichuk[20], C. Bohm[34], D. Bose[13], S. Böser[11], O. Botner[39], L. Brayeur[13], H.-P. Bretz[41], A. M. Brown[15], R. Bruijn[24], J. Brunner[41], M. Carson[22], J. Casey[5], M. Casier[13], D. Chirkin[27], A. Christov[21], B. Christy[16], K. Clark[38], F. Clevermann[19], S. Coenders[1], S. Cohen[24], D. F. Cowen[38,37], A. H. Cruz Silva[41], M. Danninger[34], J. Daughhetee[5], J. C. Davis[17], C. De Clercq[13], S. De Ridder[22], P. Desiati[27], K. D. de Vries[13], M. de With[9], T. DeYoung[38], J. C. Díaz-Vélez[27], M. Dunkman[38], R. Eagan[38], B. Eberhardt[28], J. Eisch[27], R. W. Ellsworth[16], S. Euler[1], P. A. Evenson[31], O. Fadiran[27], A. R. Fazely[6], A. Fedynitch[10], J. Feintzeig[27], T. Feusels[22], K. Filimonov[7], C. Finley[34], T. Fischer-Wasels[40], S. Flis[34], A. Franckowiak[11], K. Frantzen[19], T. Fuchs[19], T. K. Gaisser[31], J. Gallagher[26], L. Gerhardt[8,7], L. Gladstone[27], T. Glüsenkamp[41], A. Goldschmidt[8], G. Golup[13], J. G. Gonzalez[31], J. A. Goodman[16], D. Góra[41], D. T. Grandmont[20], D. Grant[20], A. Groß[30], C. Ha[8,7], A. Haj Ismail[22], P. Hallen[1], A. Hallgren[39], F. Halzen[27], K. Hanson[12], D. Heereman[12], D. Heinen[1], K. Helbing[40], R. Hellauer[16], S. Hickford[15], G. C. Hill[2], K. D. Hoffman[16], R. Hoffmann[40], A. Homeier[11], K. Hoshina[27], W. Huelsnitz[16,b], P. O. Hulth[34], K. Hultqvist[34], S. Hussain[31], A. Ishihara[14], E. Jacobi[41], J. Jacobsen[27], K. Jagielski[1], G. S. Japaridze[4], K. Jero[27], O. Jlelati[22], B. Kaminsky[41], A. Kappes[9], T. Karg[41], A. Karle[27], J. L. Kelley[27], J. Kiryluk[35], J. Kläs[40], S. R. Klein[8,7], J.-H. Köhne[19], G. Kohnen[29], H. Kolanoski[9], L. Köpke[28], C. Kopper[27], S. Kopper[40], D. J. Koskinen[38], M. Kowalski[11], M. Krasberg[27], K. Krings[1], G. Kroll[28], J. Kunnen[13], N. Kurahashi[27], T. Kuwabara[31], M. Labare[22], H. Landsman[27], M. J. Larson[36], M. Lesiak-Bzdak[35], M. Leuermann[1], J. Leute[30], J. Lünemann[28], J. Madsen[33], G. Maggi[13], R. Maruyama[27], K. Mase[14], H. S. Matis[8], F. McNally[27], K. Meagher[16], M. Merck[27], P. Mészáros[37,38], T. Meures[12], S. Miarecki[8,7], E. Middell[41], N. Milke[19], J. Miller[13], L. Mohrmann[41], T. Montaruli[21,c], R. Morse[27], R. Nahnhauer[41], U. Naumann[40], H. Niederhausen[35], S. C. Nowicki[20], D. R. Nygren[8], A. Obertacke[40], S. Odrowski[20], A. Olivas[16], M. Olivo[10], A. O'Murchadha[12], L. Paul[1], J. A. Pepper[36], C. Pérez de los Heros[39], C. Pfendner[17], D. Pieloth[19], E. Pinat[12], J. Posselt[40], P. B. Price[7], G. T. Przybylski[8], L. Rädel[1], M. Rameez[21], K. Rawlins[3], P. Redl[16], R. Reimann[1], E. Resconi[30], W. Rhode[19], M. Ribordy[24], M. Richman[16], B. Riedel[27], J. P. Rodrigues[27], C. Rott[17,d], T. Ruhe[19], B. Ruzybayev[31], D. Ryckbosch[22], S. M. Saba[10], T. Salameh[38], H.-G. Sander[28], M. Santander[27], S. Sarkar[32], K. Schatto[28], M. Scheel[1], F. Scheriau[19], T. Schmidt[16], M. Schmitz[19], S. Schoenen[1], S. Schöneberg[10], A. Schönwald[41], A. Schukraft[1], L. Schulte[11], O. Schulz[30], D. Seckel[31], Y. Sestayo[30], S. Seunarine[33], R. Shanidze[41], C. Sheremata[20], M. W. E. Smith[38], D. Soldin[40], G. M. Spiczak[33], C. Spiering[41], M. Stamatikos[17,e], T. Stanev[31], A. Stasik[11], T. Stezelberger[8], R. G. Stokstad[8], A. Stößl[41], E. A. Strahler[13], R. Ström[39], G. W. Sullivan[16], H. Taavola[39], I. Taboada[5], A. Tamburro[31], A. Tepe[40], S. Ter-Antonyan[6], G. Tešić[38], S. Tilav[31], P. A. Toale[36], S. Toscano[27], M. Usner[11], D. van der Drift[8,7], N. van Eijndhoven[13], A. Van Overloop[22], J. van Santen[27], M. Vehring[1], M. Voge[11], M. Vraeghe[22], C. Walck[34], T. Waldenmaier[9], M. Wallraff[1], R. Wasserman[38], Ch. Weaver[27], M. Wellons[27], C. Wendt[27], S. Westerhoff[27], N. Whitehorn[27], K. Wiebe[28], C. H. Wiebusch[1], D. R. Williams[36], H. Wissing[16], M. Wolf[34], T. R. Wood[20], K. Woschnagg[7], D. L. Xu[36], X. W. Xu[6], J. P. Yanez[41], G. Yodh[23], S. Yoshida[14], P. Zarzhitsky[36], J. Ziemann[19], S. Zierke[1], M. Zoll[34]

[1]III. Physikalisches Institut, RWTH Aachen University, D-52056 Aachen, Germany
[2]School of Chemistry & Physics, University of Adelaide, Adelaide SA, 5005 Australia
[3]Dept. of Physics and Astronomy, University of Alaska Anchorage, 3211 Providence Dr., Anchorage, AK 99508, USA
[4]CTSPS, Clark-Atlanta University, Atlanta, GA 30314, USA
[5]School of Physics and Center for Relativistic Astrophysics, Georgia Institute of Technology, Atlanta, GA 30332, USA
[6]Dept. of Physics, Southern University, Baton Rouge, LA 70813, USA
[7]Dept. of Physics, University of California, Berkeley, CA 94720, USA
[8]Lawrence Berkeley National Laboratory, Berkeley, CA 94720, USA
[9]Institut für Physik, Humboldt-Universität zu Berlin, D-12489 Berlin, Germany
[10]Fakultät für Physik & Astronomie, Ruhr-Universität Bochum, D-44780 Bochum, Germany
[11]Physikalisches Institut, Universität Bonn, Nussallee 12, D-53115 Bonn, Germany
[12]Université Libre de Bruxelles, Science Faculty CP230, B-1050 Brussels, Belgium
[13]Vrije Universiteit Brussel, Dienst ELEM, B-1050 Brussels, Belgium
[14]Dept. of Physics, Chiba University, Chiba 263-8522, Japan
[15]Dept. of Physics and Astronomy, University of Canterbury, Private Bag 4800, Christchurch, New Zealand
[16]Dept. of Physics, University of Maryland, College Park, MD 20742, USA
[17]Dept. of Physics and Center for Cosmology and Astro-Particle Physics, Ohio State University, Columbus, OH 43210, USA
[18]Dept. of Astronomy, Ohio State University, Columbus, OH 43210, USA
[19]Dept. of Physics, TU Dortmund University, D-44221 Dortmund, Germany
[20]Dept. of Physics, University of Alberta, Edmonton, Alberta, Canada T6G 2E1
[21]Département de physique nucléaire et corpusculaire, Université de Genève, CH-1211 Genève, Switzerland
[22]Dept. of Physics and Astronomy, University of Gent, B-9000 Gent, Belgium
[23]Dept. of Physics and Astronomy, University of California, Irvine, CA 92697, USA
[24]Laboratory for High Energy Physics, École Polytechnique Fédérale, CH-1015 Lausanne, Switzerland
[25]Dept. of Physics and Astronomy, University of Kansas, Lawrence, KS 66045, USA
[26]Dept. of Astronomy, University of Wisconsin, Madison, WI 53706, USA
[27]Dept. of Physics and Wisconsin IceCube Particle Astrophysics Center, University of Wisconsin, Madison, WI 53706, USA
[28]Institute of Physics, University of Mainz, Staudinger Weg 7, D-55099 Mainz, Germany
[29]Université de Mons, 7000 Mons, Belgium
[30]T.U. Munich, D-85748 Garching, Germany
[31]Bartol Research Institute and Department of Physics and Astronomy, University of Delaware, Newark, DE 19716, USA
[32]Dept. of Physics, University of Oxford, 1 Keble Road, Oxford OX1 3NP, UK
[33]Dept. of Physics, University of Wisconsin, River Falls, WI 54022, USA
[34]Oskar Klein Centre and Dept. of Physics, Stockholm University, SE-10691 Stockholm, Sweden
[35]Department of Physics and Astronomy, Stony Brook University, Stony Brook, NY 11794-3800, USA
[36]Dept. of Physics and Astronomy, University of Alabama, Tuscaloosa, AL 35487, USA
[37]Dept. of Astronomy and Astrophysics, Pennsylvania State University, University Park, PA 16802, USA
[38]Dept. of Physics, Pennsylvania State University, University Park, PA 16802, USA
[39]Dept. of Physics and Astronomy, Uppsala University, Box 516, S-75120 Uppsala, Sweden
[40]Dept. of Physics, University of Wuppertal, D-42119 Wuppertal, Germany
[41]DESY, D-15735 Zeuthen, Germany
[a]Physics Department, South Dakota School of Mines and Technology, Rapid City, SD 57701, USA
[b]Los Alamos National Laboratory, Los Alamos, NM 87545, USA
[c]also Sezione INFN, Dipartimento di Fisica, I-70126, Bari, Italy
[d]Department of Physics, Sungkyunkwan University, Suwon 440-746, Korea
[e]NASA Goddard Space Flight Center, Greenbelt, MD 20771, USA

**Acknowledgements**

We acknowledge the support from the following agencies: U.S. National Science Foundation-Office of Polar Programs, U.S. National Science Foundation-Physics Division, University of Wisconsin Alumni Research Foundation, the Grid Laboratory Of Wisconsin (GLOW) grid infrastructure at the University of Wisconsin - Madison, the Open Science Grid (OSG) grid infrastructure; U.S. Department of Energy, and National Energy Research Scientific Computing Center, the Louisiana Optical Network Initiative (LONI) grid computing resources; Natural Sciences and Engineering Research Council of Canada, WestGrid and Compute/Calcul Canada; Swedish Research Council, Swedish Polar Research Secretariat, Swedish National Infrastructure for Computing (SNIC), and Knut and Alice Wallenberg Foundation, Sweden; German Ministry for Education and Research (BMBF), Deutsche Forschungsgemeinschaft (DFG), Helmholtz Alliance for Astroparticle Physics (HAP), Research Department of Plasmas with Complex Interactions (Bochum), Germany; Fund for Scientific Research (FNRS-FWO), FWO Odysseus programme, Flanders Institute to encourage scientific and technological research in industry (IWT), Belgian Federal Science Policy Office (Belspo); University of Oxford, United Kingdom; Marsden Fund, New Zealand; Australian Research Council; Japan Society for Promotion of Science (JSPS); the Swiss National Science Foundation (SNSF), Switzerland.

# Results from Low-Energy Neutrino Searches for Dark Matter in the Galactic Center with IceCube-DeepCore

THE ICECUBE COLLABORATION[1],

[1] *See special section in these proceedings*

*martin.wolf@fysik.su.se*

**Abstract:** The cubic-kilometer sized IceCube neutrino observatory, constructed in the glacial ice at the South Pole, offers new opportunities for neutrino physics with its in-fill array "DeepCore". In particular, the use of the outer layers of the IceCube detector as a veto allows low-energy neutrino searches to be performed in the southern sky. This makes the Galactic Center, an important target in searches for self-annihilating dark matter, reachable for IceCube. In this contribution we present the results of the first Galactic Center analysis using more than 10 months of data taken with the 79-string configuration of IceCube-DeepCore, with a special focus on low WIMP masses reaching a sensitivity as low as 30 GeV. We also present the status of an analysis extending the sensitivity to WIMP masses up to the TeV scale.

**Corresponding authors:**
Martin Wolf[1], Samuel Flis[1], Matthias Danninger[1], Martin Bissok[2]

[1] *Oskar Klein Centre and Dept. of Physics, Stockholm University, SE-10691 Stockholm, Sweden*
[2] *III. Physikalisches Institut, RWTH Aachen University, D-52056 Aachen, Germany*

## 1 Introduction

Numerous observations imply the existence of non-baryonic cold dark matter through its gravitational interaction [1]. However, the exact nature of dark matter is still unknown. Many theories beyond the Standard Model predict stable or extremely long-lived particles that are well-motivated dark matter candidates. Weakly Interacting Massive Particles (WIMPs) are one of the most promising and experimentally accessible classes of dark matter. In super-symmetric extensions to the Standard Model, WIMPs may appear in the form of neutralinos [2]. Predicted WIMP masses range from a few GeV to a few tens of TeV. WIMP self-annihilation to Standard Model particles may produce a flux of final-state particles that include positrons, gammas, or neutrinos.

Regions of increased dark matter density are interesting from an observational point of view since the ensuing higher self-annihilation rate could result in a significant flux of detectable particles. Galaxies are believed to be embedded in halos of dark matter with a variety of models attempting to describe the density distribution, based on *e.g.* N-body simulations or observations of the motion of stars within galaxies or individual galaxies within clusters of galaxies. These models, usually assume a spherically symmetric halo where the density decreases with the distance, $r$, from the Galactic Center.

Observations of low surface brightness galaxies suggest a flat distribution in the central region [3], while fits to N-body simulations tend to show a divergent, or cusped behavior towards the center [4]. However, at large distances from the Galactic Center($\simeq 8.5$ kpc), the different models converge. A broad family of dark matter density profiles may be parametrized by [5]

$$\rho_{\mathrm{DM}}(r) = \frac{\rho_0}{\left(\frac{r}{r_s}\right)^{\gamma} \cdot \left(1 + \left(\frac{r}{r_s}\right)^{\alpha}\right)^{(\beta-\gamma)/\alpha}}, \quad (1)$$

where $r_s$ is the scale radius, $\alpha, \beta$, and $\gamma$ are profile parameters, and $\rho_0$ is the normalization. For the analyses presented here, the Navarro-Frenk-White (NFW) model [6] is used as a benchmark. It is obtained from equation 1 with the parameters $\alpha = 1, \beta = 3, \gamma = 1, r_s = 20$ kpc, and $\rho_0$ chosen so that the local dark matter density $\rho(R_{\mathrm{SC}}) = 0.3$ GeVcm$^{-3}$, where $R_{\mathrm{SC}} = 8.5$ kpc is the radius of the solar orbit.

Searching for dark matter self-annihilation in the Milky Way probes the thermal average of the annihilation rate, which is proportional to the product of the annihilation cross section and the relative velocity of WIMPs, $\langle\sigma_A v\rangle$. This is complementary to indirect Solar and Earth WIMP searches, and to direct searches, which probe the WIMP-nucleon cross-section.

The expected flux of the annihilation products depends on the integrated dark matter density squared along the line of sight, given by [7]

$$J_a(\Psi) = \int_0^{l_{max}} \mathrm{d}l \, \frac{\rho_{\mathrm{DM}}^2(\sqrt{R_{\mathrm{SC}}^2 - 2lR_{\mathrm{SC}}\cos\Psi + l^2})}{R_{\mathrm{SC}}\rho_{\mathrm{SC}}^2}. \quad (2)$$

Here $\Psi$ is the half-cone opening angle with respect to the Galactic Center, and $\rho_{\mathrm{SC}} = \rho(R_{\mathrm{SC}})$ and $R_{\mathrm{SC}}$ are scaling constants which make $J_a(\Psi)$ a dimensionless quantity. Figure 1 shows examples of $J_a(\Psi)$ for different halo models.

The expected neutrino flux at Earth is given by

$$\frac{\mathrm{d}\phi_\nu}{\mathrm{d}E} = \frac{\langle\sigma_A v\rangle}{2} J_a(\Psi) \frac{R_{\mathrm{SC}}\rho_{\mathrm{SC}}^2}{4\pi m_\chi^2} \frac{\mathrm{d}N_\nu}{\mathrm{d}E}, \quad (3)$$

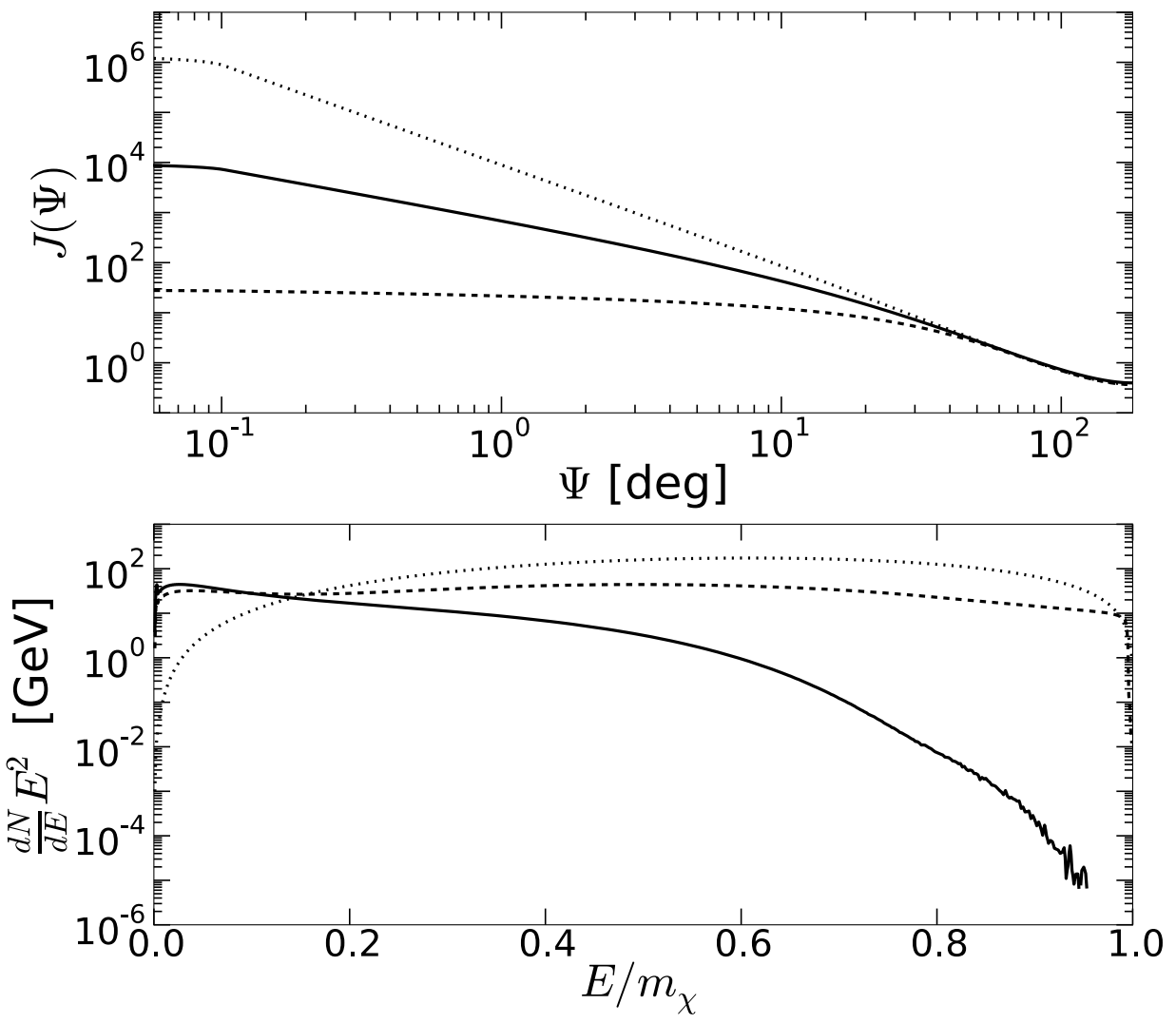

**Figure 1**: **Top:** Line-of-sight integral $J_a(\Psi)$ for three different models. The NFW [6] (solid) is the benchmark model for the analyses presented in this paper, Kravtsov [9] (dashed) and Moore [8] (dotted) represent more extremes cases of a flat-cored and a peaked profile. **Bottom:** Dark matter annihilation spectra generated with PYTHIA8 [10] for three channels, $b\bar{b}$ (solid), $W^+W^-$ (dashed) and $\mu^+\mu^-$ (dotted) and a WIMP mass of $m_\chi = 500$ GeV.

where $\mathrm{d}N_\nu/\mathrm{d}E$ is the energy dependent WIMP annihilation spectrum, and $\rho_{\mathrm{SC}}/m_\chi$, and $R_{\mathrm{SC}}$ normalize $J_a(\Psi)$ to the number density. The resulting neutrino spectra from WIMP annihilations for various signal models are simulated using PYTHIA8 [10]. Figure 1 compares three such neutrino spectra. In the case of neutrinos as final-state particles, the flux from dark matter annihilations can be probed by large neutrino telescopes such as IceCube.

IceCube is a cubic-kilometer-scale neutrino detector deployed in the ice at the geographic South Pole [11] between depths of 1450 m and 2450 m. In this work we use 320 live days of data taken from 2010 to 2011, during a period when the detector was operated in its 79-string configuration, including 6 densely instrumented strings optimized for low energies in the center of the array. Together with the 7 adjacent standard IceCube strings, these form the DeepCore subarray [12]. Neutrino detection in IceCube relies on the measurement of Cherenkov radiation produced by secondary charged leptons in neutrino interactions in the surrounding ice or the nearby bedrock.

## 2 Event Selection

Neutrinos from the direction of the Galactic Center, located in the southern hemisphere, would be down-going events within IceCube. Searches for such neutrinos must overcome a background of down-going atmospheric muons penetrating the detector. This background is reduced by selecting events with a reconstructed interaction vertex inside a fiducial volume of the detector, where the outer parts of IceCube are used as a veto.

In order to be sensitive to a wide range of possible WIMP masses, two independent analyses are performed. The DC event selection (section 2.1) is optimized to search for low WIMP mass signals (30 GeV to 500 GeV) with DeepCore. The second search uses larger parts of IceCube as a fiducial volume to improve the sensitivity to WIMP mass signals above 500 GeV, denoted by the IC event selection (section 2.2). In order to avoid confirmation bias, the right ascension in data has been scrambled during the development of the analyses.

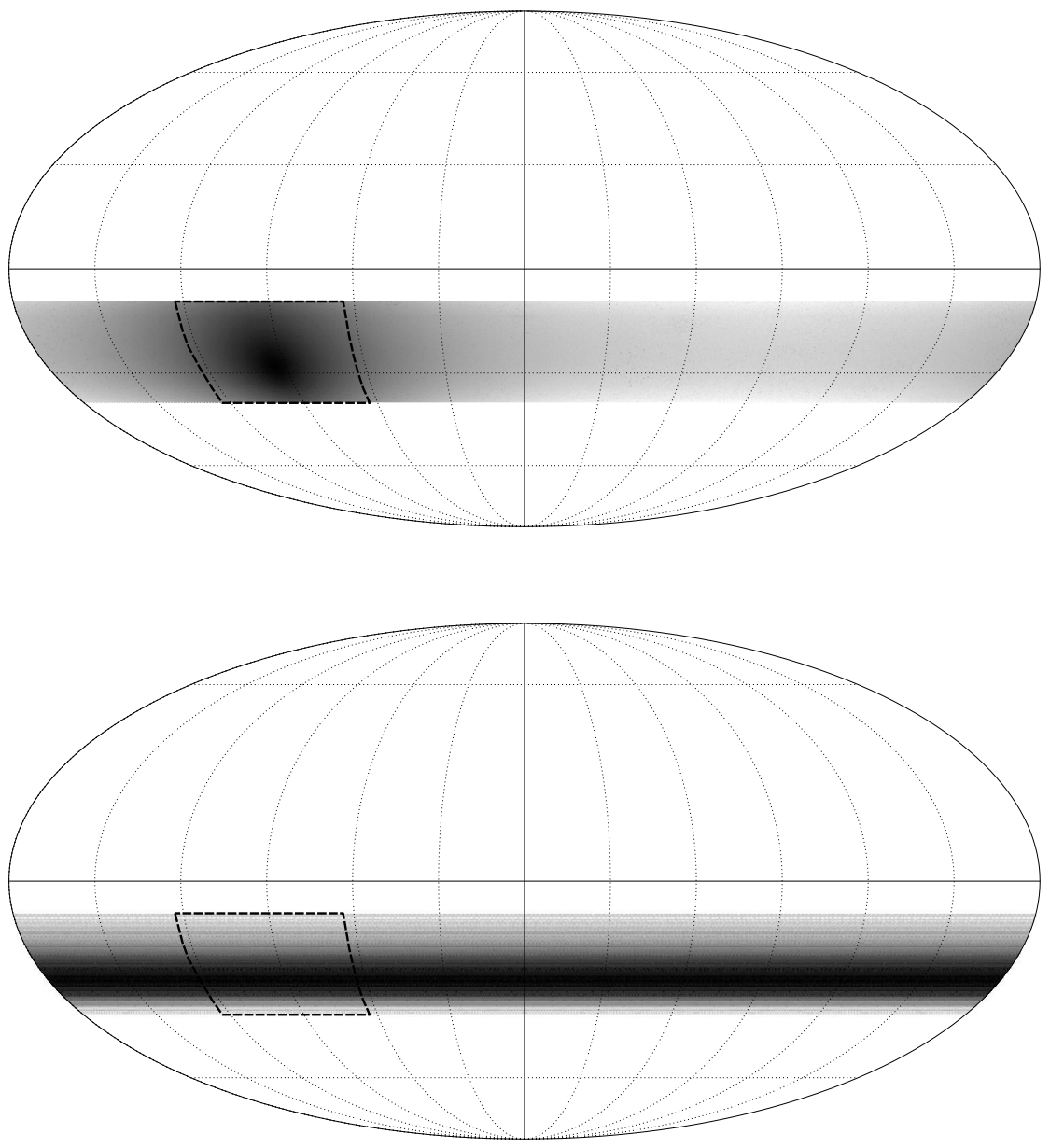

**Figure 2**: Signal and background skymap PDFs in equatorial coordinates for the DeepCore analysis. The search region around the Galactic Center at -29° dec. and 266° r.a. is indicated by a black dashed box. The black and white shading indicates the probability density. **Top:** Signal PDF for a 130 GeV WIMP annihilating into $\mu^+\mu^-$, assuming the NFW halo model. **Bottom:** Background PDF obtained from scrambled data.

### 2.1 DC event selection

To optimize for low WIMP mass signals the bottom part of the DeepCore subarray (see also [13]) is defined as the fiducial volume. This definition includes a two-DOM thick bottom veto layer, resulting in a total fiducial volume of approximately 280 m height and 260 m width.

The event selection exclusively uses data from the DeepCore filter event stream [12]. Based on distributions of event multiplicities and observables from signal simulations and experimental data, cuts are placed to reduce the content of atmospheric muon events. This is repeated for five successive linear and one multivariate cut level. The linear cuts reduce the data rate from about 200 Hz to approximately 10 mHz. These cuts rely on the expertise from the IceCube-79 Solar WIMP analysis [14], which used DeepCore for the first time in low mass WIMP searches. In addition, they incorporate newly developed muon veto methods [13].

The last cut level is implemented using a boosted decision tree (BDT). We use the TMVA software package [15] to classify events as signal or background. Two separable classes of neutrino-induced low energy signal events were identified at the linear cut levels. The first class consists of well contained starting events within the fiducial volume, while the second is dominated by partially contained events: those that start within the fiducial volume and exit it. We therefore trained two BDTs, one for the well contained (DC-contained) and a second for partially contained (DC-partial) signal events.

The cuts on the BDT scores have been optimized for each category using the maximum likelihood method described in [14]. For the optimization of the DC-contained and DC-partial event selections the $b\bar{b}$ and $W^+W^-$ WIMP annihilation spectra have been considered, respectively. The probability density functions (PDFs) for signal and background are constructed using healpix [16].

A box of size $-10°$ and $+20°$ in declination and $\pm 30°$ in right ascension around the Galactic Center position has been used as the search region. The search region is asymmetric in declination to account for the increased background rate towards lower declination angles. The final event selection (BDT cut applied) has a data rate of approximately 1 mHz. Figure 2 shows the expected signal and background PDFs at final cut level. The background PDF is found by filling a healpix map with reconstructed declination of recorded events and scrambled right ascensions. The scrambled right ascension is found by sampling fake event times from the data-taking period. The time scramble is repeated to achieve a smooth background distribution.

Sensitivities on $\langle \sigma_A v \rangle$ are derived for each signal model for both event selections, DC-contained and DC-partial. The event selection that results in a better sensitivity for a particular signal model is used in the final analysis.

### 2.2 IC event selection

The search for WIMPs with masses above a few hundred GeV benefits from the large volume of IceCube in addition to DeepCore. This increase in fiducial volume comes at a cost of a smaller veto region in addition to a higher data rate.

For the IC event selection, a dedicated data pre-selection filter has been developed, which reduces the event rate from more than 2 kHz at trigger-level to about 52 Hz. Further, this search relies on a combination of three dedicated veto techniques against incoming atmospheric muon tracks.

The veto volume contains a top layer of the upper-most 200 m (equal to 12 DOM layers). In addition, a side veto is defined, that follows the hexagonal IceCube geometry and consists of two string layers. This definition of fiducial- and veto-volumes is maintained for the following veto mechanisms.

First, we demand no reconstructed interaction vertices within the top or side veto. The vertex reconstruction is based on the projection of the first hit optical modules on the reconstructed track, taking into consideration the angle of Cherenkov light emission.

The second veto does not rely on a track reconstruction, but on the causality of hit DOMs. For each event the distance in space and time is calculated for all hits in the veto region with respect to the earliest hit inside the fiducial volume, hereby omitting any coincidence cleaning for the veto hits. Faint incoming tracks produce hits in veto DOMs that are causally connected to the earliest fiducial hit. In contrast, starting tracks (signal) produce no hits within the veto region. PDFs are constructed from the number of hits, the distance in space, and distance in time, which are used for a likelihood-ratio test.

For the third veto the reconstructed arrival direction is used to calculate the point of entry of the track into the detector volume. The distance in the $xy$-plane and the signed distance along the $z$-axis to the earliest hit DOM is calculated. Penetrating muon background produces a distinct feature in this plane, while starting signal-like tracks do not. A box cut around this feature is applied, removing a further 50% of the background while keeping more than 99% of fiducial signal events. All three veto cuts reduce the data rate from 52 Hz to 0.45 Hz.

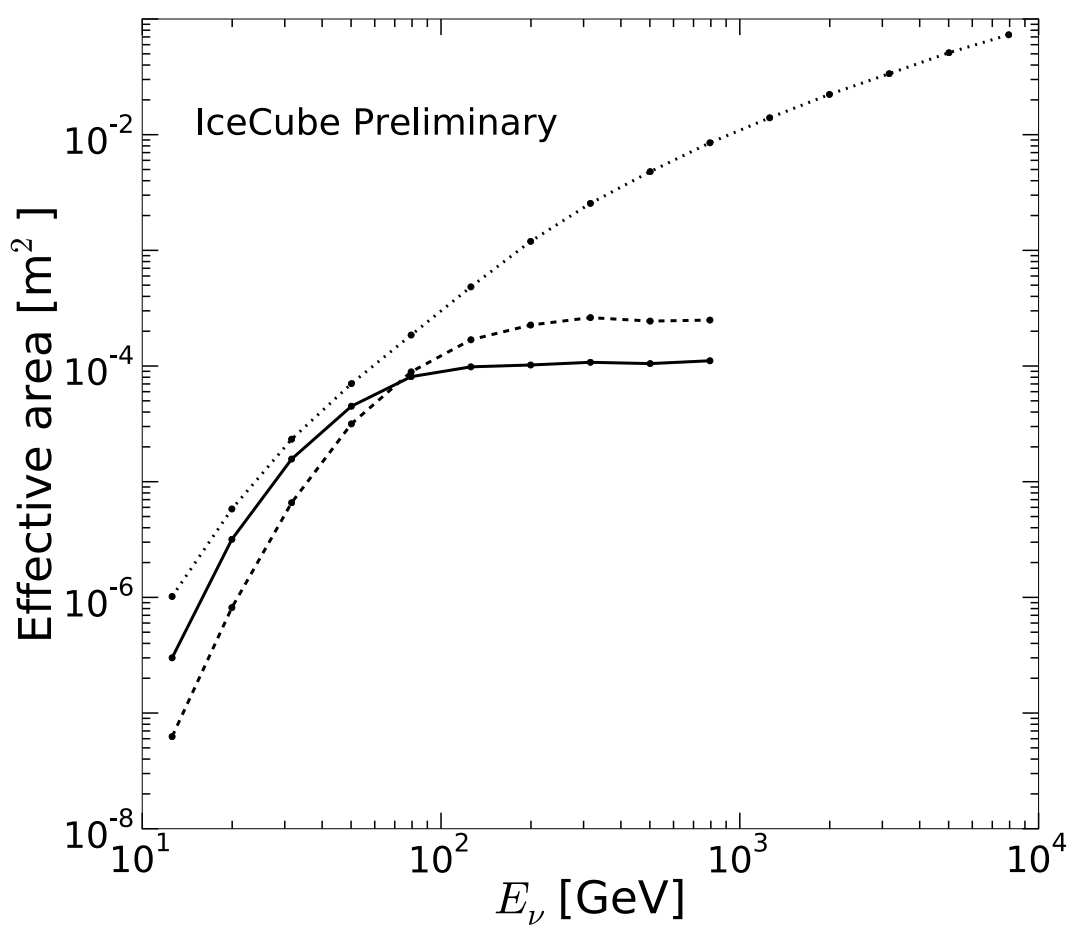

**Figure 3**: Neutrino effective area at analysis level for the IC event selection (dotted), the DC-contained (solid), and the DC-partial event selection (dashed).

The final IC event selection is also based on a BDT. It is trained on the remaining background of atmospheric showers and a starting signal event sample of 600 GeV WIMPs annihilating to $W^+W^-$. The input variables used in the BDT describe the likelihood of the event to be starting within the fiducial volume, the event quality, the event brightness, and the zenith distance to the Galactic Center. The cut on the BDT score is optimized with respect to $Signal/\sqrt{Background}$. This final acceptance criterion reduces the background by more than one order of magnitude, retaining more than 50% of the signal.

The final event sample has a data rate of about 10 mHz in the declination band $\pm 15°$ around the Galactic Center. The on-source region for a cut & count analysis is defined as $\pm 15°$ in right ascension with respect to the Galactic Center. The sensitivity to the number of signal events in the on-source region is calculated from expected background, which is found by extrapolation from the number of off-source background events in the burn sample.

## 3 Sensitivities

Figure 3 shows the effective area for the DC-contained and DC-partial event selections, as well as for the IC event selection described above. Even though the effective area for the DC event selections is smaller when compared to the IC event selection, the sensitivity to the number of signal events is higher, and the adopted on-source region is larger in the DC event selections. This is reflected in the 90% confidence level sensitivities on $\langle \sigma_A v \rangle$, which were computed using the Feldman-Cousins approach [17].

Sensitivities for the low and high WIMP mass regimes resulting from the two independent analyses, are shown in figure 4. Figure 5 shows the obtained sensitivities for annihilation to $\tau^+\tau^-$ in comparison to previous IceCube analyses: the IceCube-59 dwarf spheroid galaxy stacking and Virgo galaxy cluster search [18], the IceCube-40 Galactic Center analysis [19], and the IceCube-22 galactic halo analysis [20]. In addition, the preferred regions of the dark matter interpretation of the Pamela-excess (gray shaded region) with constraints from Fermi data (green shaded regions) [21] and the limits from Fermi [22] (green curve) are shown. The light-grey shaded band represents the natural scale at which WIMPs may appear as thermal relics.

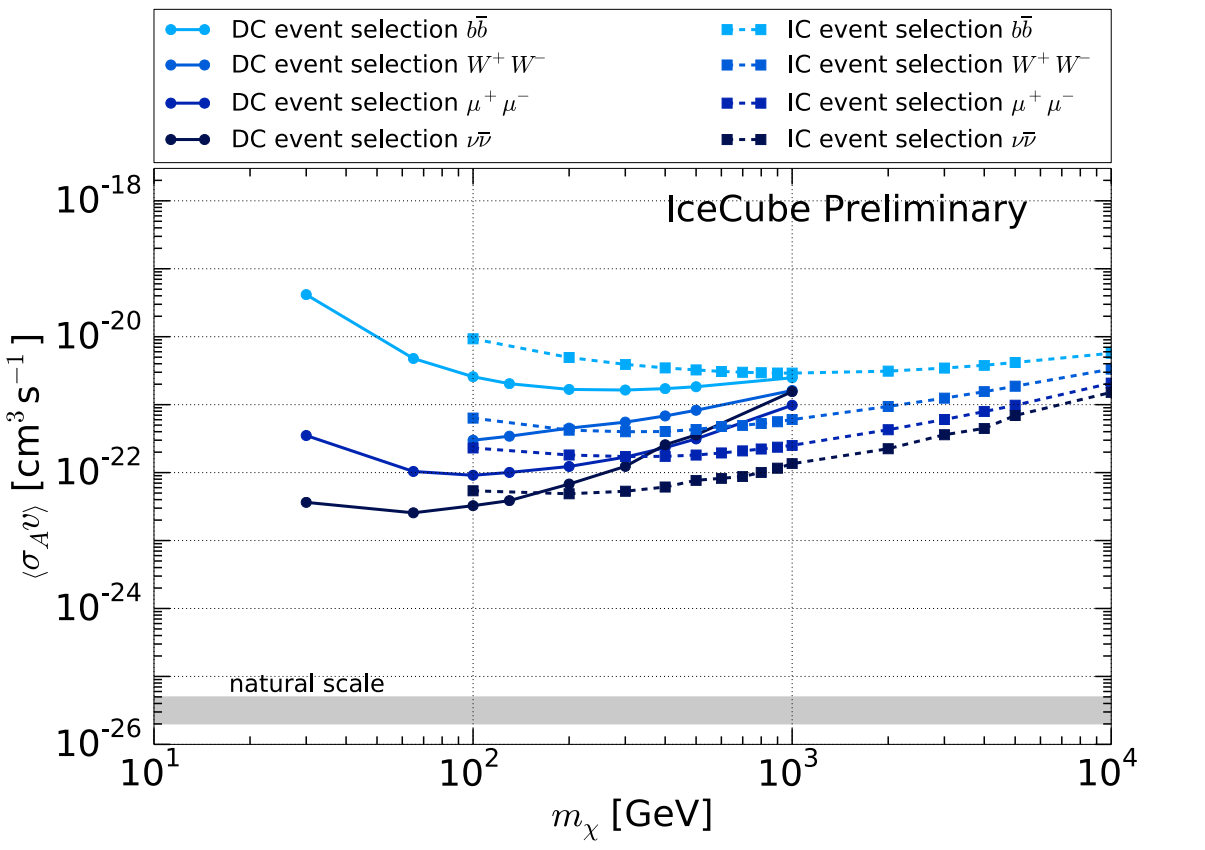

**Figure 4**: Sensitivities of the DeepCore low energy optimized (solid) and IceCube high energy optimized (dashed) independent analyses for different WIMP annihilation channels and WIMP masses.

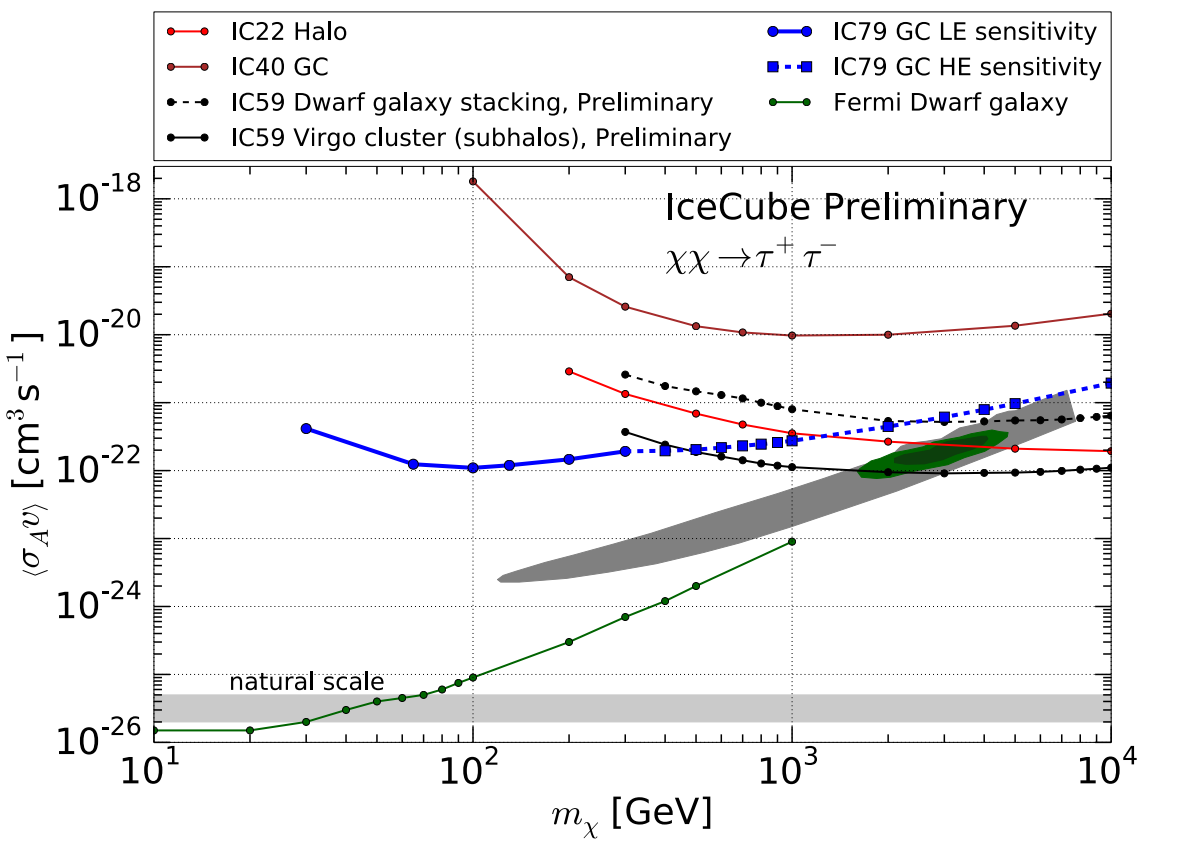

**Figure 5**: Sensitivity of this work (thick solid and dashed blue line) to the WIMP velocity-averaged self-annihilation cross-section into $\tau^+\tau^-$, compared to other analyses (see text).

For Galactic Center analyses, the largest model dependence comes from the chosen dark matter density profile. As described in section 1, the widely used NFW model has been chosen as a benchmark for the analyses presented here. Other models can be formulated based on the dark matter density profile kernel given by equation 1: the Moore and the Kravtsov model, which exhibits a cusped behavior in the central region, or has a flatter core region, respectively. The top panel of figure 1 shows the line-of-sight integral for each of the three models. The magnitude of the model-dependent variation of the presented sensitivities corresponds to the variation of the solid-angle-averaged value of the line-of-sight integral. It can be as large as one order of magnitude, compared to the NFW benchmark model.

## 4 Conclusions

Two approaches to search for annihilating dark matter in the Galactic Center have been presented, focusing on different WIMP mass regions. Both approaches use the nearly complete 79-string configuration of IceCube and 320 live days of data taken from 2010 to 2011. The sensitivity has been improved by up to 5 orders of magnitude for the $b\bar{b}$ channel of a 100 GeV WIMP mass when comparing with the previous Galactic Center analysis of IceCube-40 data [19]. Further, for the first time a Galactic Center analysis from IceCube has been extended to 30 GeV WIMP mass. The search for low-mass WIMPs benefits from the veto capabilities of large parts of the IceCube detector, while the intermediate mass range, between a few 100 GeV and a few TeV, is assisted by a larger fiducial volume definition, compared to DeepCore alone. At even higher WIMP masses, and thus higher neutrino energies, the constraint of a starting event selection as necessary veto against atmospheric muons reduces the effective area (see also dotted compared to solid and dashed lines in figure 3), which yields a weaker sensitivity.

# Search for relativistic magnetic monopoles in the IceCube Neutrino Telescope

THE ICECUBE COLLABORATION[1]
[1] *See special section in these proceedings*

*aobertacke@icecube.wisc.edu*

**Abstract:** Stable magnetic monopoles as relics of the Big Bang are a generic prediction of Grand Unified Theories (GUTs). Despite their large mass, galactic or cosmic magnetic fields may accelerate these particles to relativistic velocities. Monopoles are predicted to emit several thousand times more Cherenkov light than electrically charged particles. However, the expected flux is extremely small. Hence, large scale detectors like IceCube are required to search for relativistic monopoles. Currently, there are two IceCube analyses concerning relativistic magnetic monopoles. The first analysis covers highly relativistic velocities above the Cherenkov threshold and is based on data taken with the half completed IceCube detector in 2008. The crucial challenge is to separate a possible signal from the background of atmospheric muons which are at least six orders of magnitude more abundant. This goal is achieved by utilizing observables such as the brightness, reconstructed direction and event topology. The resulting limits are currently the best for monopole velocities between 0.76c and 0.995c. The second analysis covers mildly relativistic velocities below the Cherenkov threshold with data from 2011. While traveling through ice, the monopoles knock electrons off their atoms which produce Cherenkov light. Former analyses used the Mott cross section for the reaction between monopoles and electrons. This analysis also uses the quantum-mechanically more valid cross section by Kazama, Yang and Goldhaber and discusses the difference. The sensitivities for the monopole velocities from 0.6 to 0.75c are better than recent limits from other detectors.

**Corresponding authors:** A. Obertacke[1], J. Posselt[1],
[1] *Dept. of Physics, University of Wuppertal, 42119 Wuppertal, Germany*

## 1 Magnetic monopoles

No evidence for single magnetic poles has ever been found, although already in ancient times people tried to separate the two poles of magnetic rocks. A subtle hint on single magnetic poles is the symmetric shape of the Maxwell equations if magnetic sources were added.

The first consistent theory of magnetic monopoles was formulated by Dirac in 1933 [1]. The existence of a magnetic charge $g$ is directly related to the quantization of the electric charge $e$

$$g = N\frac{e}{2\alpha}$$

where $N$ is an integer and $\alpha$ is the fine structure constant.

Later, t'Hooft and Polyakov independently found magnetic monopoles as generic solutions of Grand Unified Theories [2]. In this context the mass of monopoles is calculable for various GUT models [3]

$$10^8\text{GeV} < M < 10^{17}\text{GeV}$$

These high masses lead to the assumption that magnetic monopoles must have been created right after the big bang by the Kibble mechanism [4]. According to this, magnetic monopoles are 1-dimensional, stable, topological defects.

An upper limit for the number of relic monopoles was calculated by Parker which is based on the requirement that the galactic magnetic field is not dissipated faster by acceleration of monopoles than it is regenerated by the dynamo action of the galactic disc [5]

$$\Phi_{Parker} \approx 10^{-15}\text{cm}^{-2}\text{sr}^{-1}\text{s}^{-1}$$

However, various experiments have already reported lower limits (see Fig. 4).

## 2 Detection using Cherenkov light

Similar to electrically charged particles in electric fields, magnetic monopoles can be accelerated by magnetic fields up to relativistic velocities, even with their high masses [6].

The phase velocity of light in a medium with refractive index $n$ is $c_P = c/n$. A monopole traveling with a velocity $v > c_P$ directly induces the emission of Cherenkov light. In ice the threshold velocity is $v \approx 0.76c$.

The number of photons d$N$ per path length d$x$ and wavelength interval d$\lambda$ can be calculated with a formula derived by Tompkins [7]

$$\frac{\mathrm{d}^2N}{\mathrm{d}x\mathrm{d}\lambda} = \frac{2\pi\alpha}{\lambda^2}\left(\frac{gn}{e}\right)^2\left(1-\frac{1}{\beta^2n^2}\right)$$

where $\beta = v/c$ as shown in Fig. 1.

Monopoles with slightly lower velocities still have very large kinetic energy, so they knock off $\delta$-electrons from the surrounding atoms. These electrons may be energetic enough to generate Cherenkov light.

For the cross section of monopoles and electrons, previous analyses in this velocity range [8] chose the Rutherford cross section [9] with the Mott form factor [10]. This cross section is calculated replacing the potential energy operator of a heavy, electrically charge by the average potential of a magnetic monopole acting on a stationary electron. This leads to the differential cross section

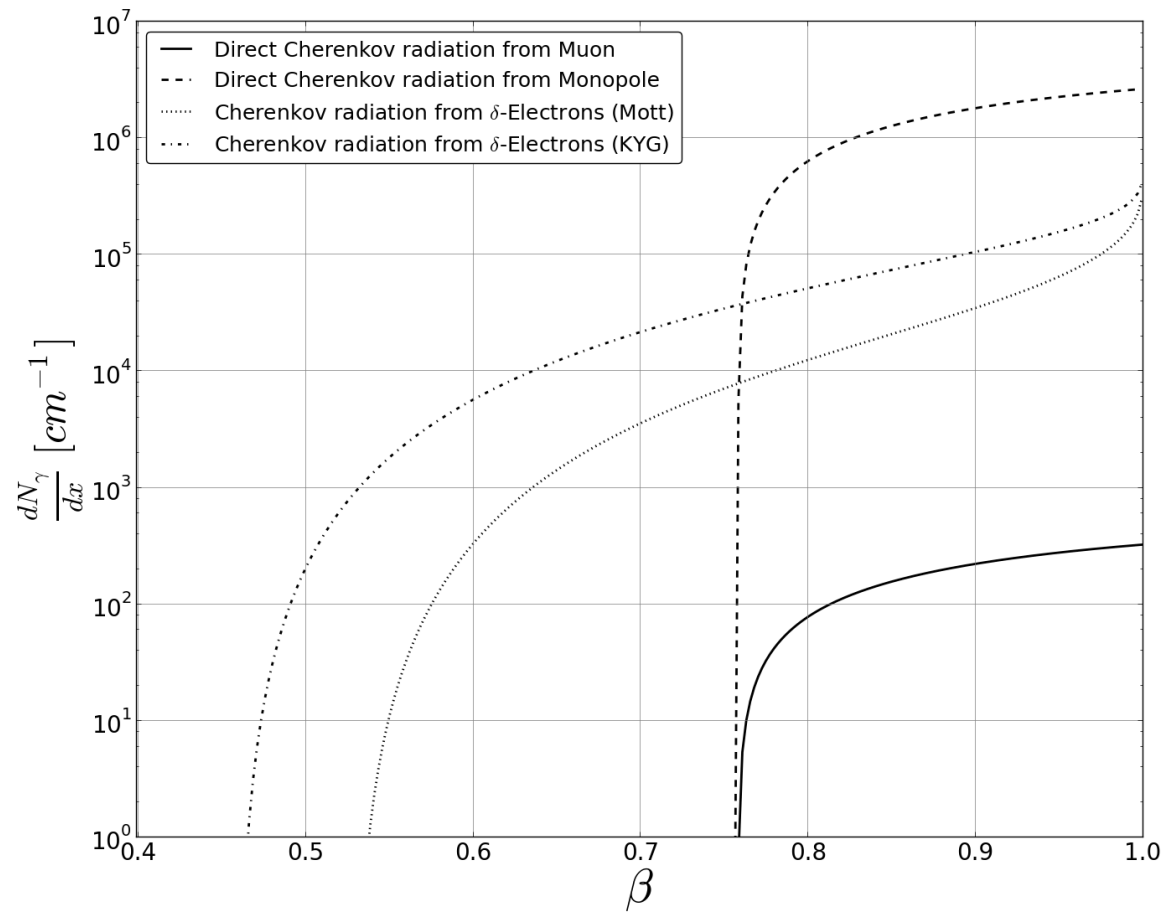

**Figure 1**: Number of photons per cm produced by a muon (solid line), a monopole by direct Cherenkov light (dashed) and monopoles by $\delta$-electrons with Mott (dotted) and KYG cross section (dash-dotted).

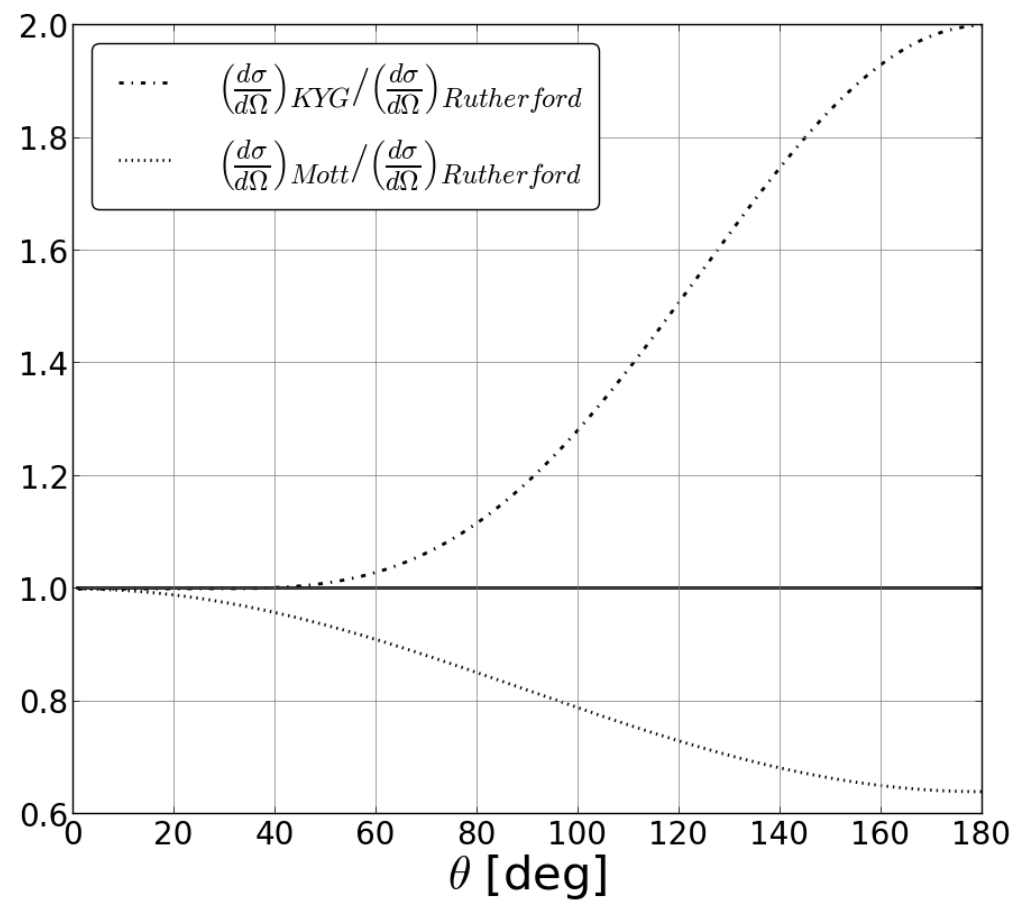

**Figure 2**: Ratio of differential cross sections to Rutherford cross section for monopoles (this is independent of $\beta$).

$$\left(\frac{d\sigma}{d\Omega}\right)^{magn.}_{Mott} = \left(\frac{g\beta \cdot Ze}{4T_0}\right)^2 \frac{1}{\sin^4\frac{\theta}{2}} \; F_{\mathrm{Mott}}(\beta,\theta)$$

$$F_{\mathrm{Mott}}(\beta,\theta) = \left(1-\beta^2\sin^2\frac{\theta}{2}\right)$$

where $\theta$ is the center-of-momentum scattering angle, $T_0$ is the initial energy of the projectile and $Ze = -e$ is the charge of the electron. This differs from the Mott cross section for electric charges by replacing the charge of the target $Z'e$ with $\beta g$.

The study presented here uses the more accurate cross section introduced by Kazama, Yang and Goldhaber (KYG cross section) [11]. This cross section differs from the Mott cross section by also taking into account the spin of the target and the vector potential $\vec{A}$.

Due to the interaction of the magnetic moment of the electron and the monopole, the spin may flip. In contrast to the Mott cross section scattering around 180 degree is favored. The helicity-flip and -nonflip amplitudes are calculated for this cross section.

By definition $\vec{A}$ has a singularity for monopoles. The problem was solved by using line bundles to define overlapping and singularity-free regions around the monopole [12]. Using that the authors of the KYG paper calculate a different form factor for the Rutherford cross section

$$F_{KYG}(\beta,T_q) = \left(\frac{g\beta}{Ze}\right)^2 \left[\frac{|T_q(\theta)|^2}{(Zeg)^2}\sin^4\frac{\theta}{2} + 2\left(\sin\frac{\theta}{2}\right)^{4|Zeg|+2}\right]$$

where $T_q(\theta)$ is a very difficult equation depending on the scattering angle.

The difference between the cross sections is shown in Fig. 2. The resulting number of photons for the Mott and KYG cross sections are shown in Fig. 1. For comparison both cross sections are considered in this analysis.

The simulation of magnetic monopole signals and background in IceCube is described in [13].

## 3 IceCube

IceCube is a cubic-kilometer neutrino detector installed in the ice at the geographic South Pole between depths of 1450 m and 2450 m [14]. Detector construction started in 2005 and finished in 2010. It now comprises 86 strings with 60 Digital Optical Modules (DOMs) each.

In the presented analyses data from 2008 and 2011 – the 40 and 86 string configurations (IC40 and IC86) – are used.

Each DOM consists of a Photomultiplier Tube (PMT) and electronics to digitize photon signals. When the PMT signal exceeds a threshold set to 0.25 photoelectrons on a DOM, it is called a hit. If also its two neighbor DOMs up and down the string are hit, it is called a hard local coincidence (HLC) hit.

The DeepCore subarray includes 8 densely instrumented strings optimized for low energies plus 12 adjacent standard strings. DeepCore is capable to detect much fainter signals than IceCube.

## 4 IC40 analysis above the Cherenkov threshold

A search for relativistic magnetic monopoles was performed on the 2008 data run with the detector operating with 40 strings representing a volume of $\sim 0.5$ km$^3$. The general procedure used is that of a blind analysis. Hence the optimization of the data selection is based on simulated data, and only $\sim 10\%$ of the experimental data, referred to as the Burn Sample, which is used for verification purposes.

Signal datasets were generated for four different speeds, $\beta = 0.995$, $\beta = 0.9$, $\beta = 0.8$, and $\beta = 0.76$, taking only direct Cherenkov light into account. The flux is assumed to be isotropic at the detector with a normalization of $5 \cdot 10^{-17}$ cm$^{-2}$ s$^{-1}$ str$^{-1}$, roughly representing the lowest limits set by BAIKAL [15] and AMANDA [16]. The principal background consists of down-going high energy atmospheric muon bundles induced by cosmic rays. These are simulated in the energy range from $10^4$ GeV to $10^{11}$ GeV, using a 2-component model with only proton and iron primaries [17].

The event selection is focused on the brightness of the monopoles and the reconstructed zenith direction. The brightness is measured as the ratio of the total number of photo-electrons (NPE), estimated by unfolding the recorded waveforms, and the number of hit DOMs (nCh). Since the number of DOMs within a radius $r$ around a particle track is proportional to $r^2$ and the photon density approximately scales as $\exp(-r/\lambda)$, where $\lambda$ is the effective absorption length, bright monopoles may trigger more DOMs than an atmospheric muon bundle. However, this increase is not enough to counter the increase in the total amount of detected light. Thus, on average NPE/nCh is expected to be larger for a monopole signal than for background.

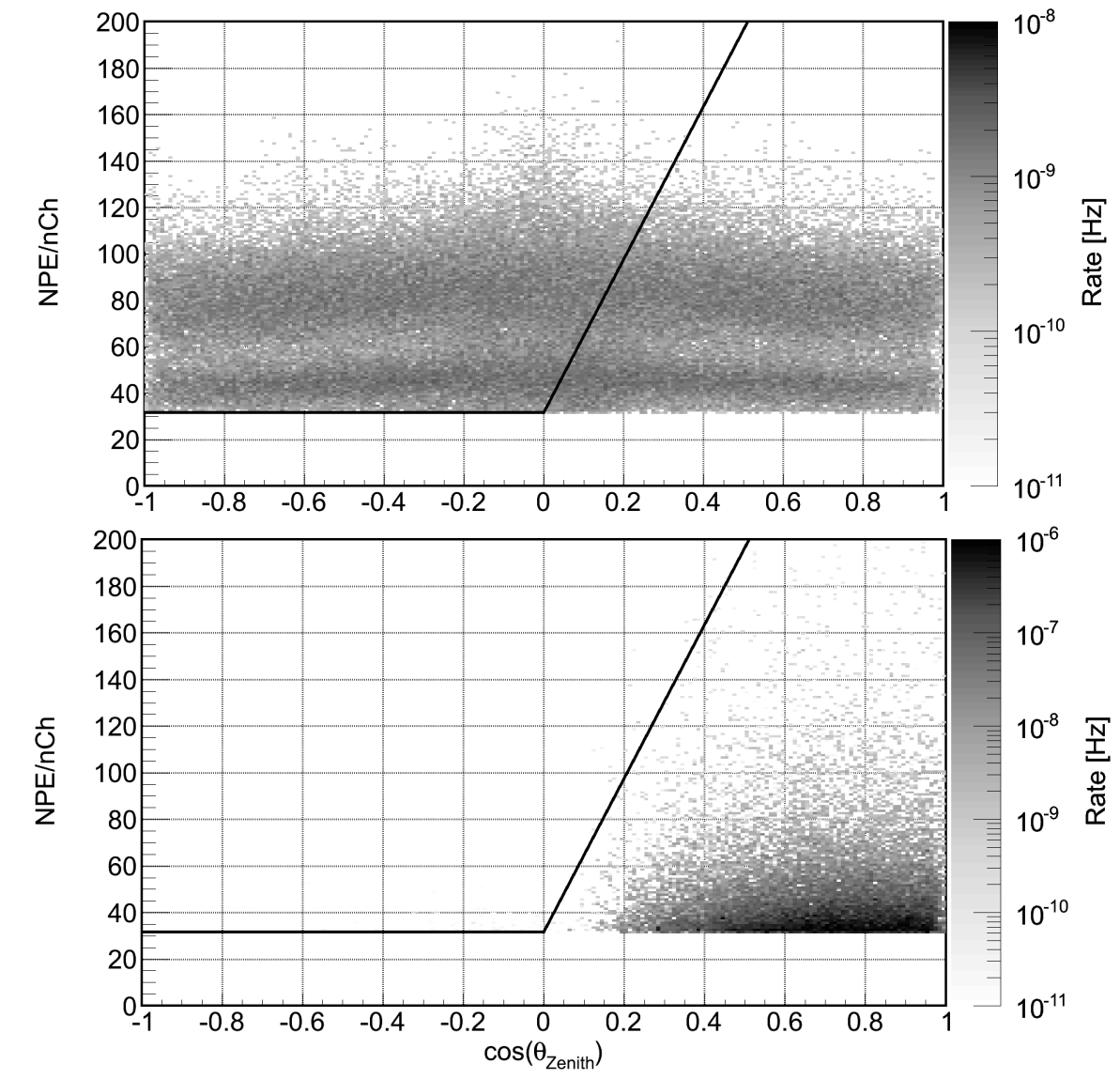

**Figure 3**: Final cut for signal (top; $\beta = 0.995, 0.9, 0.8$) and simulated background (bottom; atmospheric $\mu$ and $\nu_\mu$).

Before the final cut the data is split into a high and low NPE/nCh branch at a threshold of 31.6. This split allows to separately handle the poorly understood background near the horizon, which is contained in the low NPE/nCh branch together with the slowest considered monopoles ($\beta = 0.76$). The final cut for this branch therefore accepts only events that are clearly up-going ($\cos\theta_z < -0.2$), thus avoiding the problematic horizontal region. For the high NPE/nCh branch the final cut is in the plane of reconstructed zenith direction and NPE/nCh. In the up-going region, that is largely background free, NPE/nCh is only restricted by the split condition and no further cut is applied. For down-going directions, where muon bundles produced by high-energy cosmic rays dominate, a cut increasing linearly in strength as $\cos\theta_z$ approaches the vertical is used (see Fig. 3). The cut is optimized on background and signal Monte Carlo by considering the combination that minimizes the model rejection factor (MRF) [18] for an isotropic and monoenergetic monopole flux.

After unblinding, one event remains in the low NPE/nCh branch, which is consistent with the expected background of about two events for the considered time period. In the high NPE/nCh branch two event are observed whereas the expected background is only about 0.3 events. However, visual inspection of these two events revealed characteristics (mainly the time evolution), which are, in contradiction with simulations, more likely explained with an atmospheric muon than a magnetic monopole. In the absence of a better model for the atmospheric background, the two events are therefore treated as signal candidates. The resulting (preliminary) flux limits at the detector are given in Tab. 1 and are also shown in Fig. 4. The limits also include systematic uncertainties on parameters like the sensitivity of the DOMs or the energy spectrum sampled in simulations by averaging the MRF with an appropriate probability density function.

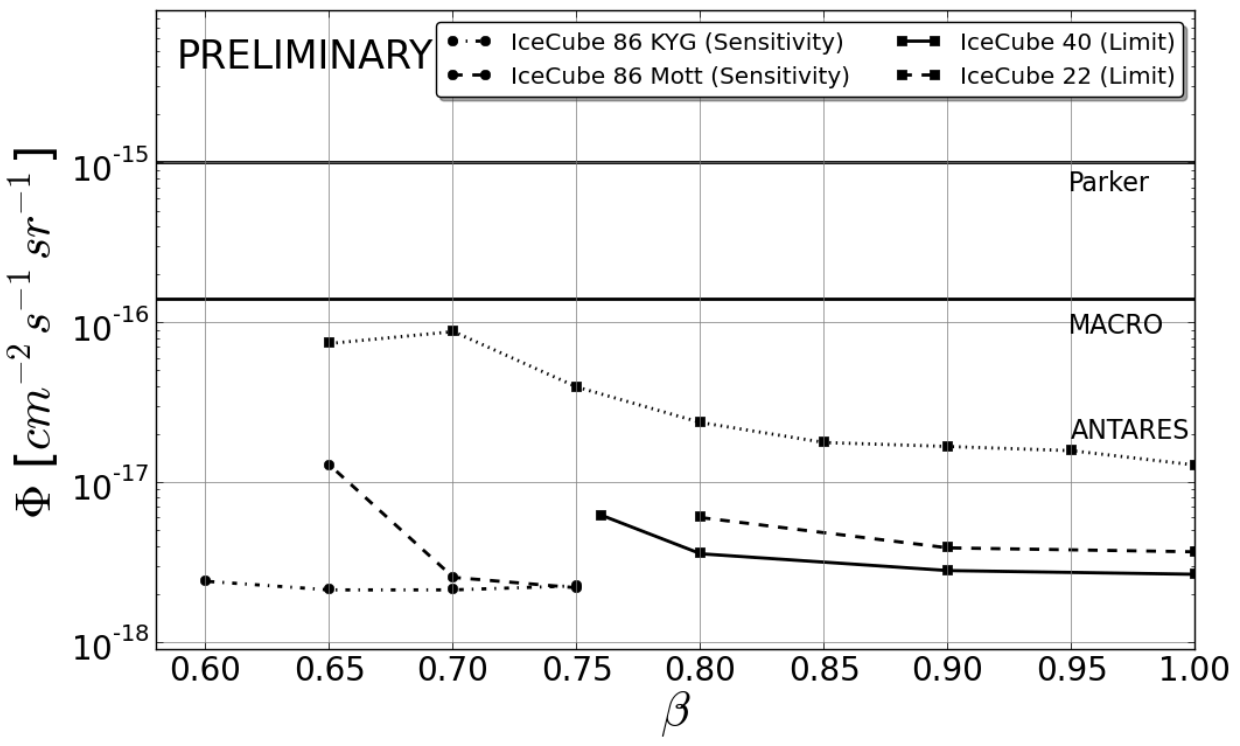

**Figure 4**: Limit and sensitivity of the introduced analyses in addition to previous analyses [8, 19, 20].

| $\beta$ | $A_{eff}$ km$^2$ | $n_{exp}$ a$^{-1}$ | $\Phi_{90}$ cm$^{-2}$ s$^{-1}$ sr$^{-1}$ |
|---|---|---|---|
| 0.995 | 0.51 | 101.9 | $2.97 \cdot 10^{-18}$ |
| 0.9 | 0.49 | 96.5 | $3.14 \cdot 10^{-18}$ |
| 0.8 | 0.38 | 76.1 | $4.00 \cdot 10^{-18}$ |
| 0.76 | 0.09 | 18.0 | $1.73 \cdot 10^{-17}$ |

**Table 1**: Effective area, expected signal and upper limit (90% C.L.) [21] at the IC 40 detector for a full year of data.

## 5 IC86 study below the Cherenkov threshold

The first IceCube study covering the velocity range below $0.76c$ used 10% of the experimental data of the season 2011/12 (the IC86 Burn Sample ) for the background study.

The monopole signal is simulated with an isotropic flux of $10^{-16}$cm$^{-2}$s$^{-1}$sr$^{-1}$ for the velocities $0.75c$, $0.7c$, $0.65c$ and $0.6c$. For these velocities the produced light exceeds the light produced by a muon with $v = c$ (compare Fig. 1). For each velocity both introduced cross sections were used for comparison reasons.

Many experimental events comprise more than one visible track when muons from different air showers pass through the detector at almost the same time. These coincident events get split by a global causality based algorithm into clusters of correlated hits.

For the track reconstruction a simple least-square fit is used which is optimized by neglecting late hits and weighting hits far-off the track less. Based on this track reconstruction the most important cuts are a zenith and a velocity cut:

Since the brightness of the simulated monopole events is similar to that of a muon, down-going monopoles would be very hard to separate from air shower muons. The number of these muons decreases much at zenith angles of 86 degree. So a cut on this value is chosen to reject this background.

The velocity cut rejects muons which travel with $v = c$ and mis-reconstructed coincident background signals with very low reconstructed velocity (see Fig. 5). The cut is softened for fainter events by using a 2-dimensional cut with NPE/nCh as a variable for event brightness.

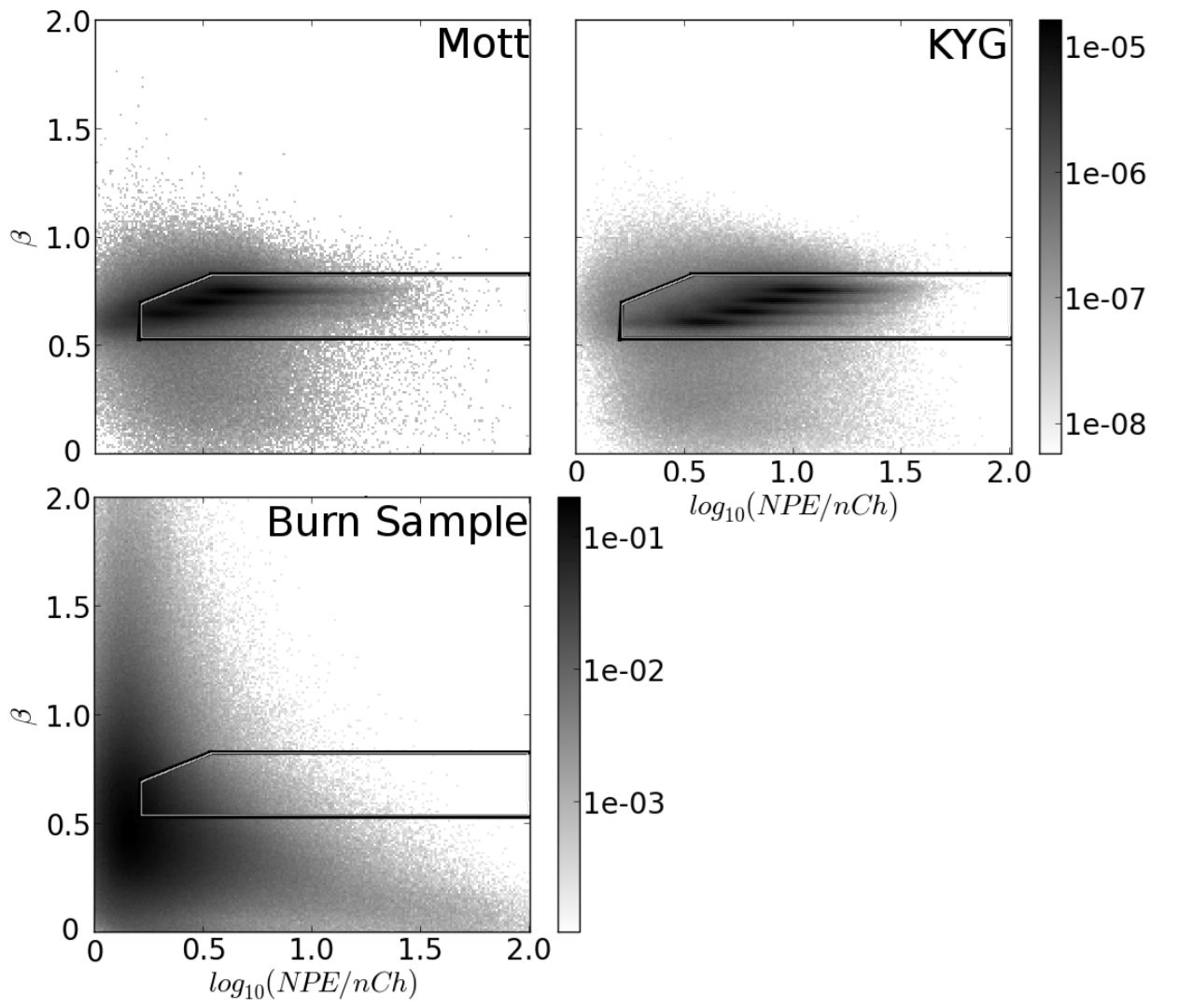

**Figure 5**: Cut on velocity and brightness (black/white line).

After these cuts, the remaining background events from the Burn Sample are coincident events. To reject those events a series of geometric variables are used. For example, the size of the maximum gap between hits which are orthogonally projected on the track and the fraction of DOMs with hits or no-hits in a radius of 50 respectively 100 meters around the track.

Every single cut is optimized so that at least 90% of the simulated signal are kept (not taking the signal, simulated with Mott cross section and $v = 0.6c$ into account).

The difference between the Mott and KYG cross section is observed for every cut variable (for example compare the maxima in Fig. 5). Cuts optimized for simulations with only one cross section remove events which are simulated with the other cross section. Therefore it is not possible to choose the Mott cross section as a conservative estimate for the light production.

After the cuts all Burn Sample events are rejected. This leads to the sensitivity given in Tab. 2 and Fig. 4.

The sensitivity using the Mott cross section is better than the ANTARES limits which are produced with the same cross section. The sensitivity using the KYG cross section is more than one order of magnitude below the previous best limits and extends to even lower velocities.

| | Mott | | KYG | |
|---|---|---|---|---|
| $\beta$ | $n_{exp}$ $a^{-1}$ | $\bar{\Phi}_{90}$ $cm^{-2}s^{-1}sr^{-1}$ | $n_{exp}$ $a^{-1}$ | $\bar{\Phi}_{90}$ $cm^{-2}s^{-1}sr^{-1}$ |
| 0.75 | 110 | $2.2 \cdot 10^{-18}$ | 107 | $2.3 \cdot 10^{-18}$ |
| 0.70 | 95 | $2.5 \cdot 10^{-18}$ | 113 | $2.1 \cdot 10^{-18}$ |
| 0.65 | 19 | $1.3 \cdot 10^{-17}$ | 113 | $2.2 \cdot 10^{-18}$ |
| 0.60 | 0.0004 | $5.4 \cdot 10^{-13}$ | 100 | $2.4 \cdot 10^{-18}$ |

**Table 2**: Expected signal and average upper limit (90% C.L.) [18, 21] at the IC86 detector for a full year of data. No Burn Sample event is left.

## 6 Conclusion and Outlook

The IC40 analysis with velocities above the Cherenkov threshold produced the currently best limits in the range of 0.995 to 0.76$c$.

The IC86 study showed that IceCube is also sensitive to faint monopoles with velocities down to 0.6$c$. Looking at Fig. 1 it seems plausible that the same analysis works for lower velocities around 0.55$c$.

Since magnetic monopoles below the Cherenkov threshold give very faint light signals there is also a good chance to reach even lower velocities using the DeepCore detector or the proposed PINGU infill.

# Multipole analysis with IceCube to search for dark matter accumulated in the Galactic Halo

THE ICECUBE COLLABORATION[1],

[1] *See special section in these proceedings*

*reimann@physik.rwth-aachen.de*

**Abstract:** Self-annihilating or decaying dark matter in the Galactic Halo may contribute to the observed flux of cosmic particles, e.g. gammas, positrons or neutrinos. High-energy neutrinos can be detected with the IceCube Neutrino Observatory, a cubic-kilometer-sized Cherenkov detector at the geographical South Pole. The additional flux from dark matter annihilation depends on the density of dark matter in the direction of sight and is expected to be larger in the direction of the galactic center and smaller in the direction of the anti-center. Given the large field of view of IceCube, such a large-scale anisotropy would leave a characteristic imprint on multipole expansion coefficients of the observed set of arrival directions in a high-purity muon neutrino event sample that can be obtained by selecting up-going muon events. This imprint can be interpreted in terms of the thermally averaged self-annihilation cross-section of dark-matter particles. We discuss the analysis of IceCube data in the 79-string configuration, using neutrinos from the Northern hemisphere. This analysis improves in sensitivity with respect to previous IceCube analyses. Resulting exclusion limits on the velocity averaged annihilation cross-section are presented.

**Corresponding authors:** René Reimann[1],
[1] *III. Physikalisches Institut, RWTH Aachen University, D-52056 Aachen, Germany*

## 1 Introduction

A variety of astronomical observations imply the existence of non-baryonic cold dark matter in the Universe. The velocity distribution of galaxy clusters suggests a mass content significantly higher than what should be expected from observations of the luminous mass [1]. Also the rotation profiles of galaxies show discrepancies to the expectation that hint at the existence of a dark matter halo reaching out far beyond the baryonic disc and bulge [1]. Based on, e.g., the anisotropy in the cosmic microwave background observed by the Wilkinson Microwave Anisotropy Probe [2] or Planck [3], a determination of cosmological parameters is possible, which yields the concordance model of cosmology with a dark matter content of about 22%. Large-scale structure formation and N-body simulations favor cold dark matter that consists of massive and non-relativistic particles. Weakly Interacting Massive Particles (WIMPs) are the generic candidates for cold dark matter in a mass range from a few GeV up to several 100 TeV [4, 5]. A promising WIMP candidate is the lightest stable particle in minimal super-symmetric extension to Standard Model.

If WIMPs are Majorana particles, they can self-annihilate and produce Standard Model particles in the final state. Observable stable messenger particles (e.g. photons, neutrinos, charged cosmic rays), could be produced as annihilation products, or in their decays. Since dark matter particles may be bound in the Galactic Halo, we expect this excess flux of neutral particles from self-annihilations to display a large-scale anisotropy that should map the dark matter density distribution in the galaxy. Due to the shape of the halo, this flux should be stronger towards the central part of our galaxy.

## 2 Neutrino flux from dark matter annihilation in the Galactic Halo

Predicted dark matter density profiles are based on observations of dark-matter-dominated galaxies and simulations of N-body systems [6]. An often used parametrization of a spherically symmetric dark-matter density profile is

$$\rho_{\mathrm{DM}}(r) = \frac{\rho_0}{\left(\frac{r}{r_s}\right)^{\gamma} \cdot \left[1 + \left(\frac{r}{r_s}\right)^{\alpha}\right]^{(\beta-\gamma)/\alpha}}, \qquad (1)$$

where $r$ is the distance from the Galactic Center (GC) in kpc and $r_s$ is a scaling radius [7]. Commonly, the Navarro-Frenk-White (NFW) profile with the parameters $r_s = 20\,\mathrm{kpc}$ and $(\alpha, \beta, \gamma) = (1, 3, 1)$ [8] is used. $\rho_0$ is chosen such that the local dark-matter density at the radius of the solar circle $R_{\mathrm{SC}}$ is $0.3\,\mathrm{GeV/cm^3}$. Other common models are e.g. the Moore profile [9] or the Burkert profile [10].

The flux from self-annihilating dark matter depends on the density squared along the line of sight

$$J_a(\psi) = \int_0^{l_{max}} \mathrm{d}l \, \frac{\rho_{\mathrm{DM}}^2\left(\sqrt{R_{\mathrm{SC}}^2 - 2lR_{\mathrm{SC}}\cos\psi + l^2}\right)}{R_{\mathrm{SC}}\rho_{\mathrm{SC}}^2}, \qquad (2)$$

where $\psi$ is the opening angle between the line of sight and the GC [11]. $R_{\mathrm{SC}}$ and $\rho_{\mathrm{SC}}^2$ are constants for scaling and $l_{max}$ is approximately the radius of the Milky Way. The line-of-sight integral for the Northern hemisphere is shown in Fig. 1. The flux is large in the direction of the galactic center and small in the direction of the anti-center. Because the galactic center is located on the Southern hemisphere at right ascension (RA) 266.3° and declination (Dec) −29.0°, the corresponding peak is not shown, but a large-scale anisotropy is clearly visible.

The differential neutrino flux is given by [11]

$$\frac{d\phi_\nu}{dE} = \frac{\langle \sigma_A v \rangle}{2} J_a(\psi) \frac{R_{SC}\rho_{SC}^2}{4\pi m_\chi^2} \frac{d\mathcal{N}}{dE}. \quad (3)$$

The factor $1/4\pi$ comes from isotropic emission of the final-state products and $m_\chi$ is the mass of the WIMP. $\frac{d\mathcal{N}}{dE}$ is the energy spectrum of final-state neutrinos, which is obtained by DarkSUSY [12] for benchmark WIMP masses and annihilation channels. $\langle \sigma_A v \rangle/2$ is the velocity-averaged self-annihilation cross-section per particle.

## 3 The IceCube Neutrino Observatory

IceCube is a Cherenkov neutrino detector located at the geographic South Pole [13]. Neutrinos interact with the clear Antarctic ice and produce secondary leptons and hadrons. These relativistic secondary particles produce Cherenkov light which is detected with Digital Optical Modules (DOMs). IceCube consists of 86 strings each instrumented with 60 DOMs, which are located at depths from 1.45 km to 2.45 km below the surface. The strings are arranged in a hexagonal pattern with an inter-string spacing of 125 m and a DOM-to-DOM distance of 17 m. A more densely instrumented sub-array called DeepCore, consisting of eight densely-instrumented strings, has been embedded to lowers the energy threshold from about 100 GeV to about 10 GeV [14]. The total instrumented volume is about one km$^3$. The detector construction was completed in December 2010.

## 4 Data Sample

This analysis uses data measured from June 2010 to May 2011, when IceCube was operating in its partial 79-string configuration. Within the live-time $T_{live}$ of 316 days, 57281 up-going muon neutrinos from the Northern hemisphere were detected. We use muon events with reconstructed declinations of 0° to 90° to reduce the atmospheric muon contamination of the sample. By means of a mixture of straight cuts and a selection by a Boosted Decision Tree (BDT) [15] the atmospheric muon contamination was reduced to $< 3\%$ [16, 17]. The resulting sample consists mostly of atmospheric neutrinos, which are background for the search of neutrinos from self-annihilating dark matter in the galactic halo. Unlike signal, the integrated atmospheric neutrino flux is nearly constant as function of RA [18].

## 5 Analysis Method

To search for the anisotropy caused by an additional flux from WIMP annihilations in the Galactic Halo, we perform a multipole expansion of the neutrino arrival directions in equatorial coordinates. This is based on spherical harmonics

$$Y_\ell^m(\theta,\phi) = \sqrt{\frac{(2\ell+1)(\ell-m)!}{4\pi(\ell+m)!}} P_\ell^m(\cos(\theta)) \exp(im\phi) \quad (4)$$

where $\theta$ is the declination and $\phi$ is the right ascension. $P_\ell^m$ are the associated Legendre polynomials and build up the canonical solution of the Legendre equation with $-\ell \leq m \leq \ell$, where $\ell$, $m$ are separation constants [19]. Because spherical harmonics build a complete set of orthonormal functions on the full sphere $\Omega$, all integrable functions $f(\theta,\phi)$ in the range $-\frac{\pi}{2} \leq \theta \leq \frac{\pi}{2}$ and $0 \leq \phi \leq 2\pi$ can be expanded in terms of spherical harmonics with the complex expansion coefficients

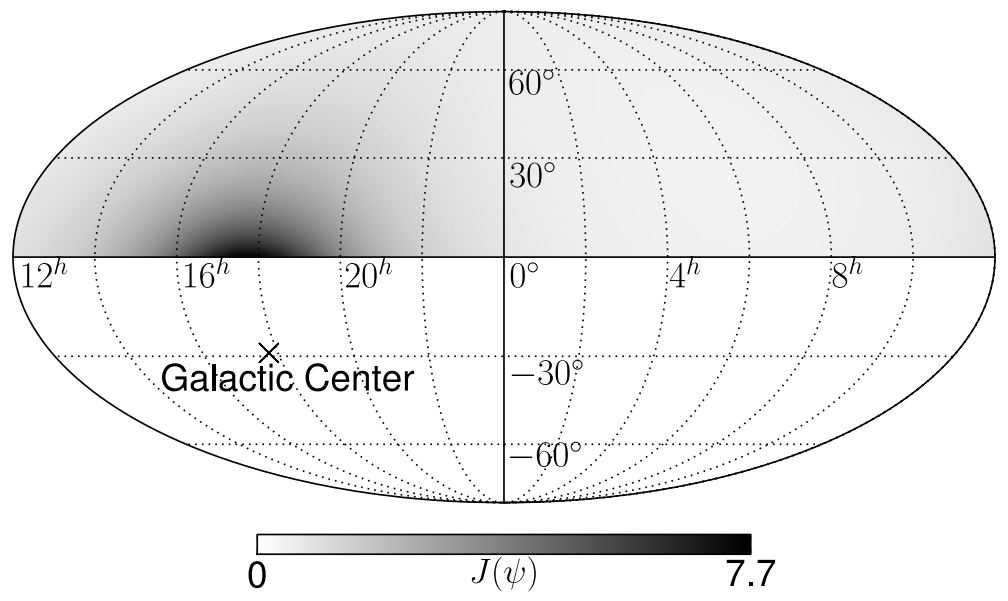

**Figure 1**: The dimensionless line-of-sight integral for the NFW model is shown for the northern hemisphere in equatorial coordinates. The anisotropy in the line-of-sight integral causes the anisotropy in the flux from self-annihilation of dark matter in the galactic halo. The position of the galactic center is indicated by the cross.

$$a_\ell^m = \int_\Omega d\Omega\, f(\theta,\phi)\, Y_\ell^{m*}(\theta,\phi). \quad (5)$$

The expansion coefficients are calculated up to a maximal $\ell$ of $\ell_{max} = 100$ using the software package HEALPix[1] [20]. The skymap is described by

$$f(\theta,\phi) = \sum_{i=1}^{N_\nu} \delta(\cos(\theta)-\cos(\theta_i)) \cdot \delta(\phi-\phi_i), \quad (6)$$

where $(\theta_i,\phi_i)$ is the measured arrival direction of each neutrino, $N_\nu$ is the total number of neutrinos on the skymap and $\delta(x)$ is the Dirac delta distribution. To generate skymaps of arrival directions, atmospheric neutrinos and missreconstructed muons are simulated as background, where declinations are taken from experimental data and right ascensions are uniformly randomized. Neutrinos from WIMP annihilation in the Galactic Halo are generated taking into account the line-of-sight integral, the angular resolution and the detector acceptance. The number of events from WIMP annihilation is varied in simulation. The number of events per skymap is fixed to the total number of measured events in the experimental sample by adding the remaining number of events from the background simulation. In Fig. 2 and Fig. 3 the mean absolute value, also called power, of pure background and signal skymaps is shown. For background, nearly all power is contained in $a_\ell^0$-coefficients, but for signal there is also power in coefficients with small $m$. The absolute value of the expansion coefficient is proportional to the signal strength, which can be found by simulation. In coefficients with large power there is also a preferred direction, indicated by the phase of the expansion coefficient. We combine phase and absolute value of the coefficient by projecting the complex expansion coefficient $a_\ell^m$ on the expected phase of the coefficient $a_{\ell,halo}^m$ from pure signal skymaps. The resulting projection is given by

$$\mathscr{A}_\ell^m = \|a_\ell^m\| \cos\left(\arg\left(a_\ell^m\right) - \left\langle \arg\left(a_{\ell,halo}^m\right)\right\rangle\right), \quad (7)$$

1. http://healpix.jpl.nasa.gov

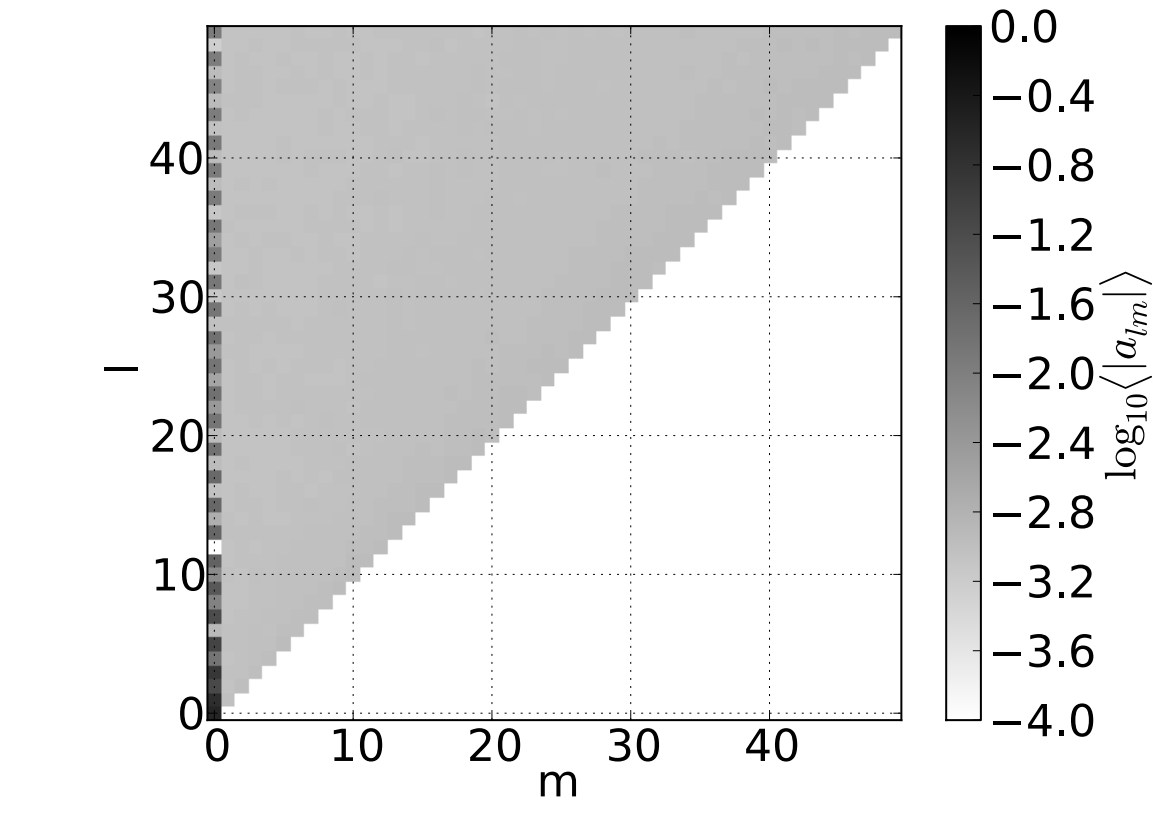

**Figure 2**: Shown is the logarithm of the absolute value for all coefficients in the $\ell$-$m$-plane. The absolute value was averaged over 1000 pure background skymaps.

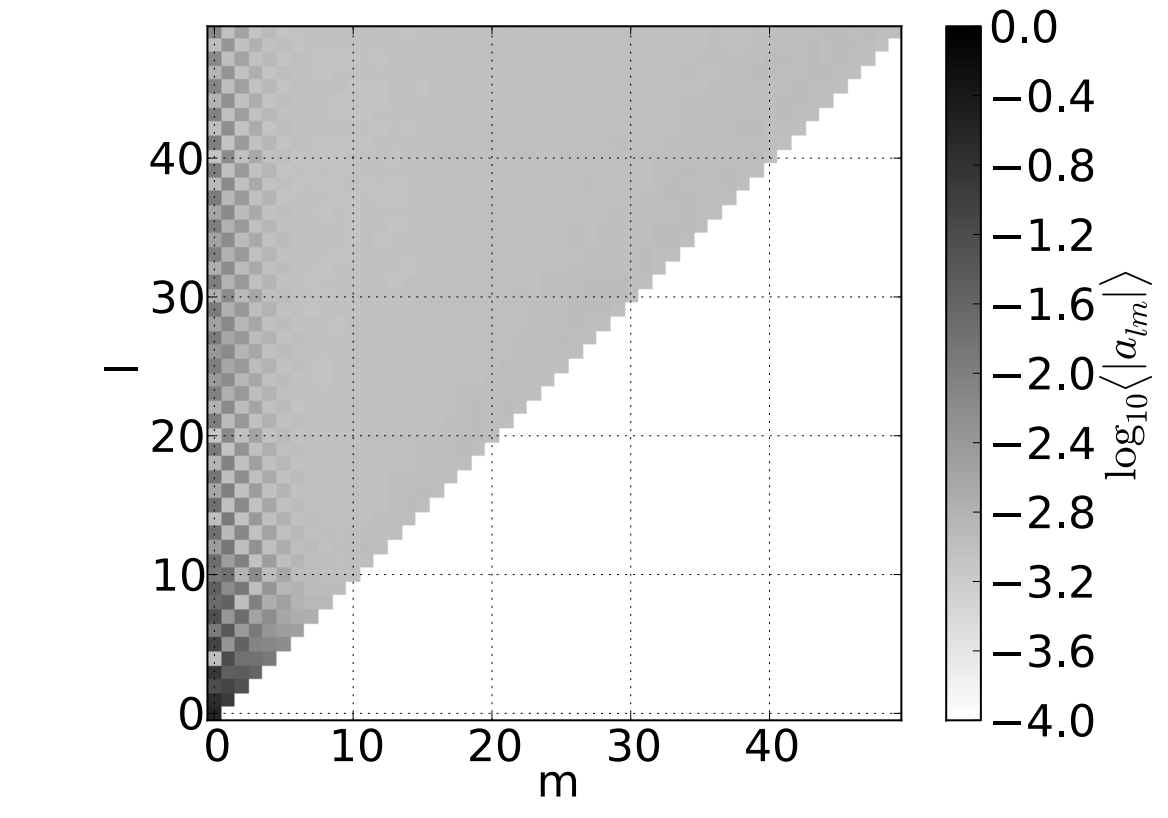

**Figure 3**: Shown is the logarithm of the absolute value for all coefficients in the $\ell$-$m$-plane. The absolute value was averaged over 1000 pure signal skymaps.

where arg $\left(a_\ell^m\right)$ gives the phase of the complex number $a_\ell^m$. To separate signal from background, we define a test statistic based on the difference of $\mathscr{A}_\ell^m$ to the background estimation from atmospheric neutrinos $\mathscr{A}_{\ell,\mathrm{atm}}^m$. The test statistic is motivated by a weighted $\chi^2$-function and is given by

$$D^2 = \frac{1}{\sum w_\ell^m} \sum_{\ell=1}^{\ell_{\max}} \sum_{m=1}^{\ell} \operatorname{sign}\left(\mathscr{A}_\ell^m\right) w_\ell^m \left( \frac{\mathscr{A}_\ell^m - \left\langle \mathscr{A}_{\ell,\mathrm{atm}}^m \right\rangle}{\sigma\left(\mathscr{A}_{\ell,\mathrm{atm}}^m\right)} \right)^2, \tag{8}$$

where $\langle x \rangle$ describes the mean and $\sigma(x)$ the standard deviation of the coefficients calculated from 1000 skymap simulations. The weights

$$w_\ell^m = \left\| \frac{\left\langle \mathscr{A}_{\ell,\mathrm{halo}}^m \right\rangle - \left\langle \mathscr{A}_{\ell,\mathrm{atm}}^m \right\rangle}{\sigma\left(\mathscr{A}_{\ell,\mathrm{atm}}^m\right)} \right\|, \tag{9}$$

are determined from pure halo signal skymaps. As can be seen from equation (4), spherical harmonics with $m = 0$ are independent of RA and just depend on declination. In the test statistic, the coefficients with $m = 0$ are excluded to reduce systematic uncertainties, which are dominated by uncertainties in the zenith dependent detector acceptance.

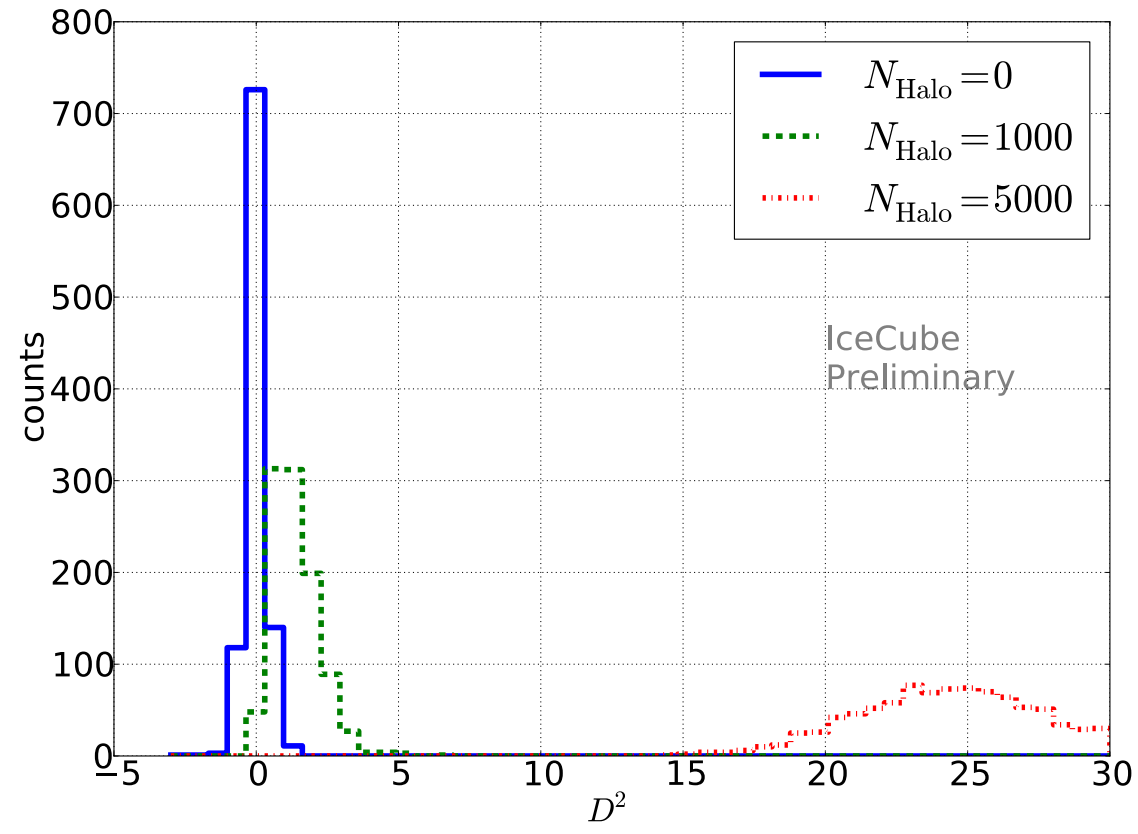

**Figure 4**: Test statistic $D^2$ for pure background simulation (solid) and simulations with small signal contributions. $N_{\mathrm{halo}}$ is the number of simulated neutrino arrival directions from WIMP annihilation in the Galactic Halo.

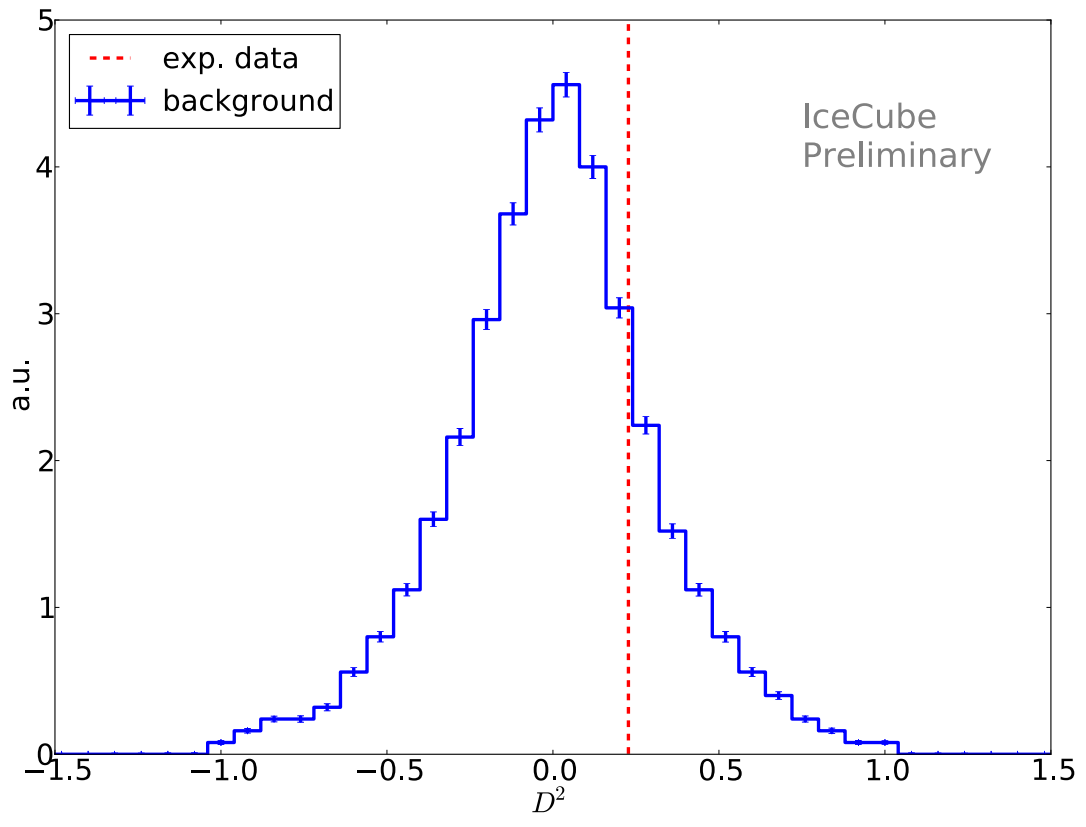

**Figure 5**: $D^2$-distribution for background expectation (solid) and the experimental observed value $D^2_{\mathrm{exp}}$ (dashed). The error of the background distribution reflects the simulation statistics.

The test statistic $D^2$ is plotted for pure background simulation and simulation with small signal contribution in Fig. 4. For each configuration 1000 skymaps were analyzed.

## 6 Results

The experimental value of the test statistic is $D^2_{\mathrm{exp}} = 0.23$. This corresponds to an overfluctuation of $0.79\sigma$. The probability to observe this value or higher from background-only case is 22%. The background $D^2$-distribution and the measured $D^2_{\mathrm{exp}}$ value are shown in Fig. 5.

Based on the Feldman-Cousins approach [21], the 90% C.L. upper limit on the number of signal events has been calculated to $N_{90} = 1061$. This number can be converted to a limit on the velocity-averaged self-annihilation cross-section by

$$\langle \sigma_A v \rangle_{90} = \frac{8\pi m_\chi^2}{R_{\mathrm{SC}} \rho_{\mathrm{SC}}^2} \frac{1}{T_{\mathrm{live}}} \frac{1}{\int\int J(\psi) A_{\mathrm{eff}} \frac{\mathrm{d}\mathscr{N}_\nu}{\mathrm{d}E}\, \mathrm{d}E\, \mathrm{d}\Omega} N_{90}, \tag{10}$$

where $T_{\mathrm{live}}$ is the live-time and $A_{\mathrm{eff}}$ is the effective area of the detector, which includes the detection efficiency. The resulting limits for the benchmark annihilation channels $b\bar{b}$, $W^+W^-$, $\mu^+\mu^-$ and $\nu\bar{\nu}$ are shown in Fig. 6. For comparison the limits of halo analyses from IceCube-22 [18] and Fermi [22] are also shown.

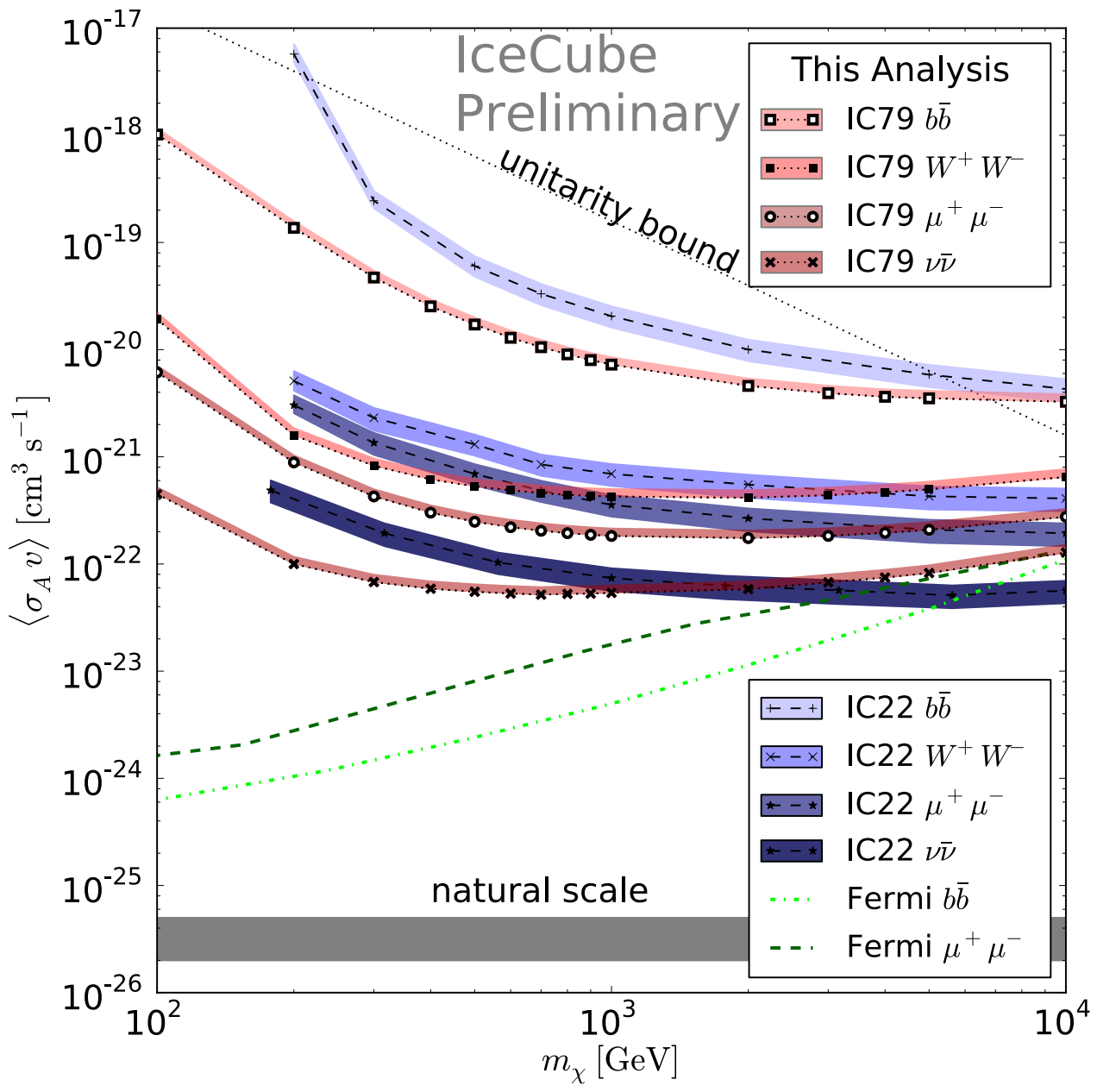

**Figure 6**: Upper limits on velocity-averaged WIMP self-annihilation cross-section from this analysis (IC79) and limits from halo analyses of IceCube-22 [18] (both 90% C.L.) and from the Fermi collaboration [22] ($3\sigma$ C.L.). The limits from Fermi-LAT [22] are scaled by a factor of 2, to account for the different assumption on the local DM density of $0.43\,\mathrm{GeV/cm^3}$. The Fermi analysis region is $\pm 15^\circ$ in Galactic latitude, excluding the central $\pm 5^\circ$. The baseline limit curves are calculated for the NFW profile. The model-dependence (bands) has been estimated from two more extreme cases of a cusped, and a flat-cored halo model. The gray band describes the natural scale if all dark matter consists of WIMPs.

| uncertainty | effect on $\langle\sigma_A v\rangle$ |
|---|---|
| zenith acceptance | $< 1\%$ |
| sky direction exposure | $\pm 1\%$ |
| cosmic ray anisotropy | $\pm 3\%$ |
| efficiency | $\pm 15\%$ |

**Table 1**: Systematic uncertainties

## 7 Systematic Uncertainties

The presented limits depend on the assumed halo density model. We used the NFW halo model as a benchmark and compare to the profiles by Moore [9] and Burkert [10] (bands in Fig. 6). The model dependency of $+12\%/-1\%$ on $\langle\sigma_A v\rangle$ is much smaller than for galactic center searches, because typical differences in the halo models are smaller for the outer halo. Also the uncertainty is reduced with respect to the previous IceCube analyses.

Systematic uncertainties resulting from the analysis method itself are listed in Table 1. For a 5% flatter or steeper zenith acceptance spectrum the value of $N_{90}$, and therefore of $\langle\sigma_A v\rangle$, increases by less than 1%. This small dependency is a result of omitting coefficients which depend on declination. A systematic uncertainty of the order of 3% on $N_{90}$ arises from a potential anisotropy of cosmic rays and the time-dependent exposure of the detector. A further systematic uncertainty is the effective area of the experiment. Our preliminary estimate based on the energy range of this analysis is $\pm 15\%$.

## 8 Conclusion

We present a search for self-annihilation of dark matter with the partially instrumented IceCube detector in its 79-string configuration based on a multipole expansion of the measured neutrino arrival directions. We search for a large-scale anisotropy in neutrino arrival directions from the Northern hemisphere, which corresponds to the outer Galactic Halo. The observed test statistic value is compatible with the null-hypothesis resulting in a limit on the velocity averaged self-annihilation cross-section for $\chi\chi \to b\bar{b}, W^+W^-, \mu^+\mu^-, \nu\bar{\nu}$ of the order of $10^{-18}\,\mathrm{cm^3/s}$ to $10^{-22}\,\mathrm{cm^3/s}$, assuming the NFW profile. Uncertainties in the detector acceptance have been largely reduced by obmitting expansion coefficients that depend on the zenith angle. Due to these smaller systematic uncertainties, as well as a larger detection efficiency the resulting sensitivity is stronger with respect to previous studies [18], especially in the low-mass region. However, due to the measured overfluctuation in this analysis the limits are about a factor two less constrained compared to the sensitivity.

# Earth WIMP searches with IceCube

THE ICECUBE COLLABORATION[1]

[1] *See special section in these proceedings*

*jkunnen@vub.ac.be*

**Abstract:** Many particle physics models predict massive and stable new particles that can solve the dark matter problem. If these particles are Weakly Interacting Massive Particles (WIMPs), they can be trapped in massive celestial bodies such as the Earth, where they may self-annihilate. The annihilation of these WIMPs into standard model particles can create neutrinos. To search for such neutrino fluxes, large scale neutrino telescopes, such as the cubic kilometre IceCube Neutrino Observatory located at the South Pole, can be used. Recent calculations indicate that the dark matter annihilation rate in the centre of the Earth, and thus the resulting neutrino flux could be within reach of a large neutrino detector. We will present the current status of the first search for neutrinos from WIMP annihilations in the centre of the Earth with the IceCube neutrino detector.

**Corresponding authors:** Jan Kunnen[1]
[1] *Vrije Universiteit Brussel, Belgium*

## 1 Introduction

Some of the most promising dark matter candidates are Weakly Interacting Massive Particles (WIMPs). In the Minimally Supersymmetric Standard Model (MSSM), the WIMP can take the form of the lightest neutralino. Dark matter particles from the galactic halo become bound in the gravitational potential of the solar system as it passes through the galaxy. These particles may then scatter weakly on nuclei in the Earth and lose energy, becoming trapped by the Earth. Over time, this leads to an accumulation of dark matter in the centre of the Earth. If the accumulated dark matter density reaches a high enough value, it may then self annihilate, generating a flux of neutrinos which is spectrally dependent on the annihilation channel and neutralino mass.

Expected neutrino event rates and energies depend on the specific model and distribution of dark matter under consideration and the chemical composition of the Earth. Taking these variables into consideration leads to a neutrino-induced muon flux from the centre of the Earth varying between $10^{-8} - 10^{5}$ per km$^2$ per year for WIMPs with masses in the GeV$-$TeV range [1] (Fig. 2), as will be discussed in Section 3. AMANDA [2] and Super-K [3] already ruled out muon fluxes above $\sim 10^3$ per km$^2$ per year. Still, the possibility of setting even more restrictive limits due to the increased size of the IceCube neutrino observatory (described in Section 2) with respect to previous detectors, motivates the continuation of searches for neutrinos coming from WIMP annihilations in the centre of the Earth. The analysis method is presented in Section 4.

## 2 The IceCube Neutrino Telescope

IceCube is located in the glacial ice at the geographic South Pole. It consists of an array of digital optical modules (DOMs), designed to collect the Cherenkov radiation produced by high energy, neutrino-induced charged leptons traveling through the detector volume. By recording the number of Cherenkov photons and their arrival times, the direction and energy of the charged lepton, and consequently that of the parent neutrino, may be reconstructed.

IceCube consists of 86 strings, each containing 60 DOMs, deployed between 1450 m and 2450 m in the ice [4]. Of these 86, 8 strings at the centre of IceCube comprise the DeepCore subarray, consisting of more densely instrumented strings and DOMs with higher light collection efficiency. This configuration lowers the energy threshold of IceCube from 100 GeV to 10 GeV. In total, approximately 1 km$^3$ is instrumented.

While the large ice overburden provides a shield against downward going, cosmic ray induced muons with energies $\lesssim$ 500 GeV at the surface, most analyses also use the Earth as a filter and focus on upward going neutrinos. Additionally, low energy analyses mainly focus on DeepCore as a fiducial volume and use the surrounding IceCube strings as an active veto to reduce penetrating muon backgrounds. Further background reduction is achieved by cuts based on topological signatures that differentiate well-reconstructed upgoing neutrino-induced muons from misreconstructed downgoing background muons.

## 3 Dark Matter from the centre of the Earth

WIMPs accumulated in the centre of the Earth will produce a unique signature in IceCube as vertically upgoing muons. The number of detected neutrino-induced muons depends on the neutralino annihilation rate $\Gamma_A$. If the capture rate $C$ is constant in time $t$, $\Gamma_A$ is given by [1]

$$\Gamma_A = \frac{C}{2}\tanh^2\left(\frac{t}{\tau}\right), \ \tau = (CC_A)^{-1/2}. \quad (1)$$

So the equilibrium time $\tau$ is defined by the time it takes for the annihilation rate and the capture rate to be in equilibrium, where $C_A$ is a constant depending on the WIMP number density. For the Earth, this equilibrium time is of the order of $10^{11}$ years if the spin-independent WIMP-proton cross section is $\sigma_p^{SI} \sim 10^{-43}$ cm$^2$ [5]. Since the age of the solar system is $t_\circ \approx 4.5$ Gyr, $t_\circ/\tau \ll 1$, such that $\Gamma \propto C^2$, i.e. the higher the capture rate, the higher the annihilation rate and thus the muon flux.

The rate at which WIMPs are captured in the Earth depends on the mass and the velocity of the WIMPs. If the WIMP mass is nearly identical to that of one of the nuclear species in the Earth, the capture rate will increase considerably, as is shown in Fig. 1. It should be noted that recent direct detection limits [7] exclude cross sections smaller than $\sigma_p^{SI} = 10^{-43}$ cm$^2$ over a wide range of WIMP masses. This implies that the normalization in Fig. 1 will be about an order of magnitude lower, as the cross section that is assumed in the calculation for the capture rate is $\sigma = 10^{-42}$ cm$^2$ [6].

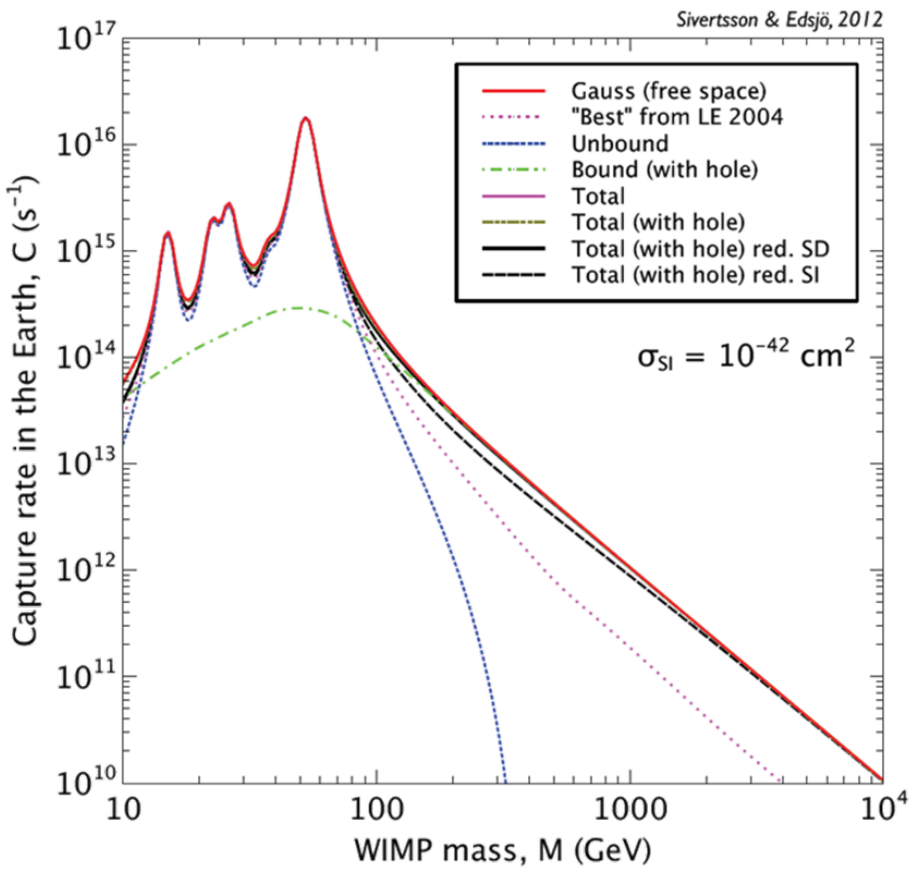

**Figure 1**: This figure shows the rate at which dark matter particles are captured to the interior of the Earth, for a scattering cross section of $\sigma = 10^{-42}$ cm$^2$. The peaks correspond to resonant capture on the most abundant elements considered in the Earth model, $^{16}$O, $^{24}$Mg, $^{28}$Si and $^{56}$Fe. A dark matter halo density of $\rho_\chi = 0.3$ GeV cm$^{-3}$ is assumed [6].

The capture rate could be higher if the WIMPs have a low velocity with respect to the Earth. This is because the escape velocity $v_{esc}$ of WIMPs inside the Earth is low. It varies from $v_{esc} \sim 11$ km/s at the mantle to $v_{esc} \sim 15$ km/s at the centre. Thus, the higher speed WIMPs will only be captured at the centre where the escape velocity is the highest, whereas the lowest speed WIMPs may be captured anywhere in the Earth.

The WIMP speed distribution is very sensitive to theoretical assumptions, as different models for the halo lead to different WIMP speed distributions. The most popular halo model is the Standard Halo Model (SHM), which is modelled as a smooth, spherically symmetric density component with a non-rotating Gaussian velocity distribution. Galaxy formation simulations including baryons indicate that there is at least one local macrostructural component beyond the SHM. These simulations show that a thick disc of dark matter, with the kinematics similar to the thick disc of stars and a mid-plane density of $0.25-1.5$ times the local dark halo density [8, 9], is formed as the baryonic disc of the Milky Way draws satellites closer to the disc plane by dynamical friction, where they are disrupted by tides [10]. The particles in this dark disc have a lower relative velocity to the Earth compared to the SHM. Therefore, less scattering is needed for the particles in the dark disc to become gravitationally captured. So the dark disk leads to a higher capture rate and thus a higher neutrino-induced muon flux at detectors such as the IceCube neutrino telescope.

Fig. 2 shows the expected [1] muon flux ($\Phi_\mu$) for $E_\mu >$ 1 GeV at the Earth's surface as a function of the WIMP mass ($M_\chi$) from neutrinos originating in the Earth. Compared to the flux from the SHM given in the left panel, the flux from the dark disc (right panel) is boosted by two to three orders of magnitude, depending on the specific model.

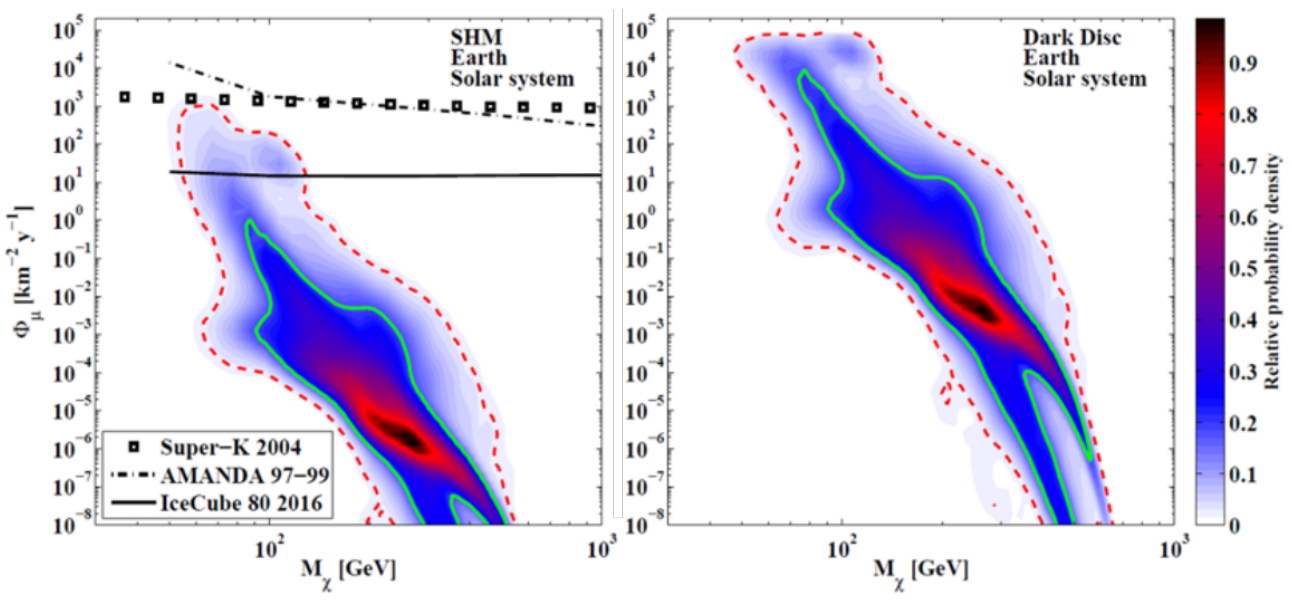

**Figure 2**: Expected muon flux at the surface of the Earth as a function of the WIMP mass for the SHM (left) and a dark disc model (right) [1].

## 4 Analysis

The last published Earth WIMP search was performed with the neutrino data recorded by the AMANDA detector over the three year period $1997-99$ [2]. No excess above the expected atmospheric neutrino background was found in this search, so an upper limit could be set on the muon flux (dashed-dotted line in Fig. 2).

A preliminary estimation of the sensitivity of the full IceCube detector was presented in [11] and is shown in Fig. 2 as the solid line. However, this estimation was performed before the construction of IceCube was started. It does not include the DeepCore array and, lacking a precise simulation of the hardware, is based on a simplified simulation of the detector response. The analysis outlined in this paper corrects all these shortcomings by using data taken with the full detector and the current state of the art IceCube simulation and data analysis techniques.

As model predictions for WIMP capture in the Earth favour low masses, the analysis will be optimised for low mass WIMPs ($m_\chi = 50$ GeV, see Fig. 1) that annihilate into $\tau^+\tau^-$. In this annihilation channel, a hard neutrino energy spectrum is produced. The expected neutrino spectrum (solid line) and the neutrino-induced muon spectrum (dashed line) for 50 GeV WIMPs annihilating into $\tau^+\tau^-$ are shown in Fig. 3. As the expected muon energy for this channel is lower than 50 GeV, the DeepCore detector will be crucial in a search for such a signal.

The signal simulations that are used in the analysis are performed using WimpSim [12]. WimpSim is a code that calculates the annihilation of WIMPs inside the Sun or the Earth, collects all the neutrinos that emerge and let these propagate out of the Sun/Earth to the detector. The code includes neutrino interactions and neutrino oscillations in a fully consistent three-flavour treatment.

The neutrino-induced muon flux in IceCube, coming from WIMPs in the centre of the Earth can not be higher than $\sim 10^3$ muons per year. This is a very low rate, compared to the 2.5 kHz rate at which IceCube is taking data. The data is dominated by atmospheric muons, which are background events, so selection cuts have to be developed to remove these events. To get to neutrino level (mHz rate), the data rate has to be lowered by ~6 orders of magnitude.

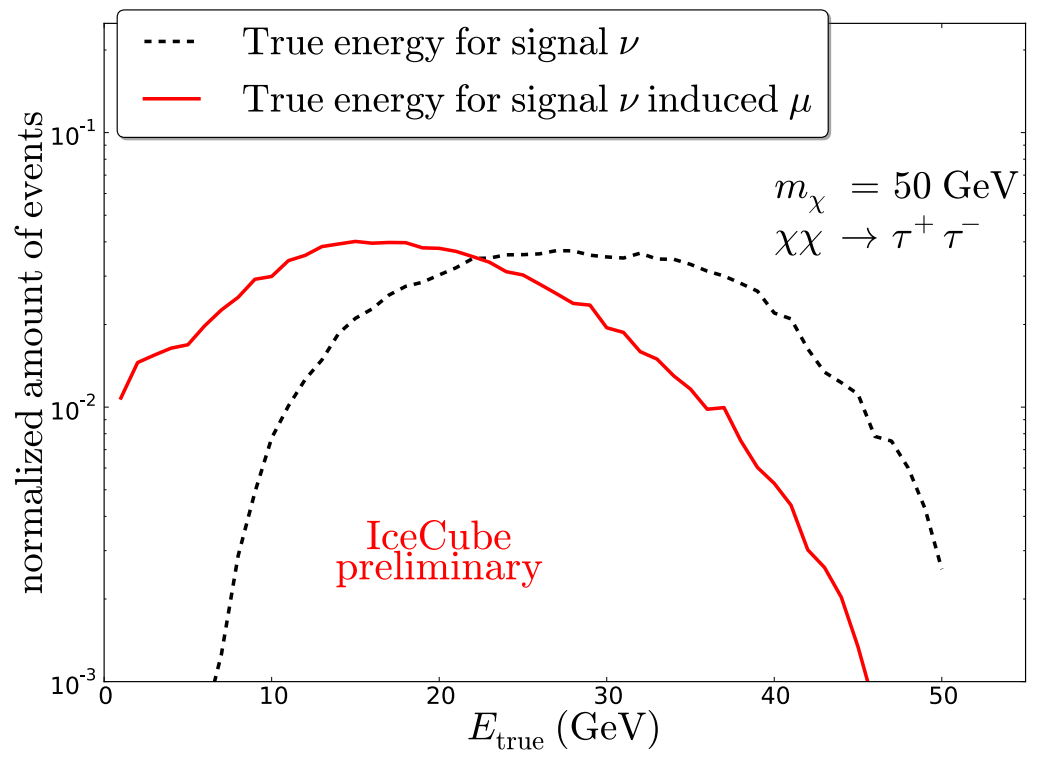

**Figure 3**: The normalized energy distributions for signal neutrinos and signal neutrino-induced muons, for a WIMP mass of 50 GeV.

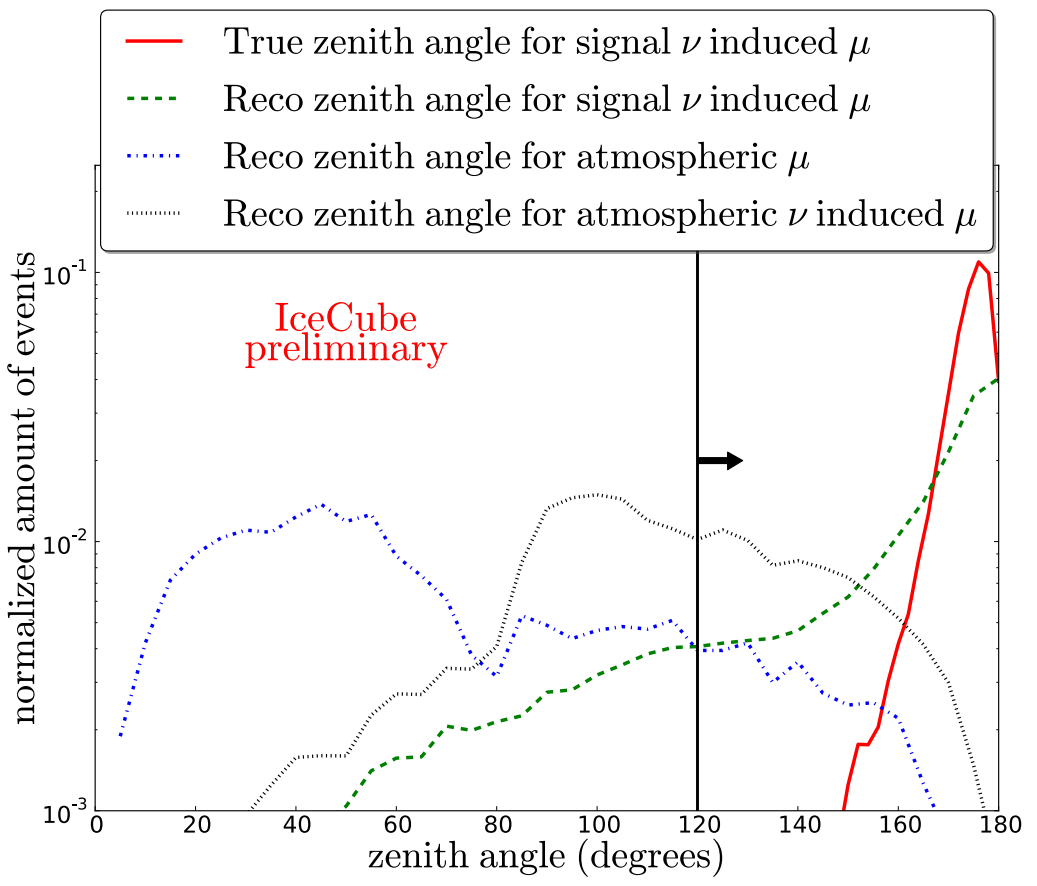

**Figure 4**: The normalized zenith angle distributions for signal (calculated with WimpSim for a 50 GeV Earth WIMP), background muons (simulated with CORSIKA) and neutrinos (simulated with NuGeN). The vertical line shows the cut value used in this analysis and the arrow indicates the parameter region that is retained.

IceCube has several online filters, that are used to *tag* different types of events. In a first step, we only select events that are tagged by online filters that look for upgoing tracks (one of the online filters that is running mainly for this analysis, optimised to tag low energy, upgoing events). After this online filter selection, the event rate is reduced to 100 Hz, while almost 100% of the signal events that triggered the detector are kept.

In a next step, linear cuts are placed to reduce the content of atmospheric muon events to ~1 Hz. These cuts are based on distributions of event multiplicities and observables from signal and background simulations.

One very straightforward selection to make is a cut on the reconstructed zenith angle. Fig. 4 shows the simulated reconstructed zenith angle distributions for signal (dashed line), atmospheric muons (dashed-dotted line) and atmospheric neutrino-induced muons (dotted line). The solid line shows the true zenith angle distribution for the signal. As the detector acceptance is zenith dependent it is not possible to define an off-source region in this analysis. We thus need to rely on simulations to understand the background, requiring a detailed understanding of systematic uncertainties. A *control region* has to be defined in which we can compare the background simulation with the data. Therefore the zenith angle cut can not be too strict, as we aim to retain a signal free control region. Based on these considerations, the zenith angle cut is chosen such that all events with a reconstructed zenith angle $< 120^\circ$ are removed.

Signal neutrino-induced muons and atmospheric muons do not only differ in direction, but also the visible starting point of the muon tracks are expected to be different. The atmospheric muons always start outside the detector (in the atmosphere) whereas muons from neutrinos can start in the detector. The positions of the starting point of the muon tracks may thus be used as a cut variable. The starting points are reconstructed using an algorithm that tries to find the range of the muon assuming no secondary processes along its path such that we only consider Cherenkov light emitted by the muon. The first step of the algorithm is to look for start and stop points by projecting all the hits (after suspected noise hits have been removed) in the event onto a given seed track. Then it calculates the probability that a DOM in the direction of this seed track is not hit by taking all non-hit DOMs within a certain radius and calculating the probability of having such an amount of non-hit DOMs in the assumption of an infinite track. The next step is to recalculate this probability, but this time with the assumption that the track has a start and/or stop point within the detector volume. The starting point of the muon is then calculated by minimizing the likelihood ratio of the two calculated probabilities.

From Fig. 5, it is clear that this reconstructed starting point is mainly concentrated in the lower central part (DeepCore) for signal events (left panel), while atmospheric muons (right panel) mainly have a reconstructed starting point at the top and side of the detector. This is because the signal is low energetic (see Fig. 3) and will thus mainly be detected by the DeepCore detector. Events with a track that has a reconstructed starting point inside the IceCube volume but well outside the DeepCore volume, either to the side ($r_{\rm vertex} > 490$ m) or above ($z_{\rm vertex} > 180$ m), are removed.

Another variable that is used for cutting is the number of direct hits in the event divided by the direct length ($N_{\rm Dir}/L_{\rm Dir}$, see Fig. 6). Direct hits are hits that are within a time residual $t_{res}$ of (-15,+25) ns according to the reconstructed track. $t_{res}$ is defined by the difference between the measured time and the expected time of the hit, where the expected time of the hit is the time it takes for a Cherenkov photon to travel the same distance without being scattered. So photons with a small $t_{res}$ have not scattered much.

Since the signal consists of vertical tracks parallel to the strings, these events should have more direct hits $N_{\rm Dir}$ than background events traversing the detector in any other direction. Also, the direct length $L_{\rm Dir}$, i.e. the distance between the first and the last direct hit, of a signal event should be small, as the signal muons have a low energy. For atmospheric muons, $N_{\rm Dir}$ and $L_{\rm Dir}$ have a broad distribution, depending on the energy and the direction of the incoming muon. Only events with $0.03\ \rm m^{-1} < N_{Dir}/L_{Dir} < 0.09\ m^{-1}$ are kept.

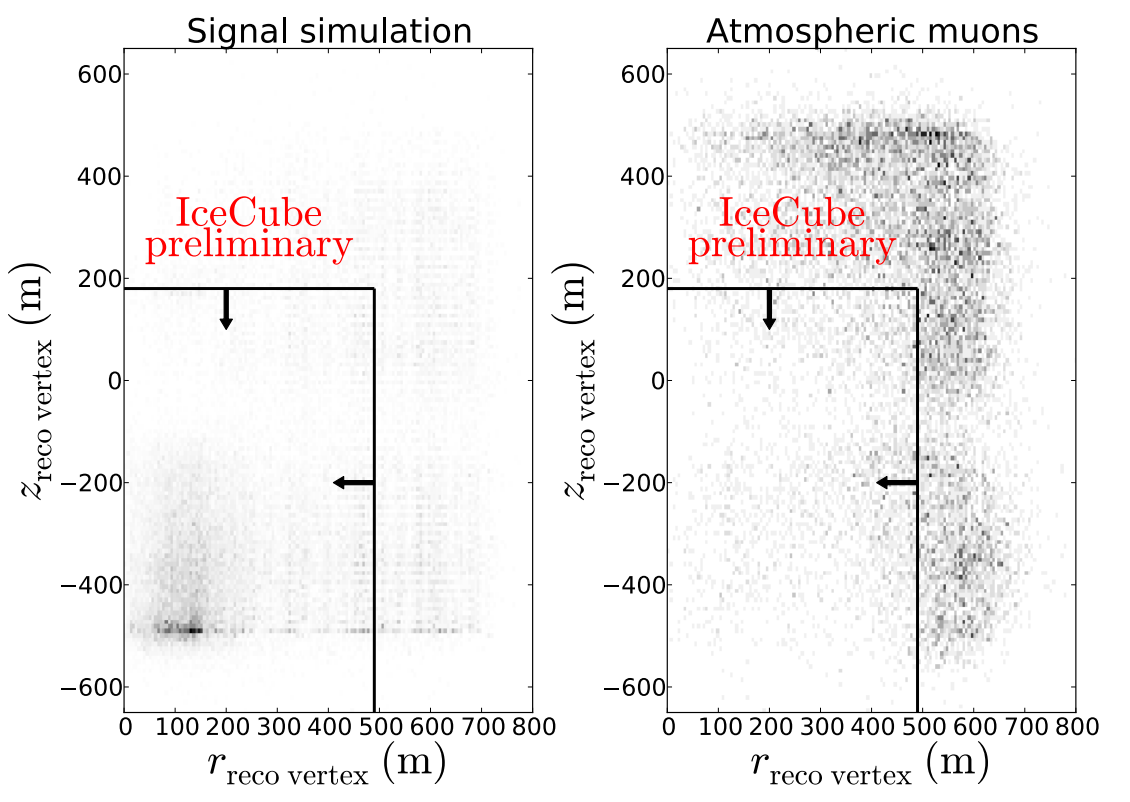

**Figure 5**: The reconstructed starting point projected in the $r-z$ plane of the detector, with $r = \sqrt{x^2+y^2}$, i.e. the distance to the central axis of the detector in the $x-y$ plane. The point density indicates the event rate. The vertical and horizontal lines show the cut values used in this analysis and the arrows indicate the parameter regions that are retained.

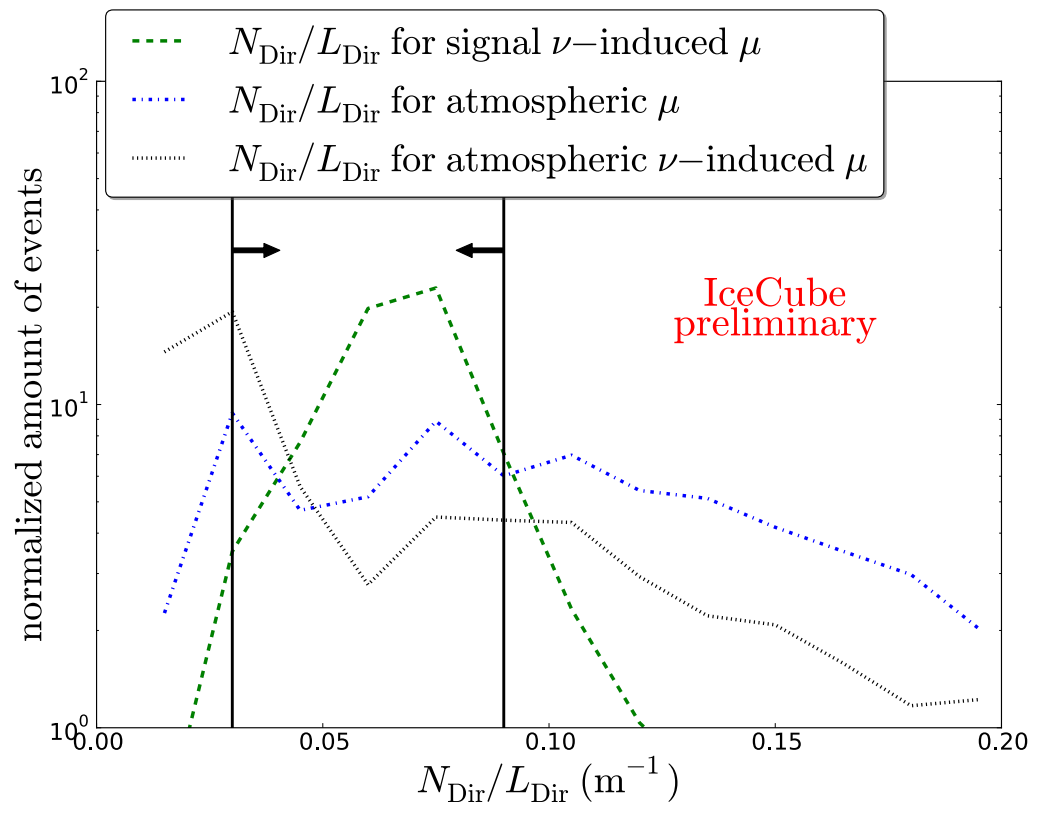

**Figure 6**: The normalized distributions for $N_{\mathrm{Dir}}/L_{\mathrm{Dir}}$. The vertical lines show the cut values used in this analysis and the arrows indicate the parameter region that is retained.

The last variable that is used at this level is $z_{\mathrm{travel}}$. This is the average drift of hits in the vertical ($z$) direction

$$z_{\mathrm{travel}} = \sum_{i=1}^{n} \frac{z_i - \langle z_{\mathrm{quartile},1} \rangle}{n}, \qquad (2)$$

where $n$ is the number of fired DOMs, $z_i$ is the $z$ position of the $i^{th}$ DOM and $\langle z_{\mathrm{quartile},1} \rangle$ is the mean $z$ position of the first quartile of hits in time. This should be negative for downgoing tracks, i.e. background, and positive for upgoing tracks, i.e. signal (see Fig. 7). Events with $z_{\mathrm{travel}} < 10$ m are removed.

The cut values (except the cut on the zenith angle) are chosen by looping over all possible combinations of cut values and checking which combination brings down the background rate below 1 Hz, while removing as little signal as possible. After this first cut level, the data rate goes down to ~0.8 Hz, while 20-40% of the signal (depending on WIMP mass and channel) is kept. The data is still dominated by atmospheric muons at this level.

Now that the data rate is below 1 Hz, we can reprocess the data using more precise (and more time-consuming) reconstructions. This is important as a good angular resolution is needed to identify tracks that come from the centre of the Earth. These reconstructions will be used to classify events as signal or background like with a boosted decision tree (BDT) from the TMVA software package [13]. We expect to reach the neutrino level (~mHz) after this BDT cut.

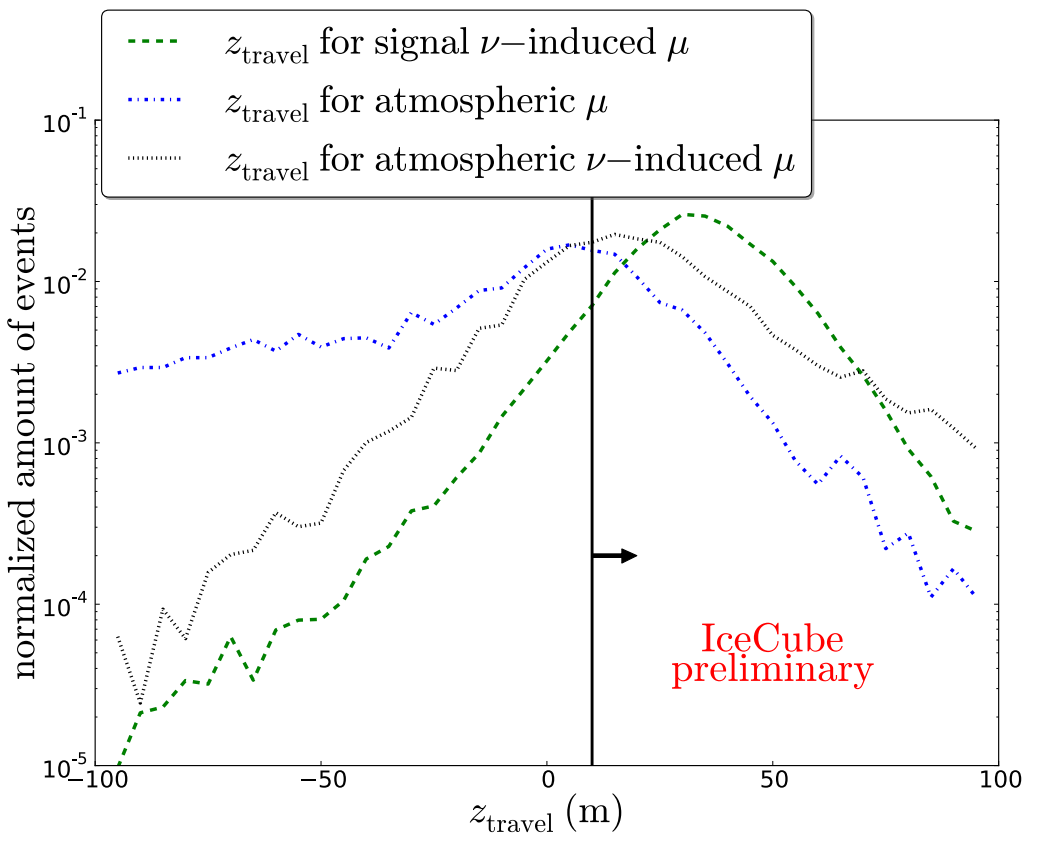

**Figure 7**: The normalized distributions for $z_{\mathrm{travel}}$. The vertical line shows the cut value used in this analysis and the arrow indicates the parameter region that is retained.

Once the atmospheric muon background has been removed, we will search for signal events in the final data set which is dominated by atmospheric neutrinos. These neutrinos come from all over the sky, while the expected signal neutrinos come from the direction of the centre of the Earth. We will use the space angle $\psi$ between the source direction (i.e. the centre of the Earth) and that of the reconstructed track to evaluate the number of signal events $\mu_s$ (similar to the method used in the search for dark matter annihilations in the Sun with the 79-string IceCube detector [14]). For signal, this space angle is expected to be close to zero, while for background, this space angle will be evenly distributed over the sky. By doing a statistical analysis on the space angle distribution, one can either detect the presence of signal neutrinos (a peak from the centre of the Earth), or set an upper limit on the dark matter induced neutrino flux.

# Extending IceCube Low Energy Neutrino Searches for Dark Matter with DeepCore

THE ICECUBE COLLABORATION[1],

[1] *See special section in these proceedings*

*samuel.flis@fysik.su.se*

**Abstract:** The cubic-kilometer sized IceCube neutrino observatory in the glacial ice at the South Pole offers new opportunities for low-energy neutrino physics and astrophysics, in particular for indirect dark matter searches. An efficient veto against atmospheric muons at an early stage of an analysis is an important tool for background rejection. For low energies, this can be achieved by using the DeepCore in-fill array of IceCube as a fiducial volume and the surrounding IceCube detector as an active muon veto. We present newly developed veto techniques for dark matter searches with DeepCore. The effective use of DeepCore and the application of these vetos are discussed, drawing on the example of two recent IceCube indirect dark matter searches for signals from the Sun and the Galactic Center, respectively.

**Corresponding authors:** Samuel Flis[1], Martin Wolf[1], Matthias Danninger[1],
[1] *Oskar Klein Centre and Dept. of Physics, Stockholm University, SE-10691 Stockholm, Sweden*

## 1 Introduction

A flux of neutrinos from Weakly Interacting Massive Particle (WIMP) [1] annihilations may be detected in large neutrino telescopes such as IceCube [2]. IceCube searches for WIMP signals from self-annihilating or decaying dark matter in the galactic halo [3] and Galactic Center [4], as well as for signals from neutrino annihilation from large celestial bodies, such as the Sun [5]. IceCube's in-fill array "DeepCore" [6] increases the low-mass WIMP sensitivity and allows the search for low-energy neutrino source candidates in the southern equatorial sky. These promising analysis prospects present new challenges to reduce the down-going atmospheric muon background. Here we report on the technical details of new analysis and muon-veto techniques that have been developed for dark matter analyses using DeepCore data.

IceCube is a cubic-kilometer neutrino detector installed in the ice at the geographic South Pole between depths of 1450 m and 2450 m. Detector construction started in 2005 and finished in 2010. Neutrino reconstruction relies on the optical detection of Cherenkov radiation, via digital optical modules (DOMs), emitted by secondary particles produced in neutrino interactions in the surrounding ice or the nearby bedrock. The completed IceCube detector consists of 86 strings with 60 DOMs each. 78 strings have a horizontal spacing of 125 m and a vertical spacing between DOMs of 17 m. In addition, the detector contains 8 strings optimized for low energies that are clustered around the center most string of IceCube. These strings feature DOMs with higher quantum efficiency photo multiplier tubes and have reduced vertical DOM spacing. The spacings are 10 m for the uppermost 10 DOMs (*DC-veto region*) and 7 m for the remaining 50 DOMs (*DC-fiducial region*). Both regions are separated by the main dust layer, which is located at a depth of about 2050 m. The horizontal distance between DeepCore strings is less than 75 m. Together with the 12 adjacent IceCube strings they form the DeepCore subarray that is optimized for neutrino energies below 100 GeV.

The work described here uses the 79-string configuration of IceCube (see figure 1). During this data taking period, DeepCore consisted of 6 densely instrumented strings together with 7 surrounding standard IceCube strings.

## 2 Methods for Efficient Event Selections

IceCube is optimized for the detection of high energy neutrinos from a few hundred GeV up to several PeV. DeepCore increases IceCube's sensitivity for neutrinos below 100 GeV and allows the detection of neutrinos with energies down to the order of 10 GeV. Newly developed veto methods against atmospheric muons, that are described here, significantly improve the sensitivities of analyses focused on low-energy events in the Southern Hemisphere. This turns IceCube into an efficient $4\pi$ detector for indirect dark matter searches.

Several methods for efficient background rejection and signal selection have been developed in order to achieve this goal. These methods include energy dependent event selection splitting, veto techniques against atmospheric muon background, and the selection of neutrino induced muon tracks starting inside a fiducial volume. In order to apply these techniques, the detector volume is divided into a fiducial and a veto volume. For the 79-string configuration of IceCube the fiducial region is chosen as the DC-fiducial volume, as shown in figure 1.

### 2.1 Event Selection Splitting

Indirect dark matter analyses with IceCube search for neutrinos originating from WIMPs with masses ranging from $\sim$ 10 GeV to $\sim$ 10 TeV. Within this mass range, signal events can have very different event topologies in the detector. To accommodate all expected event topologies in a single analysis, the full dataset may be split into multiple event selections.

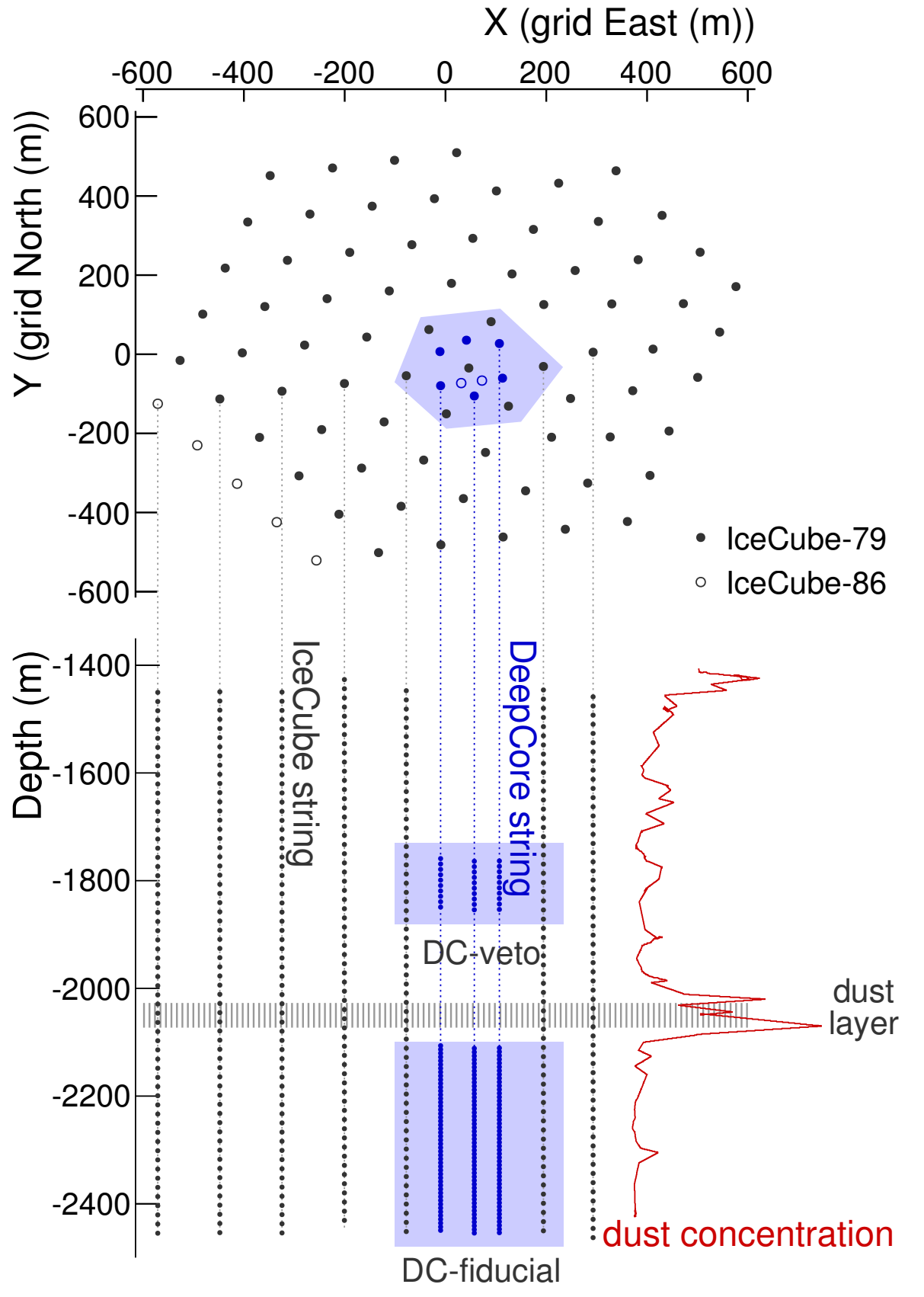

**Figure 1**: Top and side view of the IceCube detector (black) and its DeepCore subarray (blue). Strings constituting the IceCube 79-string configuration (solid) and the last deployed strings completing the IceCube detector (circles) are indicated. Also shown is the main dust layer within the detector and a dust concentration profile as measured with a dust logger [7].

Signal-efficient event selection for low-mass WIMP signals and high-mass WIMP signals is only feasible by imposing different selection criteria. This motivates splitting the event selection into a Low Energy (LE) and a High Energy (HE) event sample, which are treated independently to get the best signal sensitivity for each of the samples. Using DeepCore, the split into a LE and HE event selection can be realized in the following way: events with a larger number of hits inside the DC-fiducial region than outside are assigned to selection LE. Additionally, the number of outside hits must be less than seven. This ensures that events with a long lever arm and therefore good angular resolution are assigned to the HE selection. The step is well motivated by looking at the event topology of the two classes of muon events that are selected for each event sample. The HE selection, defined as the complement of LE, contains high energy events with no containment requirement. Associated muon tracks have accurate track reconstructions and good quality parameters. This is used to separate true up-going muon-like events from mis-reconstructed atmospheric muons, which will exhibit lower quality parameter values. The LE selection consists of shorter track-like events with, in general, lower track reconstruction quality, but good containment. In this context, creating two independent data samples by splitting the original data set is an obvious choice for a hybrid detector. Figure 2 shows the clear separation into a low and high energy event sample around 100 GeV for atmospheric neutrinos after applying the split criteria.

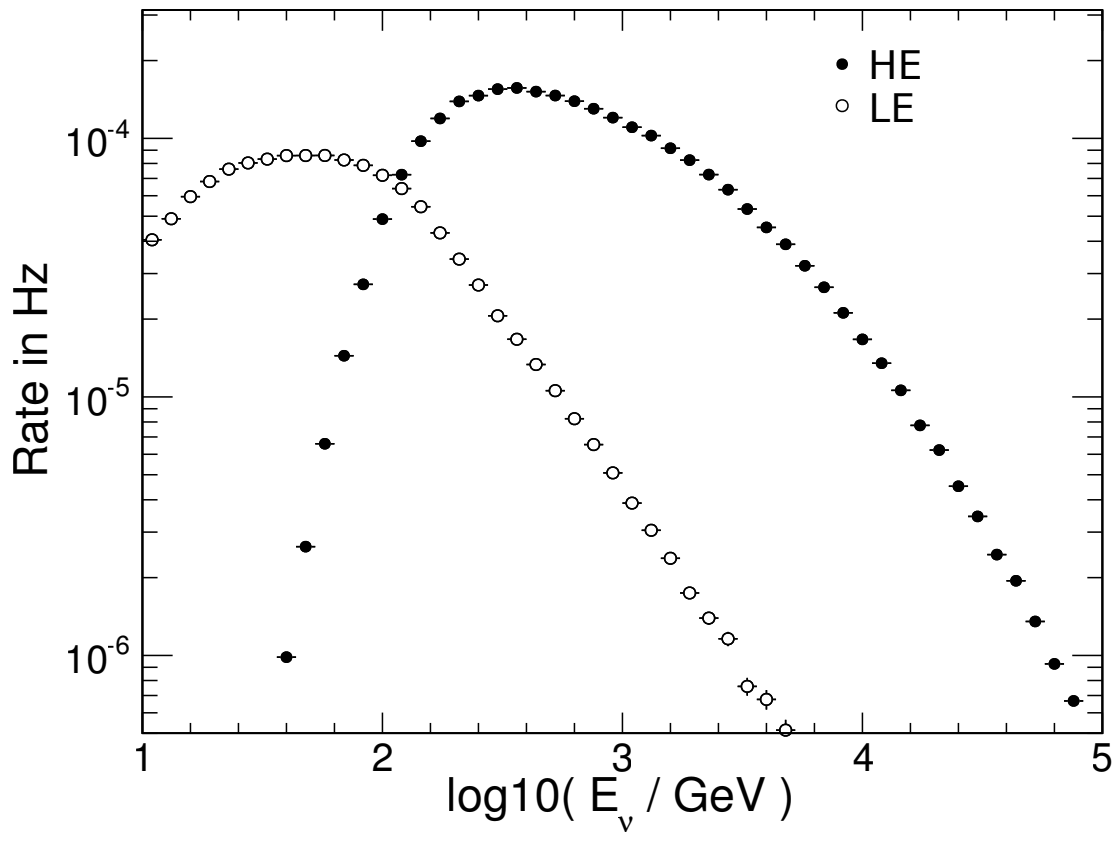

**Figure 2**: True neutrino energy ($E_\nu$) distributions from simulation for the atmospheric neutrino event rate at cut level 4 of [5]. The low-energy event selection (LE) and high-energy event selection (HE) show a clear separation in $E_\nu$ around 100 GeV.

### 2.2 Atmospheric Muon Vetos

IceCube observes a continuos stream of atmospheric muon, which are observed as down-going events and are the main background for low energy neutrino full-sky IceCube analyses. A good way to veto down-going atmospheric muons is to search for neutrino induced muon tracks starting within a fiducial volume. In this section, we discuss several such veto methods that have been developed and applied in recent dark matter searches with the IceCube-79 string configuration, detailed in Refs. [4, 5]. Events detected in IceCube consist of recorded photo multiplier tube responses from DOMs that triggered - *hit DOMs*. In the context of this article a hit DOM is denoted simply as a hit. Most reconstructions and variables used for analyses are computed using a subset of hits that has undergone cleaning to reduce the number of random hits. These subsets are defined by different cleaning algorithms based on time windows within the event and requirements on the topology of the hits. Among all veto methods to be described below, only the first uses a subset of hits, while later described veto methods consider all hits in an event.

**Veto I:** First, an attempt is made to classify an event as starting track inside the fiducial volume, based on the hit times only. This simple veto does not rely on a track reconstruction and is defined in the following way: the first $n$ hits of an event are required to occur inside the fiducial volume. By adapting $n$ to the chosen fiducial volume, backscattering from a starting event into the veto volume is accounted for. For the DC-fiducial region, defined above, $n = 4$ is found to be optimal.

**Veto II:** A second veto method, *RTVeto*, aims to tag atmospheric muons within the veto region. Faint muon tracks penetrating the detector may be identified by the sporadic clusters of hits they leave when traversing the veto region. These clusters of hits are likely to be causally connected. For a given DOM, the algorithm checks whether there are other hits within a specific time and distance in space (RT-condition). If a hit is found that fulfills the RT-condition[1], it is added to the kept set of DOMs (cluster). Furthermore, each added DOM in the cluster is also checked for additional hits that fulfill the RT-condition. We test each hit DOM in the veto region, which has a hit prior to the time of the first fiducial hit, against this condition. If a cluster of at least three hits is found that fulfill the RT-condition (an RT-cluster), the event is rejected. This cut rejects more than 80% of atmospheric muons while keeping 90% of signal neutrino events starting within the fiducial region. This is illustrated in figure 3, which shows the distribution of RT-cluster sizes for data, muon background, and signal.

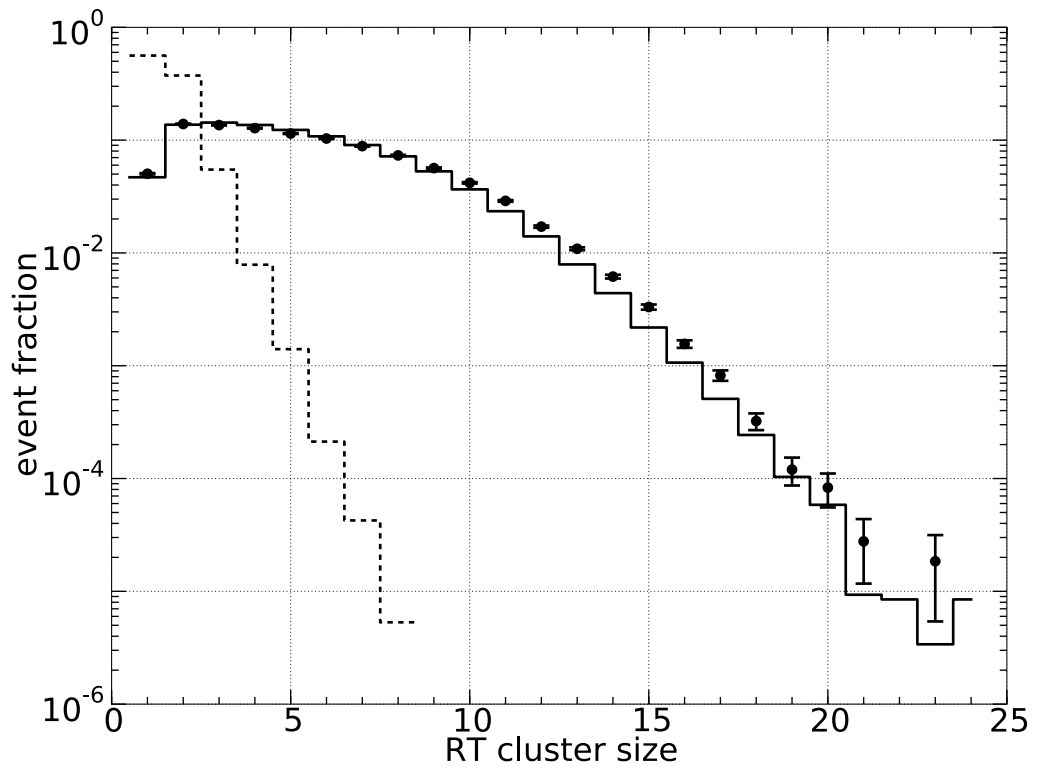

**Figure 3**: The size of the RT clusters found by RTVeto for atmospheric muon background simulation (solid line) and signal simulation (dashed line). Data is show in black dots.

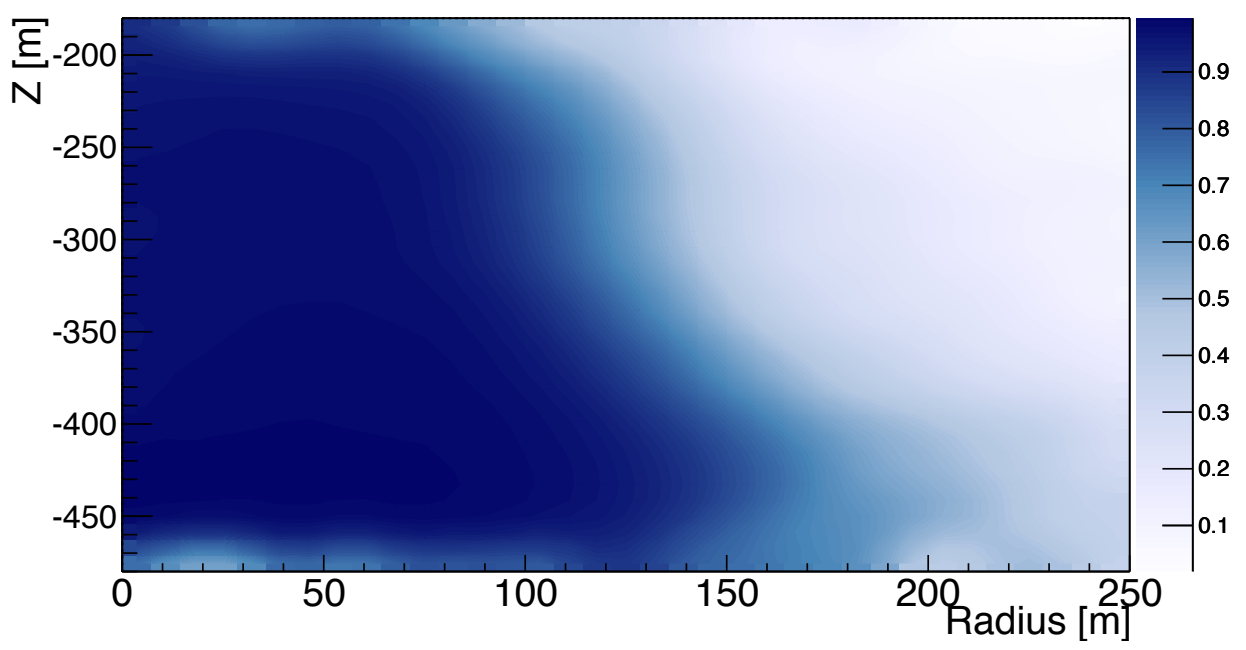

**Figure 4**: Fraction of signal per bin as a function of radial distance and $z$, defined as $f_s/(f_s+f_b)$. $f_s$ is the two dimensional probability density function of signal in this plane. $f_b$ is the corresponding probability density function for background, determined from data.

**Veto III:** The third veto method is based on the vertical position $z$ of the first hit and the radial distance in the $xy$-plane of the reconstructed vertex position[2]. The vertex is determined by a reconstruction that uses the likelihood of a muon to produce or not produce hits in the DOMs along the reconstructed track up-stream of an assumed starting point. Maximizing this likelihood gives the location of the interaction vertex. A cut can then be applied to reject background events with high $z$ positions and a reconstructed vertex at the periphery of the fiducial region. This veto, applied in sequence after the previously described veto methods, removes more than 70% background while retaining 90% signal. Figure 4 shows the fraction of signal per bin as a function of radial distance and $z$. The signal fraction is calculated as $f_s/(f_s+f_b)$, where $f_s$ and $f_b$ are the two dimensional probability density functions of signal and background in this plane, respectively.

**Veto IV:** Like the previous veto method, the *ConeHitsVeto* relies on a reconstructed track hypothesis and vertex. In this algorithm, a cone is defined around the reconstructed track direction beginning at the reconstructed staring point of the track. The cone is searched for the presence of hits at times earlier with respect to the reconstructed starting time of the track. The presence of too many hits, $n_{\text{cone}}$, inside the cone is evidence of a faint muon, and these events are rejected. To avoid events rejected due to random hits, $n_{\text{cone}}$ is optimized with respect to the opening angle, time-window and the veto region. Since the optimal opening angle is in most cases found to be rather small (typically 20° - 30°), this veto method relies to a high degree on an accurate track reconstruction to be efficient.

**Veto V:** There are two classes of faint incoming muon tracks that, while still producing hits in the veto region, are extremely hard to identify with the veto techniques discussed above. Either the incoming muon produces fewer hits along its path in the veto region than the threshold value of the veto or the hits are too far apart and at some distance from the muon track. To examine suspect hits in the veto region the likelihood for the reconstructed track hypothesis is computed given the hits in the veto region as input. The value of the likelihood indicates how likely the veto hits are compatible with the reconstructed track hypothesis and thus how likely the hits are associated with an incoming muon track. If the likelihood is normalized correctly by taking into account the number of hits considered for the likelihood calculation a threshold value on the normalized likelihood, independent of the number of hits, can be found to reject incoming muon tracks.

## 3 Application to DM Searches

The veto methods described in the previous sections have been applied by two analyses searching for Dark Matter with WIMP masses in the range of 20 GeV and 10 TeV, using the 79-string configuration of IceCube.

During a recent analysis searching for dark matter in the Galactic Center [4] several of these veto methods were developed. Since the Galactic Center is above the horizon at the South Pole, a big effort was made to reject the down-going muon background. For this purpose the RTVeto (Veto II), the RZVeto (Veto III), and the LHVeto (Veto V) were developed. Vetos I and IV, which have been used in earlier analyses, were also used. The order in which the vetos are presented in section 2.2 is also the order in which they were applied in the analysis. Inspired by the event selection splitting described in section 2.1, two different multivariate cuts were made to define two final event selections. The two event selections result in flatter sensitivity in energy. With these methods sensitivities on the velocity averaged self-annihilation cross-section, $\langle\sigma_A v\rangle$, down to $3\cdot10^{-23}\,\text{cm}^3\text{s}^{-1}$ for a 65 GeV WIMP mass were achieved.

1. For the RT condition a radius of 250 m and a time window of 1 $\mu$s between hits were applied in this work.
2. The vertex reconstruction is described in Ref. [8].

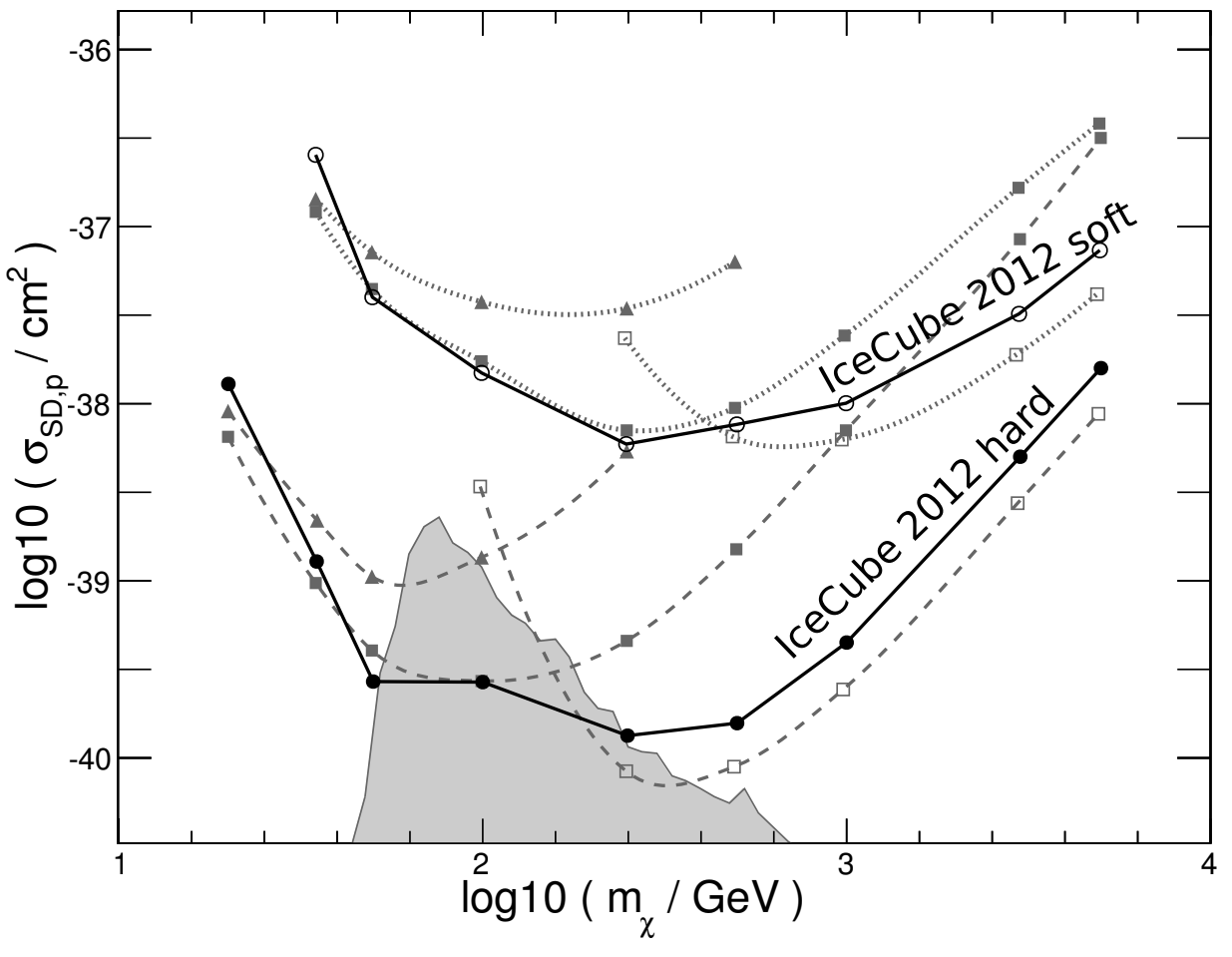

**Figure 5**: Sensitivities on $\sigma_{\mathrm{SI},p}$ from [5]. Hard channels are shown dashed and soft channels are shown dotted. The winter selection HE has empty square markers, the winter LE is shown with filled square markers, and the summer LE has filled triangle markers. The shaded region represents an allowed Minimal Supersymmetric Standard Model (MSSM) parameter space [9].

An earlier analysis searching for Dark Matter in the Sun [5] was able to set the most stringent limits to date on the spin-dependent WIMP proton cross-section for WIMPs annihilating into $W^+W^-$ or $\tau^+\tau^-$ with masses above 35 GeV. It was also the first low energy analysis in IceCube that managed to expand the search to the southern sky and set limits on WIMPs with masses as low as 20 GeV. This was possible due to the development of the Hit-Time veto (Veto I) described above, an early version of the RZVeto (Veto III) and split datasets. For this particular analysis the data was divided into a down-going and an up-going dataset corresponding in time to the summer and winter dataset, respectively, depending on the position of the Sun in the sky at the South Pole. Furthermore, the winter dataset was divided into a HE selection and a LE selection as described in section 2.1 (see also figure 2). Figure 5 illustrates how the different selections compare to each other and also how they contribute to the final limit.

## 4 Conclusions

Introducing veto methods in analyses have been crucial to transform IceCube to a $4\pi$ detector and thus extending the field of view to the southern sky. We have presented veto methods intended for rejecting muon background while retaining low energy starting events inside a fiducial region. The methods have been developed for IceCube and especially to utilize the DeepCore subarray to detect low energy events. Furthermore, the importance of differentiating between different event topologies when developing an analysis has been discussed. Using the mentioned techniques together with DeepCore, two analyses [5, 4] have been able to set limits or computed sensitivities for WIMP masses which would have been beyond reach for IceCube without DeepCore.

Since the two mentioned analyses, using the IceCube-79 string configuration, are the first to apply these techniques, there is room for improvement. In future analyses using IceCube-86 string configuration fine-tuning the veto methods might be necessary due to slight changes in detector geometry.

Neutrino oscillation analyses have successfully used vetos [10] to reject down-going muon background. With efficient veto methods it is also possible to create down-going control regions which are not affected by neutrino oscillations. The veto methods developed for dark matter searches, presented here, might make it possible to create such control regions with an adequate neutrino purity.

# Results and future developments of the search for subrelativistic magnetic monopoles with IceCube

THE ICECUBE COLLABORATION[1]

[1] *See special section in these proceedings*

*emanuel.jacobi@desy.de*

**Abstract:** The IceCube Neutrino Observatory is a large Cherenkov detector integrated into 1 km$^3$ of Antarctic ice. Besides the detection of highly energetic astrophysical neutrinos, the detector can be used to search for signatures of exotic physics. This work is the search for subrelativistic, magnetic monopoles as remnants of the GUT era (**G**rand **U**nified **T**heory) shortly after the Big Bang. These monopoles can be detected by the Cherenkov light from nucleon decays, which are catalyzed via the Rubakov-Callan effect along the trajectory of the monopole. In this paper the results of the analysis of first data taken from May 2011 until May 2012 with a dedicated slow-particle trigger for DeepCore, a subdetector of IceCube, are presented. For the brightest monopoles a different analysis technique is applied. We present the results of a corresponding analysis applied to data taken from May 2009 until May 2010 with the 59-string configuration. In both analyses no monopole signal has been observed and their flux can be constrained to a level three orders of magnitude below the Parker bound. These results improve the current best experimental limits by more than one order of magnitude for a wide parameter space of monopole velocity and catalysis cross section. Since May 2012 a dedicated trigger is running on the full IceCube detector and multiple improvements for future searches for subrelativistic magnetic monopoles have been developed. Based on this the sensitivities of a future analysis are presented.

**Corresponding authors:** M. L. Benabderrahmane[1], E. Jacobi[1], S. Schoenen[2]
[1] *DESY, 15738 Zeuthen, Germany*
[2] *III. Physikalisches Institut B, RWTH Aachen, 52056 Aachen, Germany*

## 1 Introduction

Magnetic Monopoles were first postulated by Paul Dirac in 1931 [1]. By introducing a magnetic source term the Maxwell Equations could be symmetrized. As a direct consequence, Dirac's theory would give an explanation for the quantization of the electric charge.

Grand Unified Theories (GUT) lead to predictions of magnetic monopoles once more. Since a unified theory must include the electroweak theory, a $U(1)$ factor for the electroweak theory must be left over when the unified symmetry was broken. Based upon this fact t'Hooft and Polyakov [2, 3] predicted independently of each other the existence of magnetic monopoles. Those monopoles could have been created as topological defects during the phase transition $10^{-36}$ s after the big bang.

Until now the existence of magnetic monopoles has not been proven experimentally. In the last thirty years many different experiments have been performed in order to detect magnetic monopoles. So far only upper limits on the flux could be set. The best direct flux limit on subrelativistic magnetic monopoles has been published by the MACRO collaboration [4] after 10 years of measurement.

Following the hypothesis of Rubakov [5] and Callan [6] monopoles should catalyse proton decays. This process provides a promising detection channel, especially for subrelativistic monopoles. Water Cherenkov Neutrino Detectors are sensitive to such a signal.

## 2 IceCube detector

With an instrumented volume of about 1 km$^3$ the IceCube Neutrino Observatory is well suited to search for rare particles, such as magnetic monopoles. The IceCube detector consists of 5160 digital optical modules (DOMs), deployed on 86 strings in the clear Antarctic ice at the geographical South Pole at depths between 1500 m and 2500 m. Each DOM consists of a photomultiplier tube, a high voltage generator and signal digitization hardware packed into a pressure glass vessel. The average horizontal spacing between two strings is about 125 m, the vertical spacing of DOMs along a string is 17 m.

Eight strings in the center of the detector have been deployed with a denser spacing of about 75 m between the strings and 7 m DOM spacing along one string. This infill array is the low energy extension of IceCube, named DeepCore.

### 2.1 Signature of subrelativistic magnetic monopoles

The key mechanism to detect subrelativistic, magnetic monopoles in IceCube is the Rubakov-Callan effect. Monopoles themselves, but also secondary particles produced in ionization processes of the surrounding matter, are too slow to emit Cherenkov light while propagating through the detector. The signature of these monopoles are electromagnetic or hadronic cascades from catalyzed nucleon decays along the monopole track. The energy of each cascade, and therefore the number of Cherenkov pho-tons, depends on the decay channel. For the most promising channel $p \rightarrow e^+\pi^0$ an electromagnetic cascade with almost the full proton's rest energy is produced, which results in about $10^4$ optical Cherenkov photons [7]. Therefore, the signature searched for is the Cherenkov light of consecutive cascades with mean distances $\lambda_{cat}$ along the track of a slowly moving particle. The characterizing properties of such subrelativistic, magnetic monopoles are the velocity $\beta$ and the mean free path $\lambda_{cat}$. The corresponding catalysis cross section can be calculated with $\sigma_{cat} = \frac{1}{\lambda_{cat} \cdot n}$, where $n$ is the particle density of the material.

## 2.2 Trigger

The triggers in IceCube were optimized for relativistic particles. Bright monopoles with a cross section above $\sigma = 1.7 \cdot 10^{-23}\,\text{cm}^2$ can still trigger the standard triggers, but the monopole signature may be split into subsequent events.

Since May 2011 a dedicated trigger – called Slow-Particle-Trigger (SLOP trigger) – for the search for subrelativistic magnetic monopoles is part of the IceCube Data Acquisition [8]. In the first year the trigger has been active only on the subdetector DeepCore. Since May 2012 a modified version of the trigger is running on the full 86-string detector.

The purpose of the SLOP trigger is the identification of temporarily isolated local coincidences caused by consecutive nucleon decays along the monopole track, which have to be consistent with a track from a particle moving at constant velocity $\beta \ll 1$. The trigger is based on local coincidences between two neighbouring DOMs (Fig. 1a). Using only temporary isolated coincident pairs will reject muon hits (Fig. 1b) since muons pass the detector within microseconds and therefore will produce several coincident hit pairs within a short time. The remaining pairs that fall into a given time-window are combined into groups of three (triplets) (Fig. 1c). From all possible combinations, only those triplets are kept, which fulfill two criteria that ensure their consistency with a track-like signature: (1) the time difference between the hit pairs has to be consistent with a constant velocity, (2) the hit pairs from the triplets have to be ordered along a straight line (Fig. 1d). Due to different geometries of DeepCore and the full 86-string detector, the implementation of the latter parameter differs between the trigger of the 2011 season and the 2012 season.

Finally, if the number of triplets overlapping in time, $n_{triplet}$, is greater than three (2011) or five (2012) the trigger launches and the complete detector data is read out. The trigger rate is about 2 Hz for the 2011 configuration and 13 Hz for the 2012 configuration.

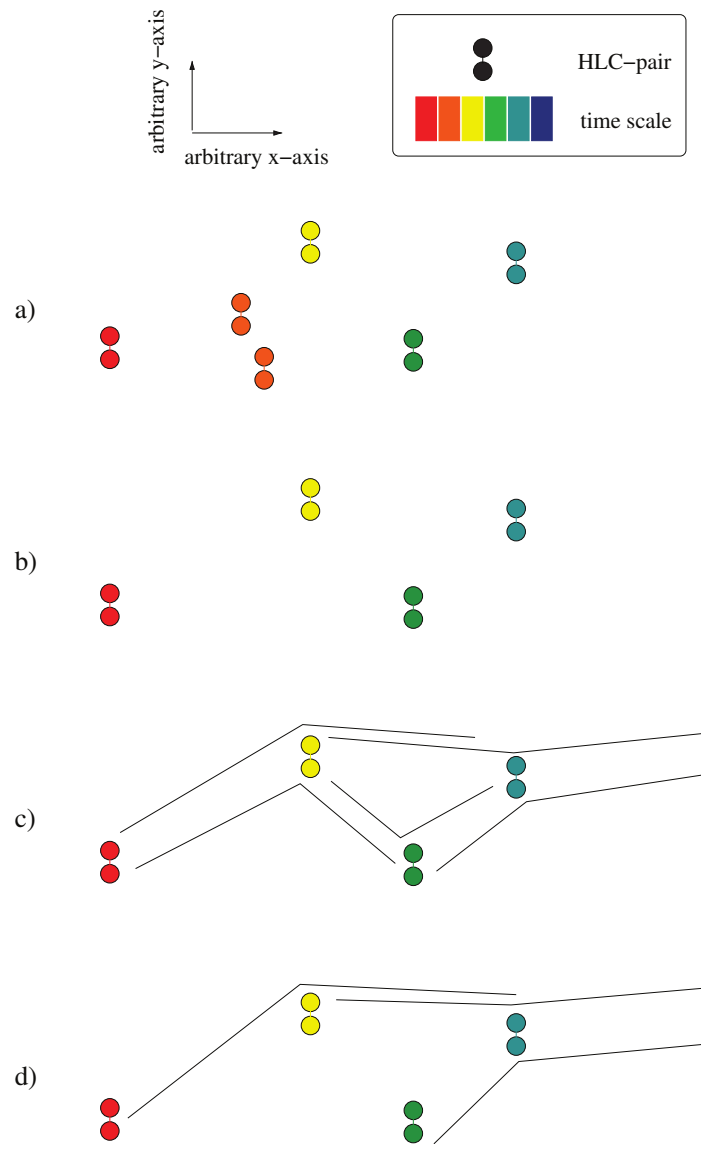

**Figure 1**: Illustration of a trigger process. The times and positions are arbitrary, the x- and y-axis correspond to spatial axes and the colorbar corresponds to a time scale [8].

# 3 Searches for bright monopole signal

The first analysis presented here uses the data from the 59 string configuration of IceCube, which was active in 2009. At this time the detector did not have any dedicated slow particle trigger. The data set has a live time of 311.25 days. The simulated signal is divided into two parts, the first one corresponds to bright monopoles with $\sigma_{cat} = 1.7 \cdot 10^{-22}\,\text{cm}^2$ (i.e. $\lambda_{cat} = 1\,\text{mm}$) and the second corresponds to a dimmer signal with $\sigma_{cat} = 1.7 \cdot 10^{-23}\,\text{cm}^2$ (i.e. $\lambda_{cat} = 10\,\text{mm}$). Simulations cover monopoles with velocities $\beta = 10^{-2}$ and $\beta = 10^{-3}$. The 2009 data set contained $2.3 \cdot 10^9$ events, vastly dominated by events due to the background of atmospheric muons. Therefore the first step of the analysis consists of a background reduction. The cut parameters are based on the time the monopole track takes to cross the whole detector and its reconstructed velocity [9]. The former parameter is calculated from the time difference of the last registered hit and the first one by a DOM. Two sets of selection criteria were optimized for the two simulated catalysis cross sections. The applied cuts on those variables reduce the data rate by a factor close to $10^{-4}$, while the signal efficiency is kept above 34% for $\lambda_{cat} = 1\,\text{mm}$ and 14% for $\lambda_{cat} = 10\,\text{mm}$. To reduce the data rate further a MultiVariate Analysis is adopted, which uses Boosted Decision Trees (BDTs). For each ($\beta$,$\lambda_{cat}$) set we train BDTs using 5% of the experimental data set. The input variables used to train the BDTs, are based on the time, position and charge of hits and their correlations. The final selection criteria, cuts on the BDT score, are optimized for every combination of ($\beta$,$\lambda_{cat}$) in order to minimize the Model Rejection Factor (MRF) [10]. The number of expected background events in the full data set is determined by extrapolating the tails of the BDT-score distributions with an exponential.

After unblinding, one event is observed, surviving the cuts for ($\beta = 10^{-3}$,$\lambda_{cat} = 1\text{mm}$). This event consists of two highly energetic vertical muons close in space (firing neighbouring strings) and in time. Figure 2 shows the BDT distribution of the corresponding parameter set after unblinding. The remaining event is compatible with background, moreover the estimated number of expected events from the fit is $4.79^{+0.65}_{-0.57}$. The flux limits obtained from this analysis are shown in Figure 6.

# 4 Search with DeepCore triggered data

The experimental data set recorded between May 2011 and May 2012 with the SLOP trigger applied on DeepCore has a live time of 351.04 days with a total number of about 50 million triggered SLOP events.

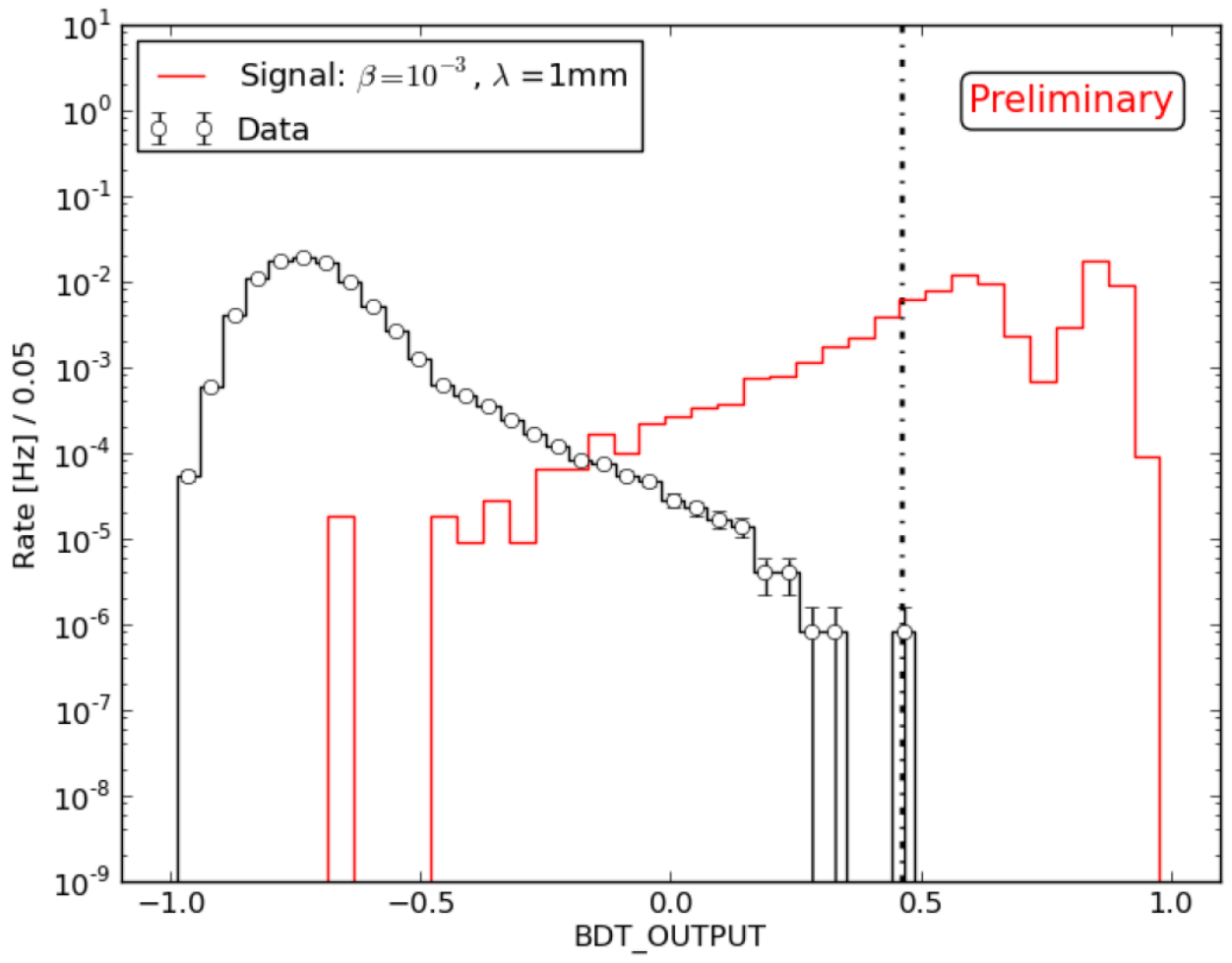

**Figure 2**: Distribution of the BDT scores, after unblinding, for data and signal with $\lambda_{\text{cat}} = 1\,\text{mm}$ and $\beta = 10^{-3}$. The dot-dashed line shows the optimized cut on the BDT score obtained from the Model Rejection Factor method. One event survived the BDT cut, this event is compatible with background.

For the analysis of this data set a robust approach is chosen, which uses a two-dimensional final selection criterion based on the number $n_{\text{triplet}}$ and the reconstructed velocity. While the background distribution decreases rapidly for high $n_{\text{triplet}}$, the signal distribution is almost flat (Fig. 4). The background for this analysis mainly consists of uncorrelated random noise. Since the analysis takes into account only coincident hit pairs selected by the SLOP trigger, which has an integrated muon veto, the influence of atmospheric muons can be neglected. The cuts on $n_{\text{triplet}}$ are optimized and defined before unblinding the full experimental data [10].

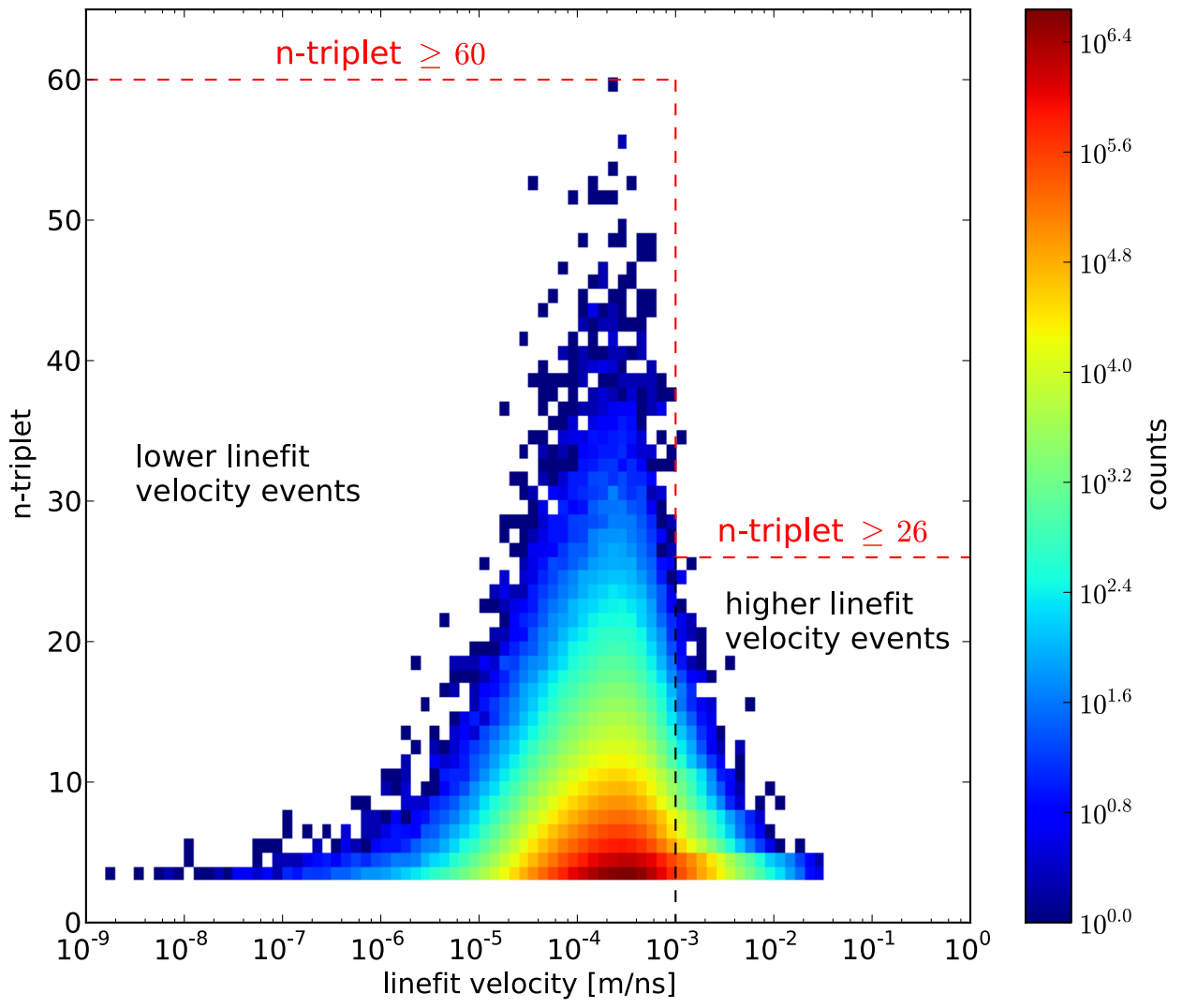

**Figure 3**: $n_{\text{triplet}}$-linefit-velocity distribution of one year experimental data. Additionally, the final cuts on $n_{\text{triplet}}$ (dashed red) and the two velocity regions (dashed black) are shown.

Using the *linefit* [9] to estimate the velocity of a particle, the data can be separated into two velocity regions. Since the reconstructed velocity of most of the background events is slower than $\beta = 10^{-3}$ (Fig. 3) a lower $n_{\text{triplet}}$ cut for $\beta = 10^{-2}$ monopoles can be chosen. Therefore, the sensitivity for faster monopoles has been increased. Figure 3 shows the $n_{\text{triplet}}$-linefit-velocity distribution of one year experimental data including the final cuts. After those cuts only one experimental event with $n_{\text{triplet}} = 34$ and $v_{\text{linefit}} = 1.15 \cdot 10^{-3}\,\text{m/ns}$ remains. Thus, upper limits on the flux of subrelativistic magnetic monopoles can be derived. For the background estimation a model based on a generic ansatz [11] has been fitted to the $n_{\text{triplet}}$ distributions (Fig. 4) of both velocity regions. The deviations at small $n_{\text{triplet}}$ could be explained by not taking into account the correlated noise. Using this fit the total expected number of background events can be estimated by $n_{\text{b}}^{\text{median}} = 3.18^{+1.74}_{-1.09}$, which is consistent with the one observed event.

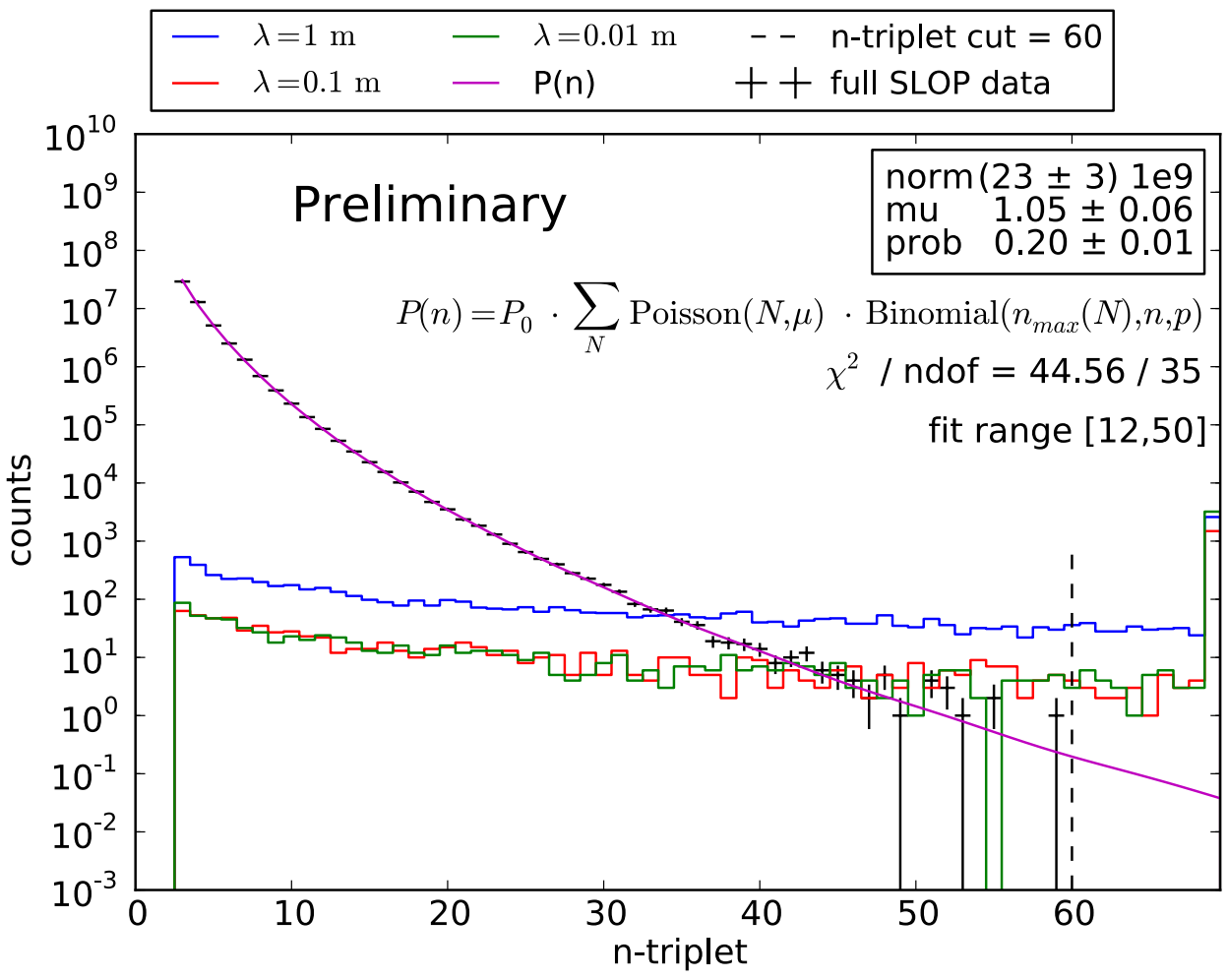

**Figure 4**: $n_{\text{triplet}}$ distribution for events with a slower reconstructed velocity. The black data points show the distribution of one year experimental data. The signal distributions are shown with decreasing $\lambda_{\text{cat}}$ in blue, red and green. Also the fitted background model $P(n\,|\,\mu, p)$ is shown (purple) [11]. Additionally, the final cut on $n_{\text{triplet}}$ is shown (dashed black).

## 5 Search with the full detector

The data set from the 2012 season, taken from May 2012 until May 2013 with a live time of 352 days, consists of about 400 million triggered SLOP events. The analysis of this data is currently ongoing. By using the full 86-string configuration of IceCube instead of the DeepCore subdetector only, the effective area of the trigger increased by one order of magnitude.

With an average event length of 1.3 ms SLOP data are dominated by uncorrelated noise. At the first step a hit cleaning is performed, where only hits are chosen which are accompanied by another hit in the same DOM within a given time. Contrary to relativistic particles, slowly moving particles will create subsequent hits in the same DOM, because they remain in the DOM's vicinity long enough.

In the next step a common object tracking algorithm, the Kalman Filter [12], is applied to the cleaned hits. The algorithm is seeded with a linefit along the triplets from the trigger. In three iterations, track-like hits are selected.

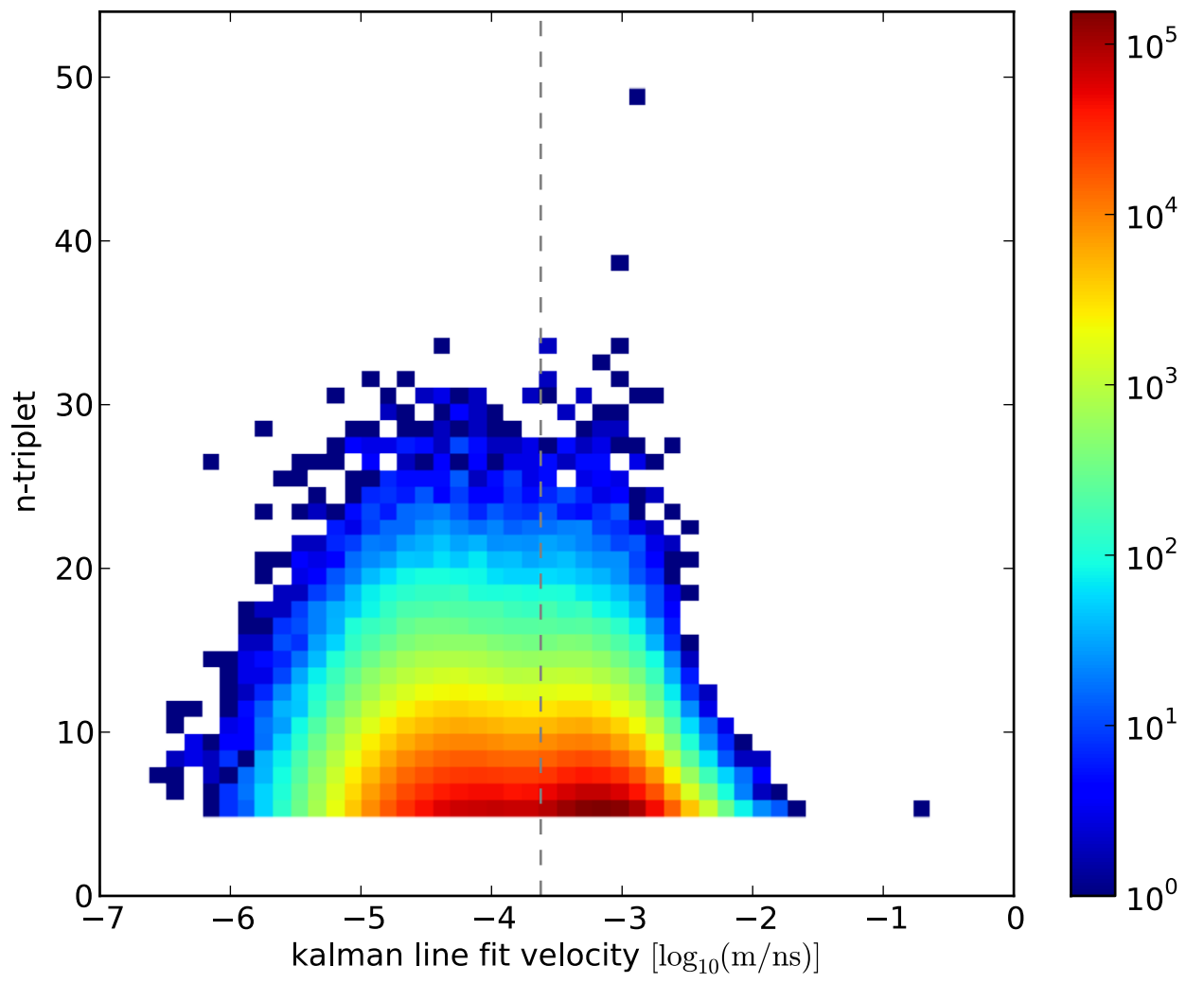

**Figure 5**: $n_{\mathrm{triplet}}$ - Kalman Linefit velocity distribution of 2.8 days experimental data. The data is split at $\beta = 10^{-3.1}$.

A new linefit along the Kalman-selected hits provides an improved direction and velocity reconstruction [13].

To demonstrate the potential of the new trigger configuration and the Kalman Filter, a sensitivity has been derived using a similar approach as in the analysis of the DeepCore data. The reconstructed velocity can be separated into two regions of equal number of events, a slow region with velocities of $\beta < 10^{-3.1}$ and a fast region with $\beta \geq 10^{-3.1}$ (Fig. 5).

The bigger size of the full detector leads to an increase of the number of triplets due to combinatorics. For velocities in the slow region a cut on $n_{\mathrm{triplet}}$ is not efficient anymore. A new analysis technique is currently being developed, using improved cleaning and a multi-variate approach.

The tail of the $n_{\mathrm{triplet}}$ distribution is fitted with an exponential function with $\chi^2/\mathrm{ndof} = 1.57$. The final cut on $n_{\mathrm{triplet}}$ for the fast and the slow samples are optimized using the Model Rejection Factor (MRF).

## 6 Results

The flux limits on subrelativistic magnetic monopoles depend on the monopole velocity $\beta$ and the catalysis cross-section $\sigma_{\mathrm{cat}}$. They are based on the Rubakov-Callan effect and the assumed proton decay channel. Also an isotropic monopole flux is assumed.

The limits are calculated by the Feldman and Cousins approach [14] and take into account all uncertainties in background and signal prediction. The sensitivity is derived using the same method, but does not take systematic uncertainties into account.

Figure 6 shows the flux limits of the IC59 and the DeepCore analysis, together with the sensitivity of the analysis of the 2012 data and the MACRO limits as comparison. Except for $\lambda_{\mathrm{cat}} = 3\,\mathrm{m}$ ($\sigma_{\mathrm{cat}} = 5.5 \cdot 10^{-26}\,\mathrm{cm}^2$) the flux limits are improved by an order of magnitude or more (Fig. 6). Moreover, the monopole flux can be constrained to a level up to three orders of magnitude below the Parker bound [15]. The ongoing analysis is expected to be sensitive to monopole fluxes one order of magnitude below the flux limits presented here.

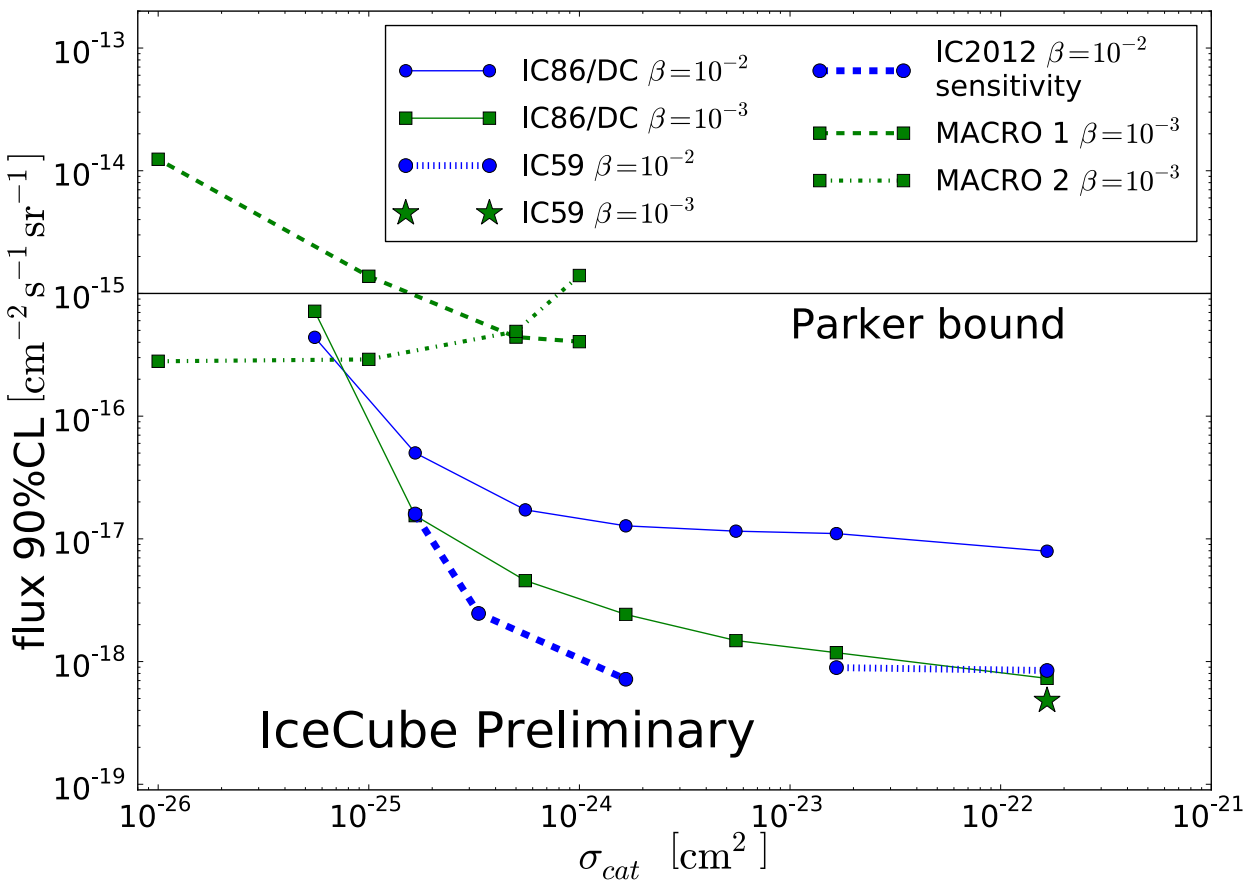

**Figure 6**: Limits and sensitivity of the IceCube slow monopole analyses in comparison with MACRO [4].

# Exotic signatures from physics beyond the Standard Model in IceCube - Signal and background simulations

THE ICECUBE COLLABORATION[1]

[1] *See special section in these proceedings*

*soldin@uni-wuppertal.de*

**Abstract:** Cosmic rays reach the atmosphere with energies up to $10^{11}$ GeV and high energy neutrinos are expected to reach similar energy regions. The incident particle may interact with a nucleon within the atmosphere or inside the Earth to produce exotic particles, thereby probing physics beyond the Standard Model (SM). Attractive scenarios are, for example, supersymmetric extensions of the SM (SUSY) or Kaluza-Klein models. Under favorable conditions meta-stable, charged SUSY or Kaluza-Klein particles give rise to well-separated, parallel, minimum ionizing tracks in IceCube. Since background events from charm and bottom decays and Drell-Yan processes inside the air shower as well as coincident neutrino events could lead to similar signatures, these events are expected to form a relevant background. As the production of exotic particles is highly suppressed, the understanding of this di-muon background seems to be crucial for an analysis in IceCube.

**Corresponding authors:** D. Soldin[1], L. Gerhardt[2,3], K. Helbing[1], S. Klein [2,3], S. Kopper[1], D. van der Drift[2,3]

[1] *Dept. of Physics, University of Wuppertal, D-42119 Wuppertal, Germany*
[2] *Dept. of Physics, University of California, Berkeley, CA 94720, USA*
[3] *Lawrence Berkeley National Laboratory, Berkeley, CA 94720, USA*

## 1 Introduction

The cosmic ray spectrum is known up to energies of $10^{11}$ GeV [1] and recently IceCube reported the first observation of high-energy neutrinos in the PeV range [2]. These high-energy particles may interact with a nucleon within the atmosphere or inside the Earth to produce exotic particles, thereby probing physics beyond the Standard Model. Although large neutrino-telescopes like IceCube [3] are focused on the detection of single neutrino events, they are able to look for more exotic event signatures. One signature of great interest consists of two parallel tracks going upward through the detector, that could be produced by new physics.

Attractive scenarios are, for example, supersymmetric extensions of the SM where R-parity is conserved. Under favorable conditions parallel double-tracks can be produced when a neutrino interacts inside the Earth [4, 5] or cosmic rays interact in the atmosphere [6, 7], generating a pair of supersymmetric particles. These SUSY particles may then decay immediately into a pair of charged next-to-lightest SUSY particles (NLSPs), typically long-lived staus[1]. If these particles have a lifetime $\sim \mu$s, they can live long enough to travel large distances ($\sim$ 1000 km) and enter the IceCube detector with track separations of several hundred meters. Other theories beyond the SM, e.g. Kaluza-Klein models (KK), could lead to similar predictions [8].

We expect a very low event rate of these particles, thus the understanding of background events is crucial. Previous studies have considered SM backgrounds from processes producing di-muons directly that can cause parallel double-tracks in the detector [4, 9, 10]. The decay of charmed/bottom hadrons, for example, as well as muons from Drell-Yan processes inside an air shower or coincident neutrino interactions could lead to similar signatures.

In this paper we show the signature of exotic double-track events in IceCube as well as the expected event rates for different benchmark SUSY *'toy-models'* based on Monte Carlo simulations. We also discuss relevant background events and show their expected signatures based on simulations as well as recent results from our analysis of laterally separated muons in IceCube cosmic ray events [11].

## 2 Exotic signatures in IceCube

The spatial distribution of exotic double-track events depends on the underlying theory. In Supersymmetry, for example, the distributions depend on the given particle spectrum and SUSY breaking mechanism. To simulate exotic event signatures in IceCube we consider different SUSY models and use Monte Carlo simulations for the production, decay, propagation and detector simulation of supersymmetric particles.

We consider minimal supersymmetric models (MSSM) and use the benchmark models from [12] shown in Table 1, assuming different SUSY breaking mechanisms: Minimal supergravity (mSUGRA) and gauge-mediated symmetry breaking (GMSB). These mechanisms can be characterized by the sign of the SUSY Higgs mass parameter $\mu$ and four parameters: The scalar mass parameter $m_0$, the gauging mass parameter $m_{1/2}$, trillinear coupling $A_0$ and the ratio of the Higgs vacuum expectation values $\tan\beta$ in mSUGRA scenarios and the messenger mass $M_{Mess}$, messenger index

1. The stau is the superpartner of the SM tau.

| mSUGRA | $M_{1/2}$ | $m_0$ | $\tan\beta$ | $A_0$ |
|---|---|---|---|---|
| I | 280 GeV | 10 GeV | 11 | 0 GeV |
| $\varepsilon$ | 440 GeV | 20 GeV | 15 | -25 GeV |
| GMSB | $M_{Mess}$ | $\Lambda$ | $\tan\beta$ | $N_{Mess}$ |
| II | 70 TeV | 35 TeV | 15 | 3 |
| SPS 7 | 80 TeV | 40 TeV | 15 | 3 |

**Table 1**: SUSY models used for simulation [12]. $\text{sgn}(\mu) = +$ for all models.

$N_{Mess}$, the universal soft SUSY breaking mass scale $\Lambda$ as well as $\tan\beta$ and $\text{sgn}(\mu)$ in GMSB scenarios respectively [13]. Although these models seem to be disfavoured with respect to the squark mass limits of the LHC data and due to implications of the Higgs mass on supersymmetry breaking scenarios [14], we use these benchmark points to develop SUSY simulations, since inserting different models beyond the SM is straight forward.

In order to estimate the rate of production of SUSY particles due to high-energy neutrino interactions with a nucleon $N$ inside the Earth, we assume an $E^{-2}$ primary neutrino flux normalized to current experimental limits [2]:

$$\frac{d\Phi_\nu}{dE} \approx 10^{-8}\, E^{-2} \frac{\text{GeV}}{\text{cm}^2\ \text{s sr}} \qquad (1)$$

To generate the event and handle fragmentation and decay of the involved particles into the NLSP we use PYTHIA [15] in connection with Madgraph [16].

To simulate SUSY events from cosmic ray interactions (hadronic interactions) the flux of primary nucleons can be approximated by

$$\frac{d\Phi_N}{dE} \approx 1.8\, E^{-\gamma} \frac{\text{nucleons}}{\text{cm}^2\ \text{s sr GeV}}, \qquad (2)$$

where E is given in GeV. At $E \approx 10^6$ GeV the spectral index changes from $\gamma = 2.7$ to $\gamma = 3$ [1]. We assume that the interaction typically takes place at a height $H \approx 25$ km. The production of SUSY particles is simulated using PYTHIA and the further shower development of the SM particles is simulated with CORSIKA [17]. We assume energy losses in the atmosphere to be negligible for SUSY particles.

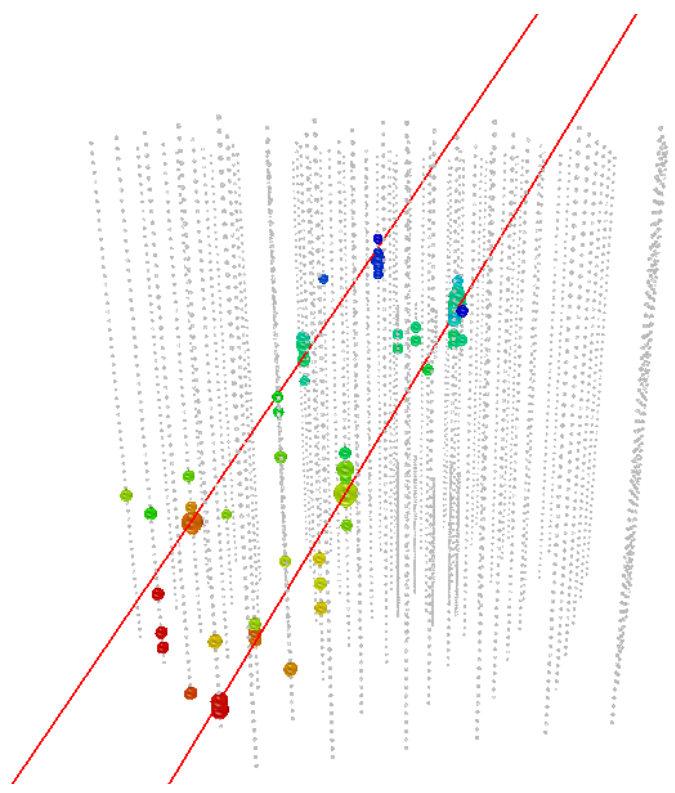

**Figure 1**: Simulated SUSY (SPS 7) double-track event and reconstructed tracks in the IceCube. The colored dots indicate hits in the detector, where the hits are time-ordered from red to blue.

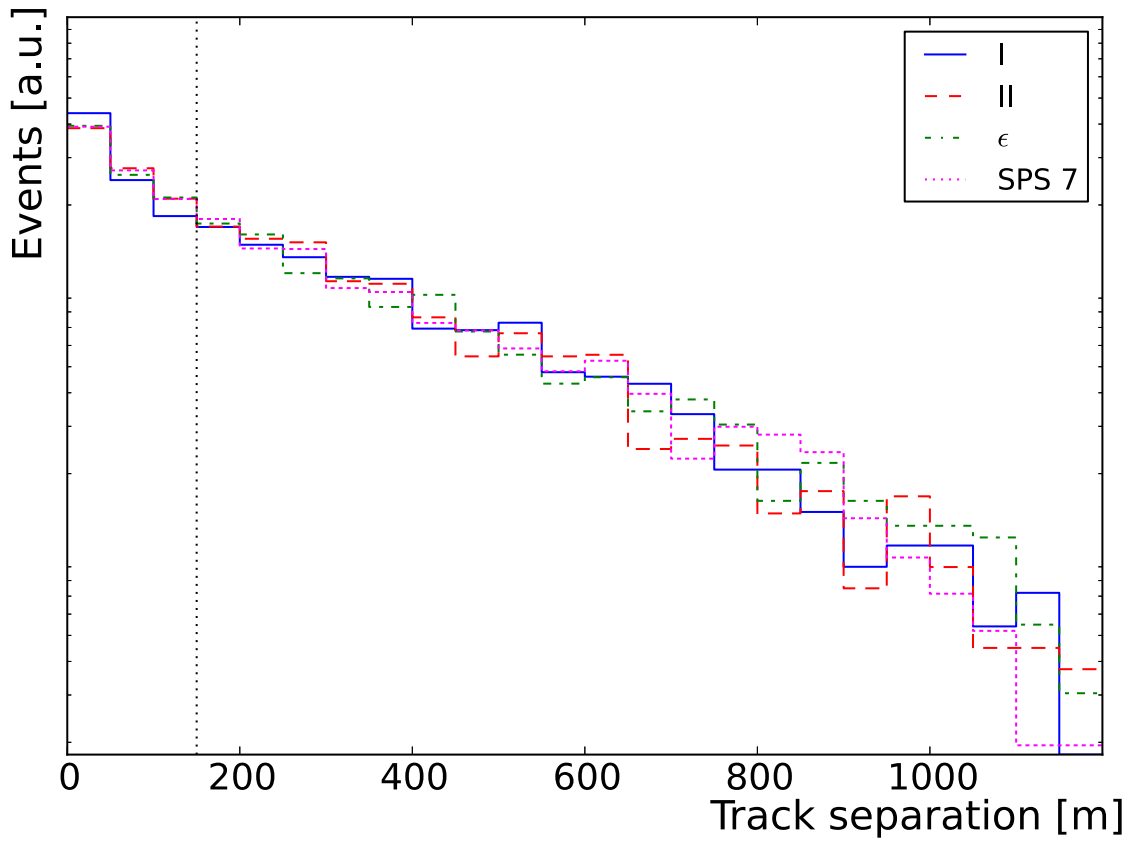

**Figure 2**: Distribution of track separation at the detector obtained from simulations of double-tracks for all MSSM models shown in Table 1.

The propagation of SUSY particles in the antarctic ice is done using the Muon Monte Carlo simulation package [18], where we treat staus like SM taus, but replacing the masses accordingly and switching off the decay. The light propagation, detector response, and trigger are simulated with IceCube software in both cases.

Figure 1 shows the distinctive double-track signature of a simulated SUSY event (SPS 7) from a $\nu N$ interaction in the IceCube. As expected, we observe two well-separated, parallel, minimum ionizing tracks. The track separation distributions from $\nu N$ interactions for different SUSY models are shown in Figure 2. We expect track separations of roughly 150 m $< d <$ 1000 m to be resolvable in IceCube[2], thus making a direct detection of supersymmetric double-tracks feasible.

## 3 SUSY event rates

Using cross sections obtained from PYTHIA and the incoming fluxes mentioned in section 2 we estimate the SUSY event rates at the detector. We assume that every event triggers, which corresponds to an effective area of $\approx 1$ km$^2$. We also assume that every SUSY particle decays into the NLSP, which does not decay on the way to the detector. Thereby we over-estimate the SUSY rates in IceCube. For the hadronic case we assume a uniformly distributed stau flux up to zenith angles of 115° [7]. The expected event rates under these assumptions are shown in Table 2.

Since in several cases these rates are below the order of 1 event/year, consistent with [12], finding a signal will be - at best - very challenging. Current LHC data shows that the simplest SUSY models are disfavored. As one moves away from minimal SUSY models (e.g. MSSM), the phase space expands and there may be many more models that produce exotic double-track signatures in IceCube. Thus an excess of double-tracks would however be a clear signature of new physics. To detect such an excess independently of the assumed exotic physics model, the understanding of the relevant SM background is essential.

2. Due to the cubic detector volume of $\sim 1$ km$^3$ and the horizontal string spacing of 125 m.

| Model | $\nu$N interactions | hadronic interactions |
|---|---|---|
| I | 1.7 | $1.5 \cdot 10^{-3}$ |
| $\varepsilon$ | 0.5 | $1.1 \cdot 10^{-4}$ |
| II | 0.82 | $6.3 \cdot 10^{-4}$ |
| SPS 7 | 0.5 | $2.9 \cdot 10^{-4}$ |

**Table 2**: Estimated SUSY event rates in IceCube given in events/year.

## 4 Backgrounds

Previous studies have considered charm production in SM processes as a major background in exotic double-track searches [5]. In [9] it was shown that other SM processes could produce double-track signatures as well. We consider the following SM background processes.

### 4.1 Charm/Bottom hadrons

If a neutrino interacts with a nucleon inside the Earth, a charmed hadron may be produced, which can produce di-muons via

$$\nu N \to \mu^- H_c \to \mu^- \mu^+ H_x \nu, \quad (3)$$

where $H_c$ is the charmed hadron that decays to another hadron $H_x$ [5]. These background events are simulated using the same method as described in section 2, with processes that produce charmed hadrons, instead of SUSY particles. In our simulations, we do not observe double-tracks from neutrino interactions with separations above 100 m; this is different from [4] and [8]. Since we are able to resolve track separations above $\sim 150$ m, these events would be reconstructed as single tracks and therefore do not form a relevant background in double-track analyses.

The decay of heavy hadrons, mostly from charmed decays, in air showers can also produce laterally separated, high-energy muons (LS muons) with high transversal momentum $p_T$. These muons, together with the associated muon bundle traveling along the core direction, may produce a double-track signature in IceCube. The muon $p_T$ is related to the muon separation from the shower core by

$$p_T \simeq \frac{d \cdot H}{E_\mu \cos(\theta)}, \quad (4)$$

where $d$ is the perpendicular separation of the muon from the shower core, $E_\mu$ its energy at generation, $H$ the interaction height, and $\theta$ the zenith angle of the shower direction. The transition from soft interactions with $p_T < 2$ GeV, that are not describable in pQCD (perturbative Quantum Chromodynamics), to hard interactions is visible as a spectral change in the $p_T$ spectrum. This spectral change should be visible in lateral separation distributions [11].

In [11] we obtained the lateral distribution of muons produced in cosmic ray events from data taken with IceCube in its 59-string configuration. Figure 3 shows the separation distribution of LS muons where, after applying all selection criteria, 34,754 events remain in the data. The expectation from simulated data, as well as coincident double showers estimated from off-time data are also shown.

The simulated di-muons seem to be in good agreement with the data and a linear fit to the ratios of simulation to data did not show any statistically significant difference in the slopes. But we have also shown that the zenith angle distributions are in significant disagreement between simulation and data. We observe that QGSJET [21] and Sibyll [19] overpredict the event rate at high zenith angles and underpredict the rate for more vertical events, while DPMJET [20] shows a better match to data, but underpredicts the rate for more vertical showers. Note that none of these interaction models include bottom quark production and DPMJET is the only model that includes a hard component of charmed particles. It was previously shown that the hard components of charm and bottom quarks, as well as hard muons from Drell-Yan processes become relevant in double-track searches in neutrino telescopes, especially when including events from horizontal directions [9]. Therefore future simulations with more sophisticated $p_T$ modeling, such as a modified *'heavy-quark'* CORSIKA version proposed in [22], may reduce the disagreement in simulations.

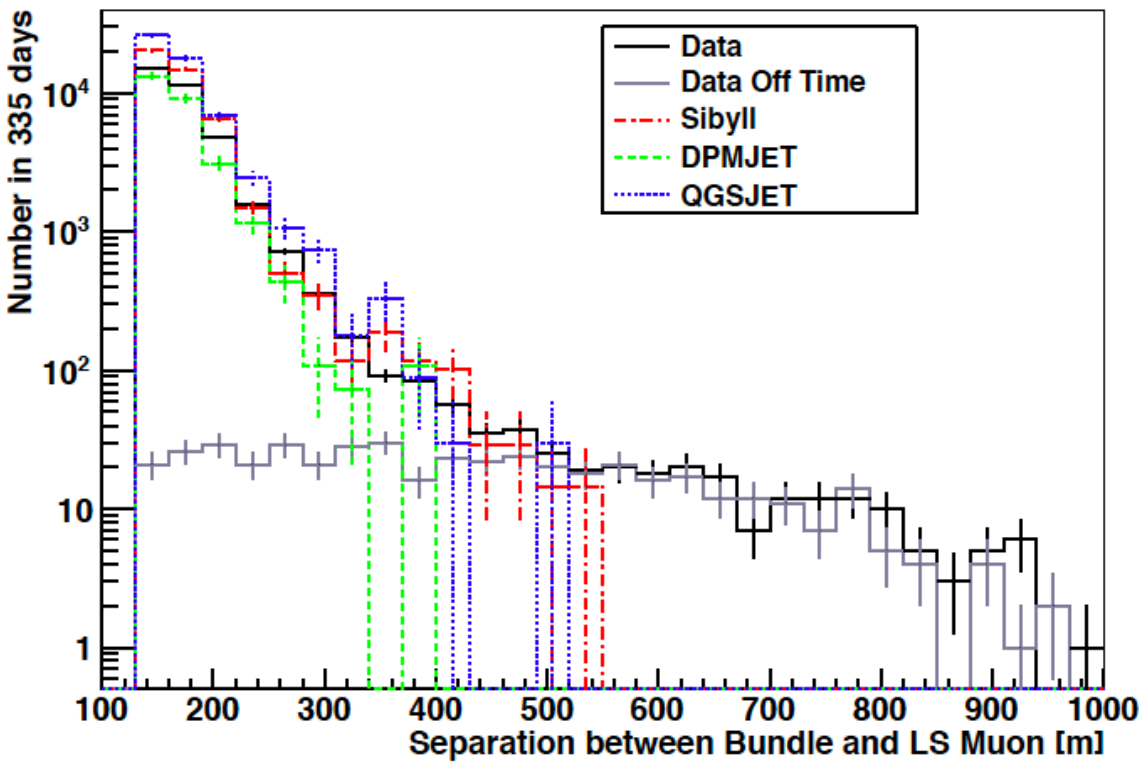

**Figure 3**: Lateral separation of muons after applying all selection criteria for data and simulation, as well as double showers estimated from off-time data [11]. Three different interaction models, Sibyll, DPMJET, and QGSJET, were used in the background simulations.

### 4.2 Drell-Yan processes in air showers

In [10] it was shown that Drell-Yan processes in air showers could contribute to the background in double-track searches. As far as we know there is no interaction model available including Drell-Yan processes in air showers. We simulate these processes using the method shown in section 2. Unfortunately, PYTHIA does not produce Drell-Yan pairs below 2 GeV invariant mass, since this region can not be described in pQCD. The obtained lateral separation of muons in air showers including Drell-Yan processes is shown in Figure 4, where we assumed a cosmic ray spectrum from Eq. 2 with primary energies in the range of $10^5$ GeV $< E_{Prim} < 10^9$ GeV up to zenith angles of 89.99°. Although rate calculations are not representative of the actual physics, the events with track separations up to $\sim 400$ m and above could form a relevant background. Moreover we think that these processes are an interesting topic by itself and may be considered in future analyses.

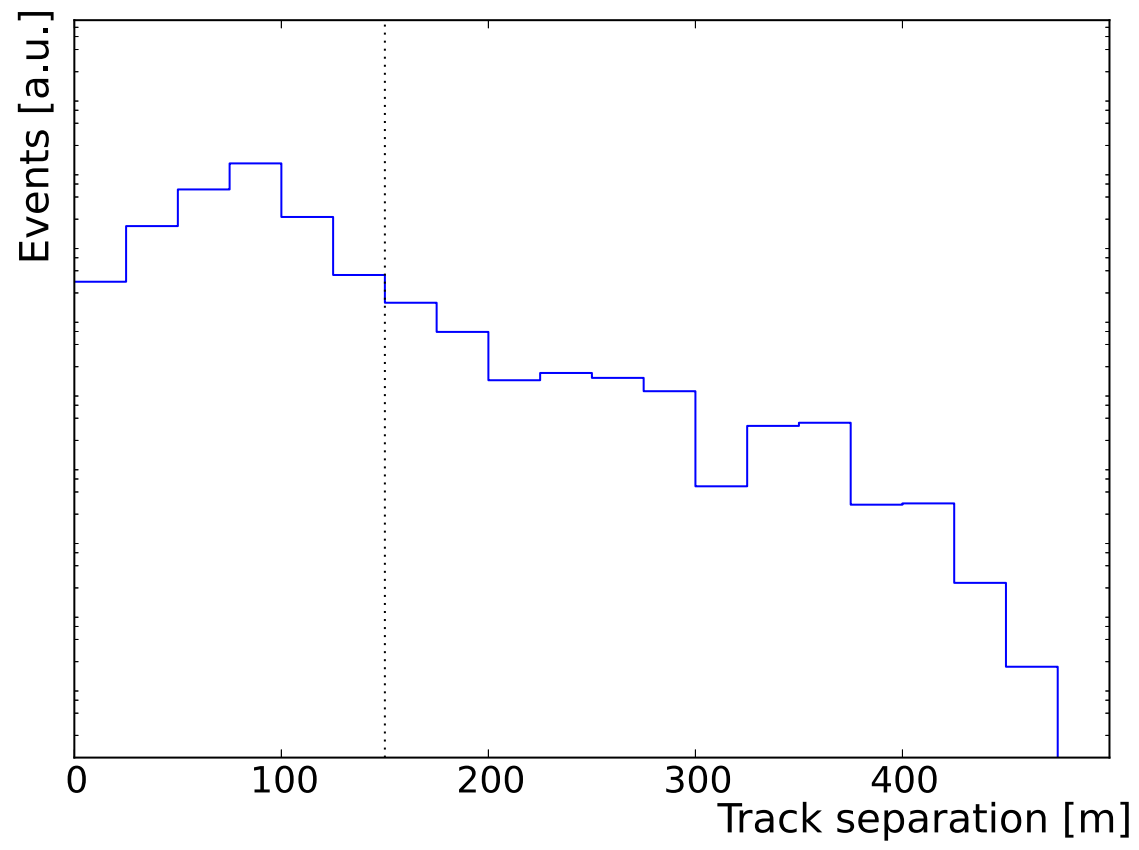

**Figure 4**: Lateral separation of muons from air showers including Drell-Yan processes obtained from simulations.

### 4.3 Double-neutrino events

Another possible source of background events arises from two muons from a pair of neutrino interactions, produced in the same cosmic ray air shower. These muons will be nearly parallel and thus mimic a double-track signature in IceCube. In [23] we estimate the event rates using CORSIKA simulations where DPMJET and QGSJET interaction models and the cosmic ray spectrum in Eq. 2 are used. All the neutrinos in each event are paired with all of the other neutrinos in each event and the separation distance is computed. The neutrino detection probability $P(E_\nu)$ depends on the neutrino interaction probability and the probability of observing the produced muon, and is given by [24]

$$P(E_\nu) = \begin{cases} 1.3 \cdot 10^{-6} E_\nu^{2.2} & \text{if } E_\nu \leq 1 \text{ TeV} \\ 1.3 \cdot 10^{-6} E_\nu^{0.8} & \text{if } E_\nu > 1 \text{ TeV} \\ 0 & \text{if } d > d_{max} \end{cases} \quad (5)$$

where $d$ is the track separation and $d_{max} \approx 1$ km the maximal resolvable track distance in IceCube and $E_\nu$ in units of TeV. Table 3 shows the overall atmospheric neutrino rates from cosmic ray interactions in events/year under two extreme assumptions of the cosmic ray composition, for all-proton and all-iron. The expected event rate to observe two upward-going neutrinos from the same air shower is about one in 14 years, thus being a highly suppressed background and negligible in IceCube SUSY analyses.

## 5 Conclusions

We studied signatures from supersymmetric models in IceCube proposed in [5, 6] using the simulation methods shown in section 2. The simulated MSSM track separations are in agreement with those obtained in [4]. The expected event rates are of the order of less than 1 event/year for currently allowed SUSY. By moving away from minimal SUSY models the phase space expands and there may be many more models that produce exotic signatures in IceCube. Other theories beyond the SM (e.g. Kaluza-Klein models) can lead to double-track signatures in IceCube that may produce more events per year [8]. Hence, this could be a signal of new physics. Therefore we studied any possible SM background events. Decays from heavy quarks produced in neutrino interactions or air showers, as well as muons produced in Drell-Yan processes inside the shower can produce double-tracks in IceCube with similar track separation distributions. Comparison between experimental data and simulations of laterally separated muons from cosmic rays shows a disagreement in the zenith angle distributions. When improved simulations become available future analyses could improve the understanding of these events and estimate the muon ratios, as well as measure the transverse momentum spectrum in air showers. Furthermore the understanding of background events may make it possible to find an excess of double-tracks which would be a signature of physics beyond the Standard Model.

| | QGSJET | DPMJET |
|---|---|---|
| Protons | 0.068 | 0.070 |
| Iron | 0.065 | 0.056 |

**Table 3**: Calculated double atmospheric neutrino event rates in IceCube given in events/year.